%% file: acmsmall.tex
\documentclass[format=acmsmall, review=false, screen=true]{acmart}

\usepackage{booktabs} 
\usepackage{array,multirow,graphicx}
\usepackage[normalem]{ulem}
\useunder{\uline}{\ul}{}
\usepackage{longtable}
\usepackage{afterpage}
\usepackage{float}
\usepackage{subcaption}
\usepackage{ragged2e}
\usepackage{wrapfig}

\usepackage[ruled]{algorithm2e} 

\SetAlFnt{\small}
\SetAlCapFnt{\small}
\SetAlCapNameFnt{\small}
\SetAlCapHSkip{0pt}
\IncMargin{-\parindent}

\acmJournal{TWEB}
\acmVolume{1}
\acmNumber{1}
\acmArticle{1}
\acmYear{2020}
\acmMonth{5}
\copyrightyear{2020}

\setcopyright{acmlicensed}

\acmDOI{0000001.0000001}

\received{February 2007}
\received[revised]{March 2009}
\received[accepted]{June 2009}

\begin{document}
\title[``I Cannot Do All of This Alone'': Exploring Instrumental and Prayer Support in OHCs]{``I Cannot Do All of This Alone'': Exploring Instrumental and Prayer Support in Online Health Communities}



\author{C. Estelle Smith}
\orcid{0000-0002-4981-7105}
\affiliation{%
  \institution{GroupLens Research, Dept. Computer Science, University of Minnesota (UMN)}
  \city{Minneapolis}
  \state{MN}
  \postcode{55455}
  \country{USA}}
\email{smit3694@umn.edu}

\author{Zachary Levonian}
\affiliation{%
  \institution{GroupLens Research, Dept. Computer Science, UMN}
  \city{Minneapolis}
  \state{MN}
  \postcode{55455}
  \country{USA}}
\email{levon003@umn.edu}

\author{Haiwei Ma}
\orcid{0000-0002-6621-9131}
\affiliation{%
  \institution{GroupLens Research, Dept. Computer Science, UMN}
  \city{Minneapolis}
  \state{MN}
  \postcode{55455}
  \country{USA}}
\email{maxxx979@umn.edu}

\author{Robert Giaquinto}
\affiliation{%
  \institution{Dept. Computer Science, UMN}
  \city{Minneapolis}
  \state{MN}
  \postcode{55455}
  \country{USA}}
\email{smit7982@umn.edu}

\author{Gemma Lein-Mcdonough}
\affiliation{%
  \institution{Simmons University}
  \city{Boston}
  \state{MA}
  \postcode{02115}
  \country{USA}}
\email{gemmaleinmcdonough@gmail.com}

\author{Zixuan Li}
\orcid{0000-0002-6820-7687}
\affiliation{%
  \institution{GroupLens Research, Dept. Computer Science, UMN}
  \city{Minneapolis}
  \state{MN}
  \postcode{55455}
  \country{USA}}
\email{lixx4110@umn.edu}

\author{Susan O'Conner-Von}
\affiliation{%
  \institution{Earl E. Bakken Center for Spirituality \& Healing, UMN}
  \city{Minneapolis}
  \state{MN}
  \postcode{55455}
  \country{USA}}
\email{ocon0025@umn.edu}

\author{Svetlana Yarosh}
\affiliation{%
  \institution{GroupLens Research, Dept. Computer Science, UMN}
  \city{Minneapolis}
  \state{MN}
  \postcode{55455}
  \country{USA}}
\email{lana@umn.edu}

\renewcommand\shortauthors{Smith, C.E. et al}

\begin{abstract}
Online Health Communities (OHCs) are known to provide substantial emotional and informational support to patients and family caregivers facing life-threatening diagnoses like cancer and other illnesses, injuries, or chronic conditions. Yet little work explores how OHCs facilitate other vital forms of social support, especially instrumental support. We partner with CaringBridge.org---a prominent OHC for journaling about health crises---to complete a two-phase study focused on instrumental support. Phase one involves a content analysis of 641 CaringBridge updates. Phase two is a survey of 991 CaringBridge users. Results show that patients and family caregivers diverge from their support networks in their preferences for specific instrumental support types. Furthermore, ``prayer support'' emerged as the most prominent support category across both phases. We discuss design implications to accommodate divergent preferences and to expand the instrumental support network. We also discuss the need for future work to empower family caregivers and to support spirituality.
\end{abstract}

\newcommand{\irrtable}{
\begin{table}[H]
\begin{tabular}{rccc}
\hline
\multicolumn{1}{c}{\textbf{Help Type}} & \textbf{Acknowledged} & \textbf{Direct Request} \\ \hline \hline
Prayer (PR) & 0.84 & 0.89 \\
Food (FO) & 0.75 & 0.29 \\
Medical (MED) & 0.71 & 0.07  \\
Remote Emotional (EMO\textsubscript{R}) & 0.7 & 0.7 \\
Other/Non-Specified (OH) & 0.7 & 0.34  \\ 
Transportation (TR) & 0.62 & 0.25 &  \\
Chores (CH) & 0.6 & 0  \\
Co-located Emotional (EMO\textsubscript{CO}) & 0.57 & 0.65 \\
Shelter (SH) & 0.49 & 0.25 \\
CaringBridge (CB) & 0.48 & 0.41 \\
Exercise (EX) & 0.42 & 0 \\
Personal Care (PC) & 0.34 & 0  \\
Personal Donation (PD) & 0.33 & 0.42 \\
Practical Item (PI) & 0.27 & 0.18  \\
General Donation (GD) & 0.13 & NA \\
Information (INFO) & 0.08 & 0 \\
Family Care (FC) & 0.05 & 0.5  \\
\hline
\end{tabular}
\caption{Interrater Reliability scores (Krippendorff's alpha) averaged across all rounds of iterative codebook development with 5 coders. ``NA'' indicates that no cases of this category were identified, while ``0'' indicates a lack of agreement due to very few cases of this category appearing in the number of cases initially classified.}
\label{tab:IRR}
\end{table}
}


\newcommand{\prelimcodes}{
\begin{figure}[H]
    \centering
    \includegraphics[width=\textwidth]{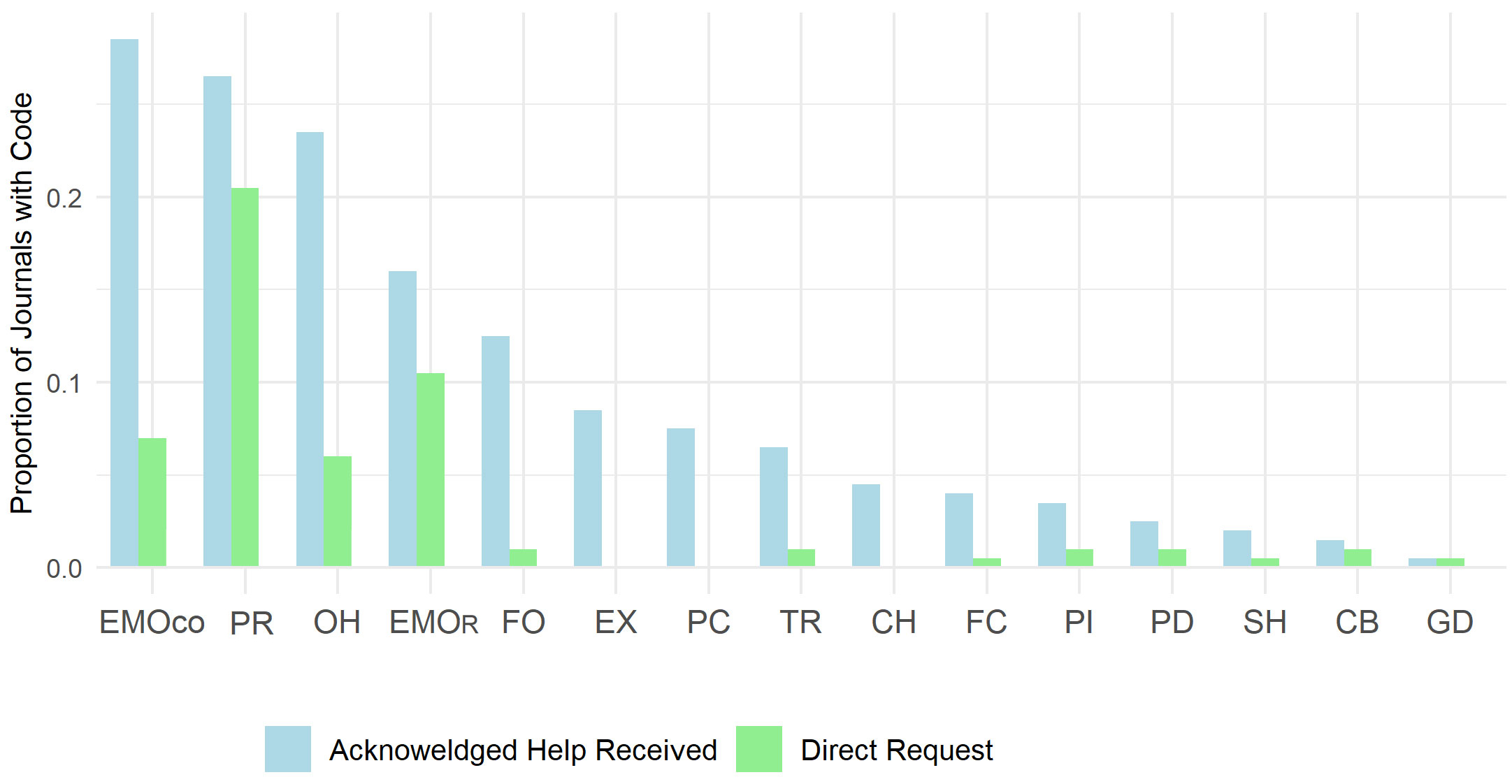}
    \caption{Distribution of help type codes across 200 journal updates during preliminary codebook development. Since IRR scores were low in categories that did not appear frequently, this figure should not be interpreted as formal data. Rather, we include it to validate our methodology of choosing to code expressions of appreciation for acknowledged help received rather than direct requests.}
    \label{fig:prelimcodes}
\end{figure}
}

\newcommand{\cbfig}{
\begin{figure}
  \frame{\includegraphics[width=\textwidth]{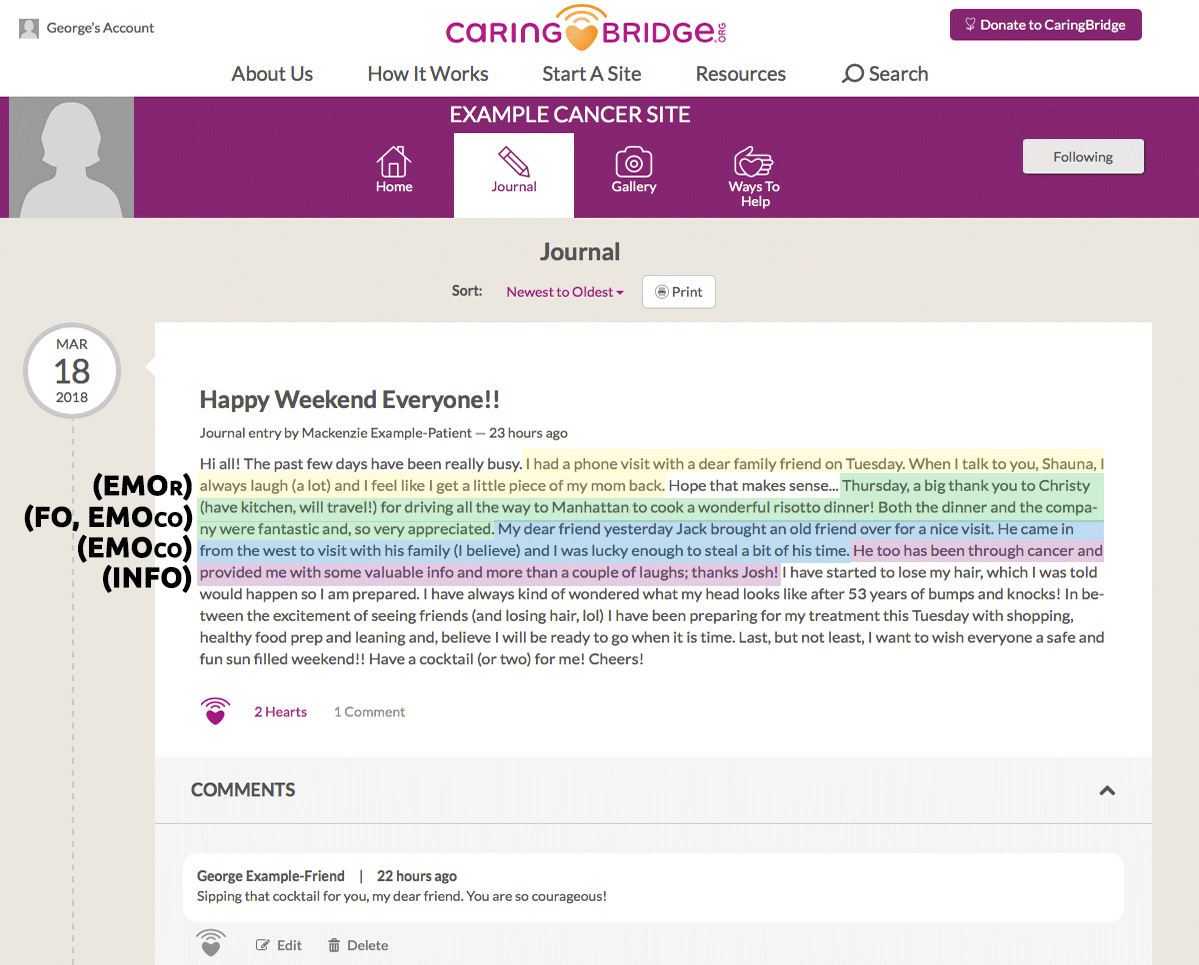}}
  \caption{Screenshot of CaringBridge Journal Update from 2018. Text in this example has been anonymized with fictional names/dates/locations from real user data and annotated according to the appreciation coding protocol described in section~\ref{coding}. (Table~\ref{tab:codebook} defines code abbreviations.)}~\label{fig:CB}
\end{figure}
}

\newcommand{\pcguseful}{
\begin{figure}[H]
  \includegraphics[width=\textwidth]{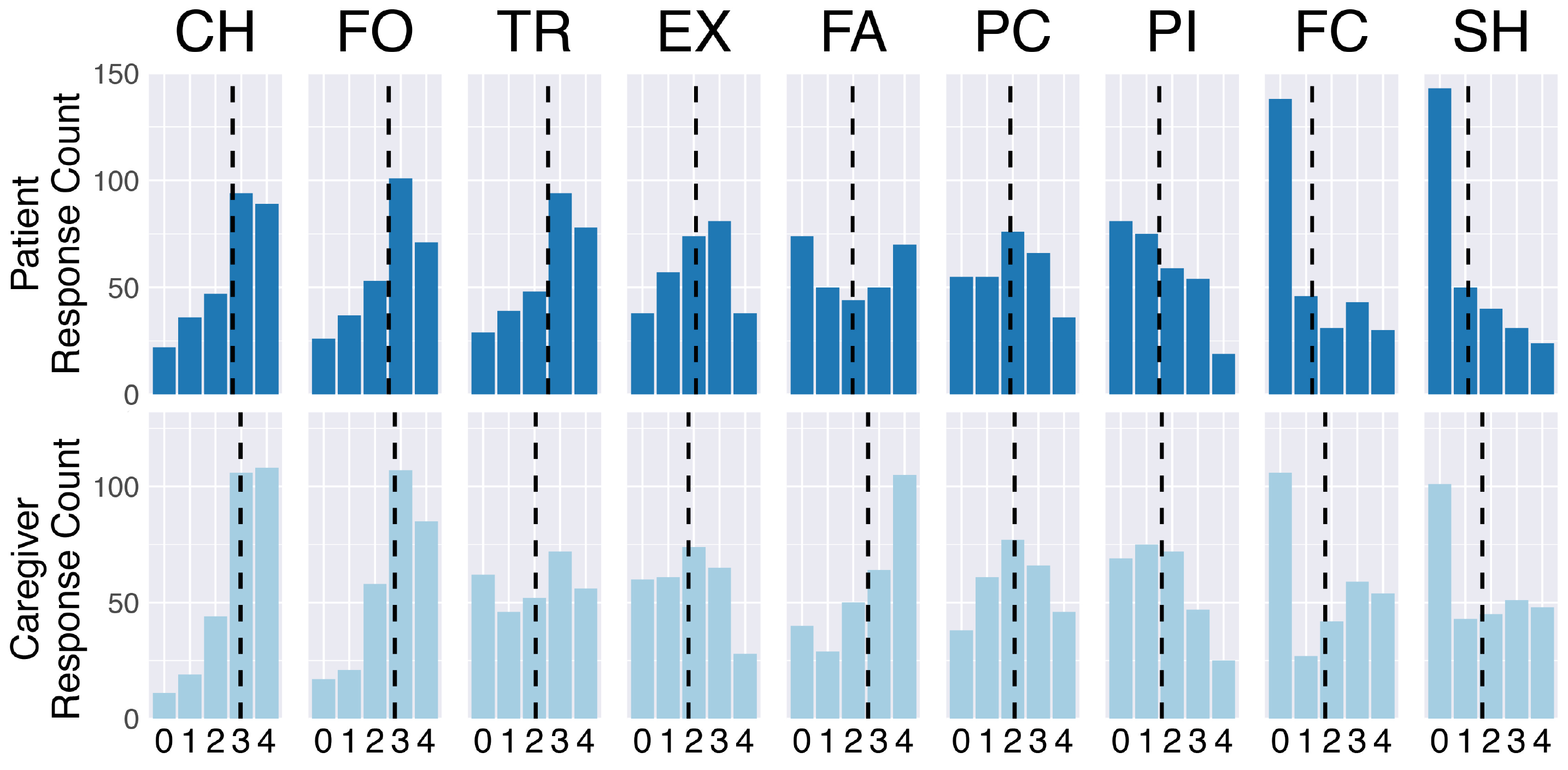}
  \caption{Distributions of Usefulness Ratings of Instrumental Support Types by Patients and Caregivers. 5-point unipolar Likert responses on a scale of 0: "Not at all" to 4: "Extremely" useful, question ~\ref{useful_help}. Vertical dashed lines indicate mean value of the ratings; see Figure~\ref{fig:usefulplot} above for numerical value of the means.}~\label{fig:useful}
\end{figure}
}

\newcommand{\usefulplot}{
\begin{figure}[H]
  \includegraphics[width=\textwidth]{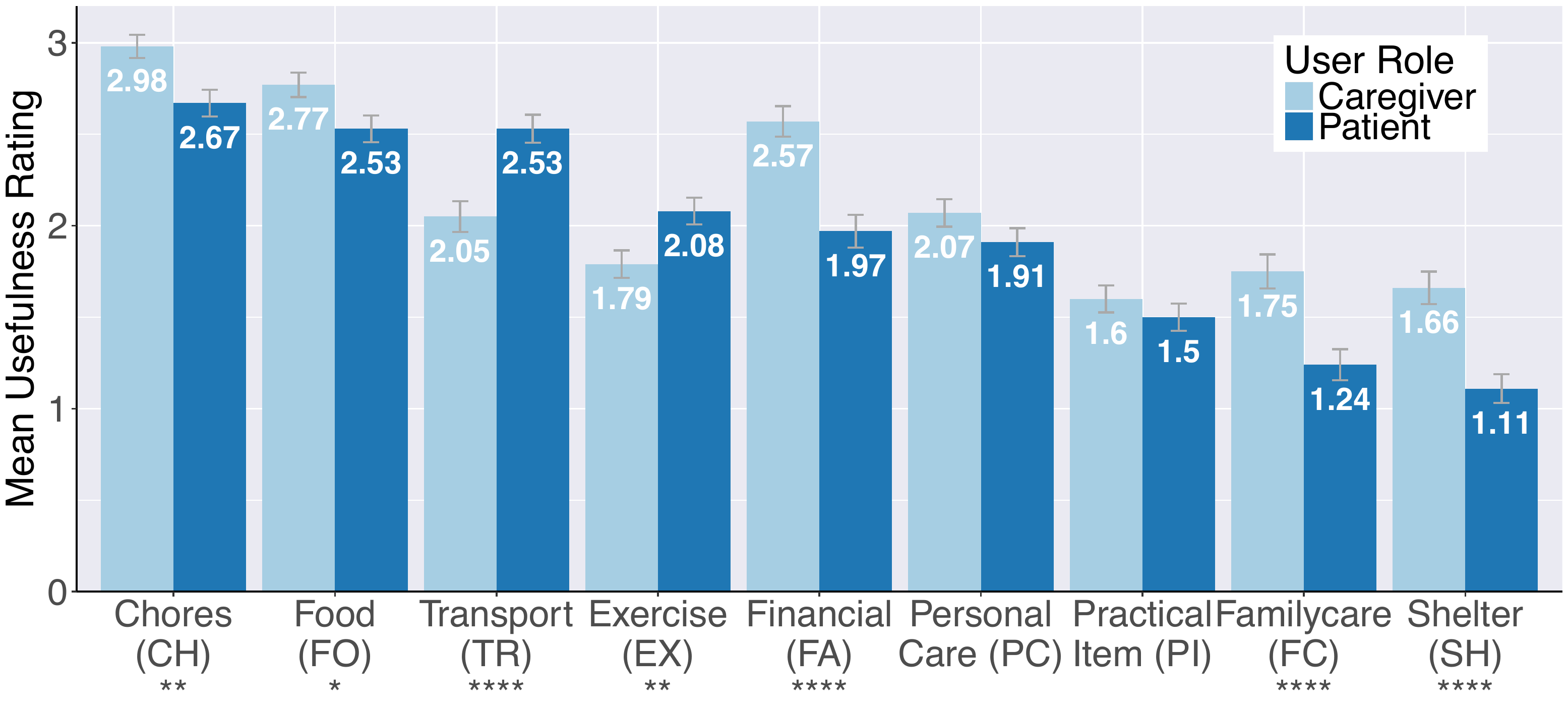}
  \caption{Mean Usefulness Ratings of Instrumental Support Types. 5-point unipolar mean Likert responses on a scale of 0: "Not at all" to 4: "Extremely" useful, question ~\ref{useful_help}. Error bars show standard errors. Asterisks below X-axis labels indicate significance levels of the difference of means between patient and caregiver ratings. (Note: Table~\ref{tab:useful} in the appendix represents this information in table format for accessibility and reference.)
  }~\label{fig:usefulplot}
\end{figure}
}

\newcommand{\interestplot}{
\begin{figure}[H]
  \includegraphics[width=\textwidth]{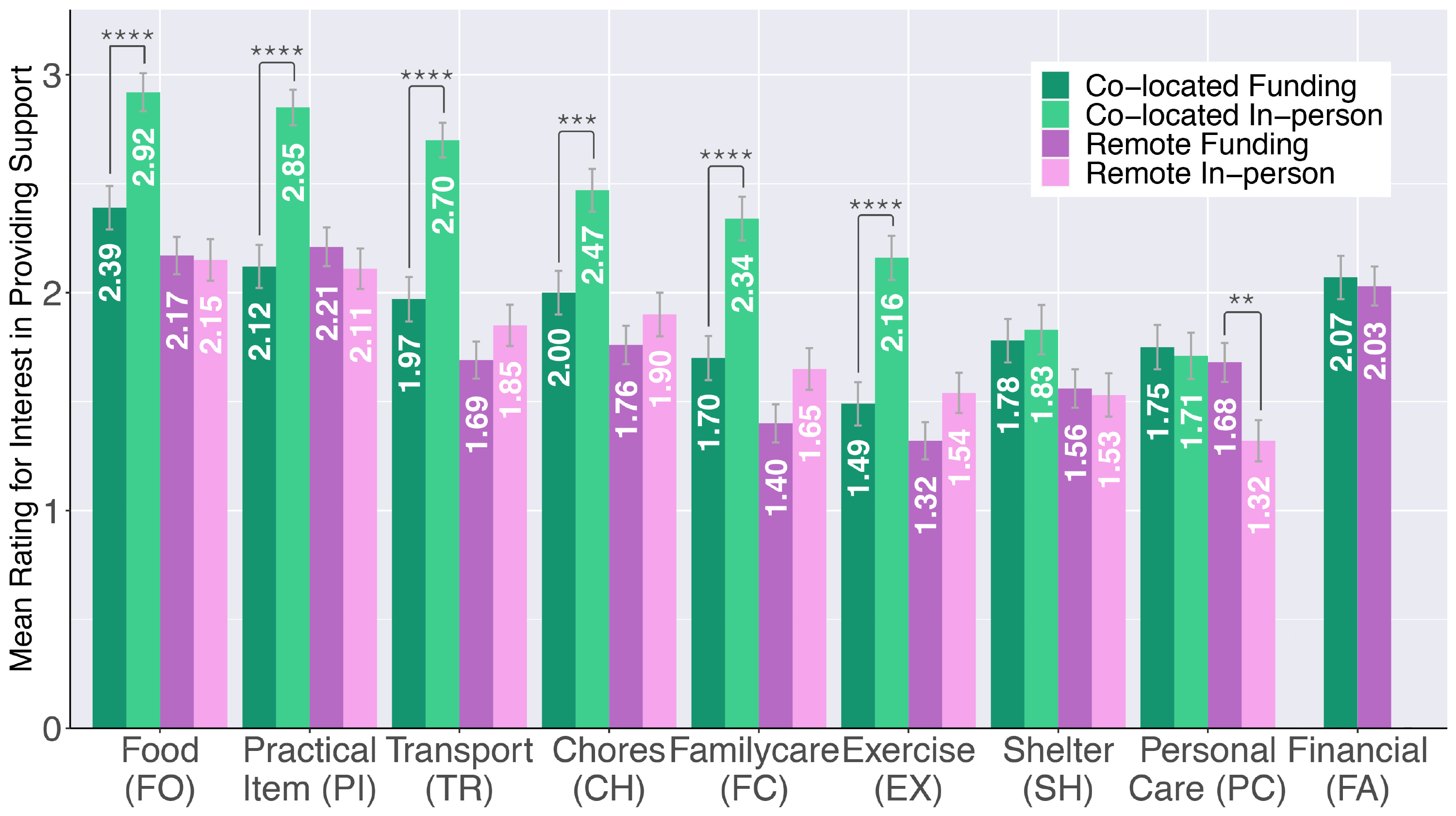}
  \caption{FFA Interest in Providing Instrumental Support Types. 5-point unipolar mean Likert responses on a scale of 0: "Not at all" to 4: "Extremely" interested. Error bars show standard errors. Two-sided t-tests were used to test difference of means between between funding services (question~\ref{funding_ffa}) versus providing services in-person (question~\ref{in_person_ffa}) separately for co-located FFA and remote FFA. Intervals are marked only for categories where significant differences were detected; asterisks indicate significance levels. \\ (Note: Table~\ref{tab:ffainterest1} in the appendix represents this information in table format for accessibility and reference.)
  }~\label{fig:interestplot}
\end{figure}
}


\newcommand{\ffadistributions}{
\begin{figure}[H]
  \includegraphics[width=5.1in]{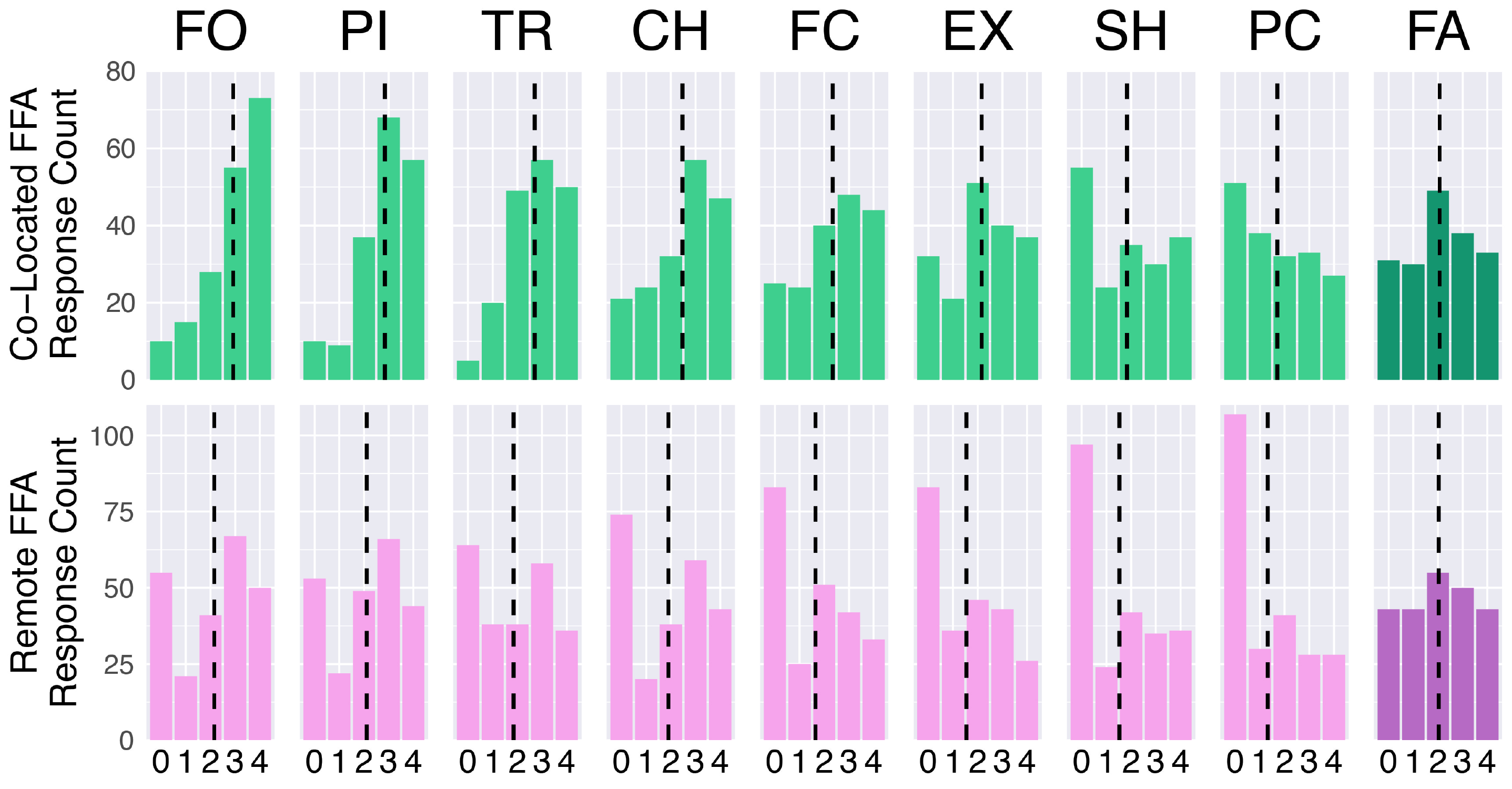}
  \caption{Distributions of Ratings of FFA Interest in Providing Support In-Person. 5-point unipolar Likert responses on a scale of 0: "Not at all" to 4: "Extremely" interested in providing help, question ~\ref{in_person_ffa}. Vertical dashed lines indicate mean value of the ratings; see Figure~\ref{fig:interestplot} above for the numerical value of the means.}~\label{fig:interest}
\end{figure}
}

\newcommand{\codebook}{
\begin{table}
\centering
\begin{tabular}{|l|c|m{2.5cm}|m{5.9cm}|}
\hline
\multicolumn{2}{|c|}{\textbf{Help Types}} & \multicolumn{1}{c|}{\textbf{Description}} & \multicolumn{1}{c|}{\textbf{Example}} \\ \hline
\multicolumn{1}{|c|}{\multirow{4}{*}{\rotatebox[origin=c]{90}{\textbf{NON-INSTRUMENTAL}}}} & \textbf{\begin{tabular}[l]{@{}c@{}}PRAYER\\ SUPPORT (PR)\end{tabular}} & \begin{tabular}[l]{@{}l@{}}  Prayer, blessings, \\ energetic support    \end{tabular} & \textit{"Thank you for all the prayers and happy thoughts. We appreciate and feel it all."} \\ \cline{2-4} 
\multicolumn{1}{|c|}{} & \textbf{\begin{tabular}[l]{@{}c@{}}REMOTE EMOTIONAL\\ SUPPORT (EMO\textsubscript{R})\end{tabular}} & \begin{tabular}[l]{@{}l@{}}    Messages, phone \\ calls, cards  \end{tabular} & \textit{"Thanks for all the well wishes and encouraging words. I've heard how comforting those words are to families when I have posted on sites, but now I really know!"} \\ \cline{2-4} 
\multicolumn{1}{|c|}{} & \textbf{\begin{tabular}[l]{@{}c@{}}CO-LOCATED EMO.\\ SUPPORT (EMO\textsubscript{CO})\end{tabular}} & \begin{tabular}[l]{@{}l@{}}    In-person social \\ interaction  \end{tabular} & \textit{"We had an awesome visit from X. It was great to catch up and relive old memories."} \\ \cline{2-4} 
\multicolumn{1}{|c|}{} & \textbf{\begin{tabular}[l]{@{}c@{}}INFORMATIONAL\\ SUPPORT (INFO)\end{tabular}} & \begin{tabular}[l]{@{}l@{}}    Health resources, \\ information   \end{tabular} & \textit{"I was so thrilled to finally have some guidance with how to start working in developmental milestones."} \\ \hline
\multirow{9}{*}{\rotatebox[origin=c]{90}{\textbf{INSTRUMENTAL SUPPORT}}} & \begin{tabular}[l]{@{}c@{}}Food \\ (FO)\end{tabular} & \begin{tabular}[l]{@{}l@{}} Meals, and/or \\ treats, drinks \end{tabular} & \textit{"X cooked and delivered an awesome steak dinner."} \\ \cline{2-4} 
 & \begin{tabular}[l]{@{}c@{}}Transportation\\ (TR)\end{tabular} & \begin{tabular}[l]{@{}l@{}}    Rides via car, \\ train, or plane  \end{tabular} & \textit{"Thank you to cousin X for coming with me and driving me to radiation yesterday!"} \\ \cline{2-4} 
 & \begin{tabular}[l]{@{}c@{}}Chores \\ (CH)\end{tabular} & Household tasks, errands & \textit{"Thanks to X and Y for helping with my yard and patio."} \\ \cline{2-4} 
 & \begin{tabular}[l]{@{}c@{}}Shelter \\ (SH)\end{tabular} & \begin{tabular}[l]{@{}l@{}}   Overnight \\ accommodations   \end{tabular} & \textit{"X was very happy to move into the Ronald McDonald house yesterday instead of Mon."} \\ \cline{2-4} 
 & \begin{tabular}[l]{@{}c@{}}Exercise \\ (EX)\end{tabular} & \begin{tabular}[l]{@{}l@{}}   Help with \\ physical activity   \end{tabular} & \textit{"Also thankful for exercise, and people who "push" others into it, like Coach X today, convincing Y to try the gym."} \\ \cline{2-4} 
 & \begin{tabular}[l]{@{}c@{}}Personal \\ Care (PC)\end{tabular} & \begin{tabular}[l]{@{}l@{}}    Services to \\ improve wellbeing  \end{tabular} & \textit{"Thanks to X and her nice friend, she was able to get her hair cut into an easy style."} \\ \cline{2-4} 
 & \begin{tabular}[l]{@{}c@{}}Financial \\ Assistance (FA)\end{tabular} & \begin{tabular}[l]{@{}l@{}}   Cash, gift cards, \\ bill payments   \end{tabular} & \textit{"We are all so grateful for the generous donations to X."} \\ \cline{2-4} 
 & \begin{tabular}[l]{@{}c@{}}Practical \\ Item (PI)\end{tabular} & \begin{tabular}[l]{@{}l@{}}    Tangible goods \\ and gifts  \end{tabular} & \textit{"Thanks to everyone for their gifts--they are all much appreciated."} \\ \cline{2-4} 
 & \begin{tabular}[l]{@{}c@{}}Family Care \\ (FC)\end{tabular} & \begin{tabular}[l]{@{}l@{}}   Caretaking for \\ dependents   \end{tabular} & \textit{"Special thank you to my mom and sister for taking such good care of our girl so we are able to focus on X during this time."} \\ \hline
\multirow{5}{*}{\rotatebox[origin=c]{90}{\textbf{OTHER SUPPORT*}}} & \begin{tabular}[l]{@{}c@{}}Medical\\ (MED)\end{tabular} & \begin{tabular}[l]{@{}l@{}}   Medical help \\ from people \end{tabular} & \textit{"The support that X oncology gave us today was tremendous."} \\ \cline{2-4} 
 & \begin{tabular}[l]{@{}c@{}}General \\ Donation (GD)\end{tabular} & \begin{tabular}[l]{@{}l@{}}   Charity donations \\ or fundraising   \end{tabular} & \textit{"The X 5K was a huge success thanks to the best family friends ever!"} \\ \cline{2-4} 
 & \begin{tabular}[l]{@{}c@{}}CaringBridge \\ (CB)\end{tabular} & \begin{tabular}[l]{@{}l@{}}    Help maintaining \\ CaringBridge site  \end{tabular} & \textit{"Our daughter X set up this account, which at least some of you have thankfully found."} \\ \cline{2-4} 
 & \begin{tabular}[l]{@{}c@{}}Generic \\ Help (GH)\end{tabular} & \begin{tabular}[l]{@{}l@{}}  Other or non- \\ specified support \end{tabular} & \textit{"I'm speechless to how kind everyone has been!"} \\ \cline{2-4} 
 & \begin{tabular}[l]{@{}c@{}}Non-Help \\ Related (NH)\end{tabular} & \begin{tabular}[l]{@{}l@{}}  Appreciation un- \\ related to help    \end{tabular} & \textit{"Today is a very happy friday for me! Today is my last day of radiation!"} \\ \hline
\end{tabular}
\caption{Abbreviated Codebook for Help Types in CaringBridge Journal Updates. (Supplementary material, sec. ~\ref{sec:complete}, shows complete codebook.) ``Example'' contains real user data with proper names replaced by `X' or `Y.' \textbf{BOLDED UPPERCASE} indicates high-level support type. Asterisk(*) indicates "other" categories not included in survey development; all other categories were present in at least one survey question.}
\label{tab:codebook}
\end{table}
}

\newcommand{\firstresults}{
\begin{figure}
\centering
\begin{subfigure}[t]{0.5\textwidth}
  \includegraphics[height=2.74in]{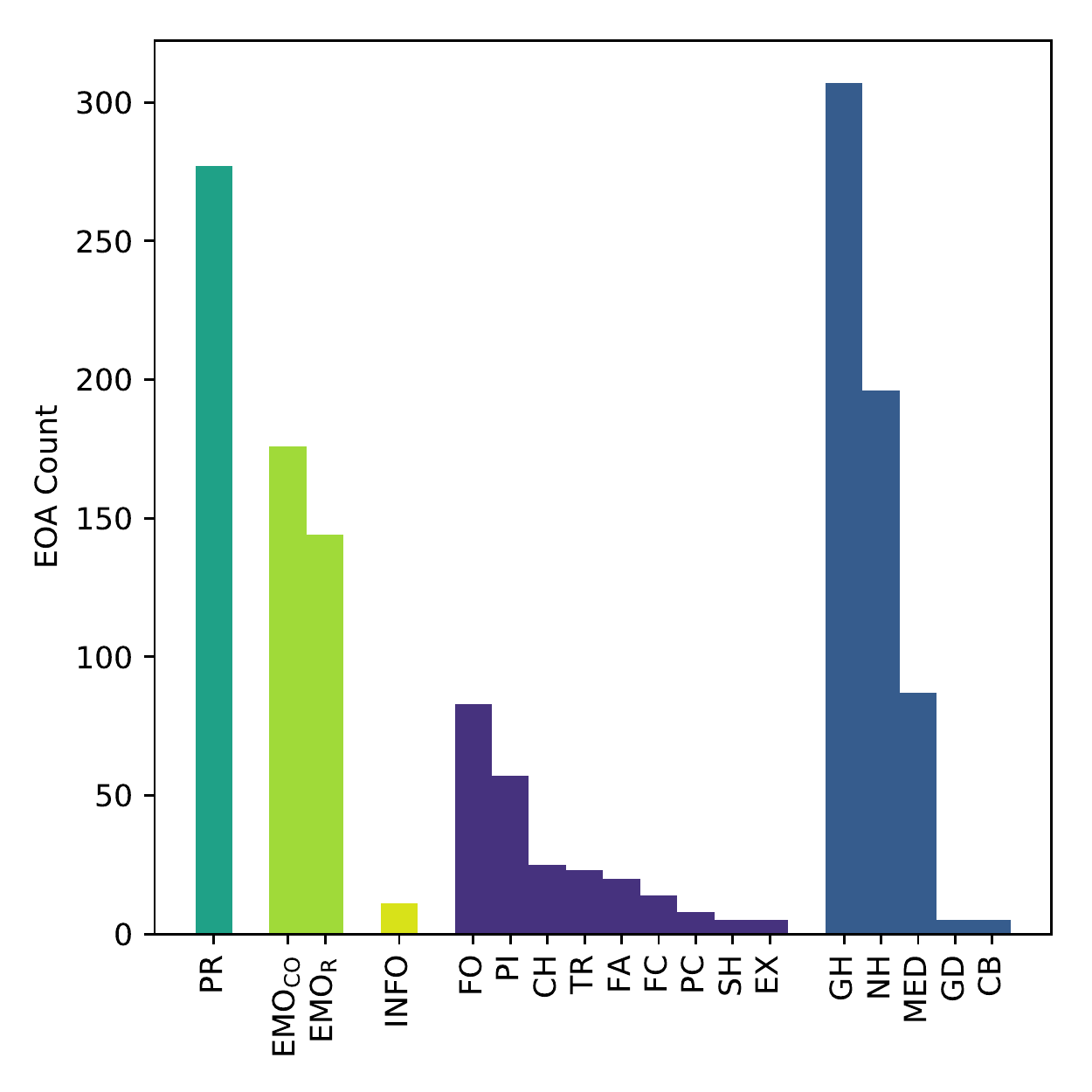}
\caption{Total counts of help types acknowledged by \\ either patients or caregivers.}
\label{fig:counts_of_help}
\end{subfigure}%
\begin{subfigure}[t]{0.5\textwidth}
  \includegraphics[height=2.74in]{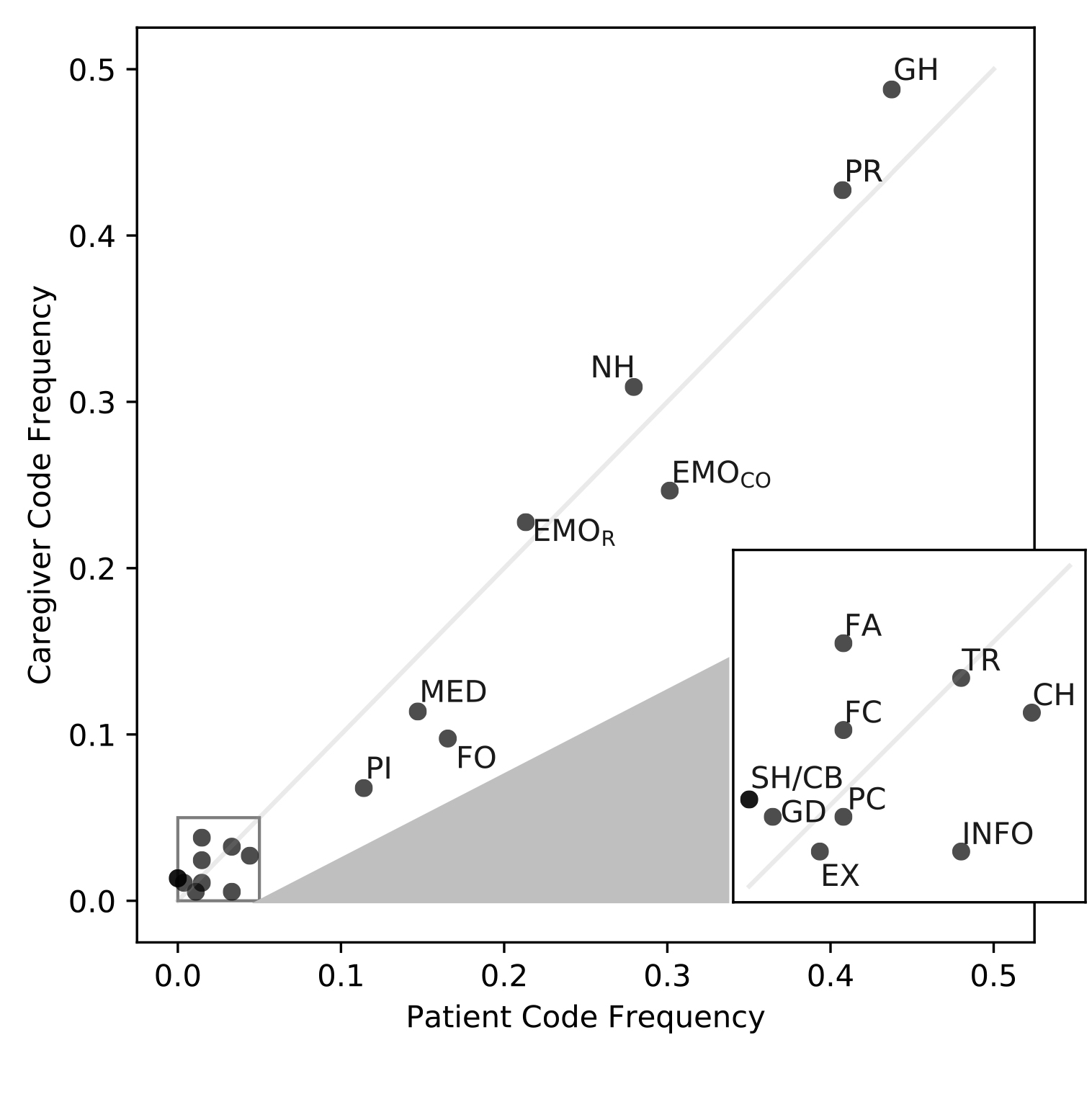}
\caption{Comparison of frequencies of help types acknowledged by caregivers (y-axis) and patients (x-axis).}
\label{fig:freq_of_help}
\end{subfigure}
\caption{Help Types Positively Acknowledged in CaringBridge Journal Updates. Data presented above are out of 641 total CaringBridge journal updates that contain at least one expression of appreciation (EOA).}
\label{fig:app_category_counts}
\end{figure}
}

\newcommand{\MLauthors}{
\begin{figure}
\centering
\begin{subfigure}[t]{0.5\textwidth}
  \includegraphics[height=2.74in]{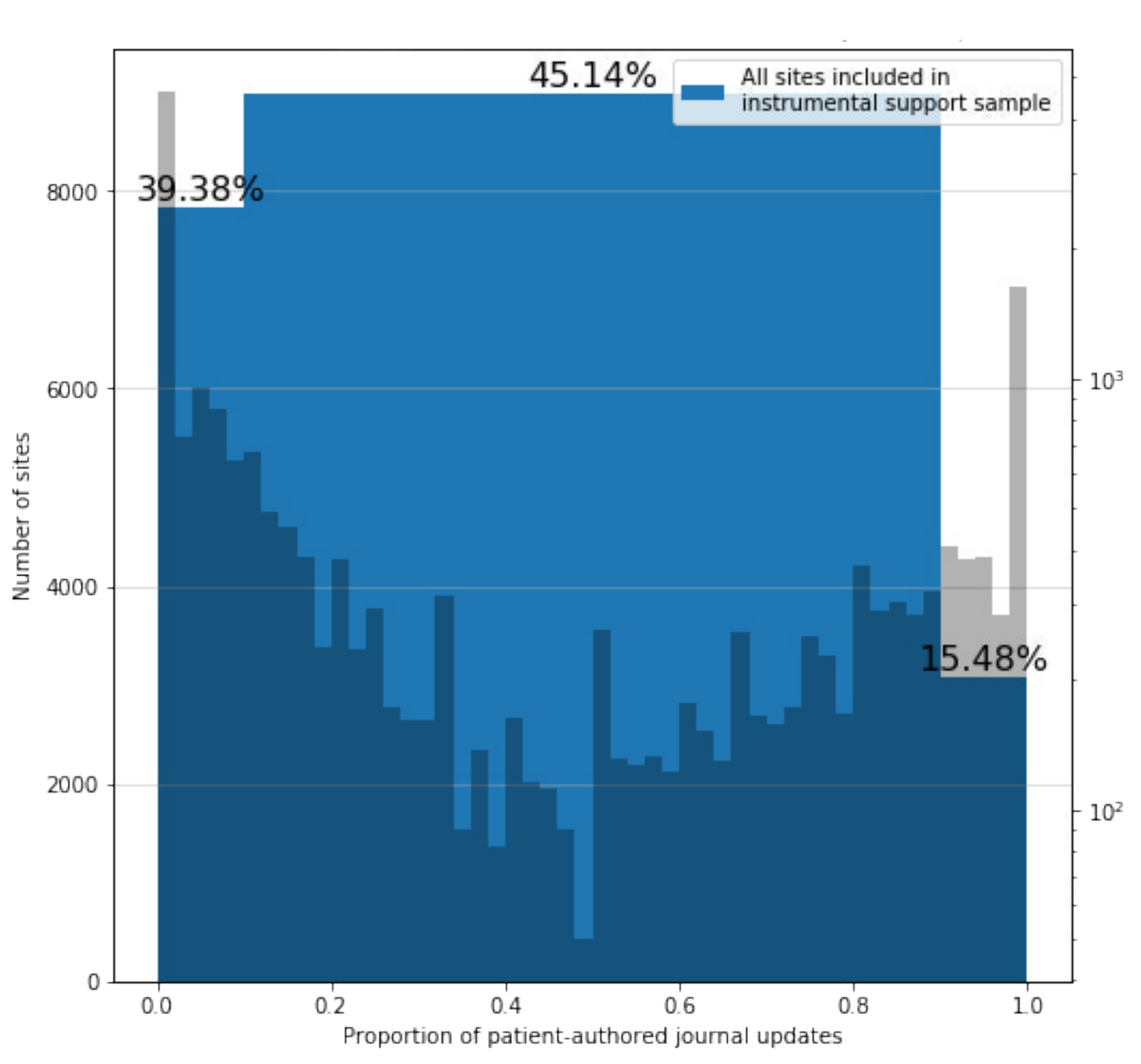}
\caption{Thresholds in the distribution where 15.5\% of \\ sites are independently authored by patients, \\45.1\% are collaboratively edited, and 39.4\% are \\independently authored by non-patients.}
\label{fig:counts}
\end{subfigure}%
\begin{subfigure}[t]{0.5\textwidth}
  \includegraphics[height=2.74in]{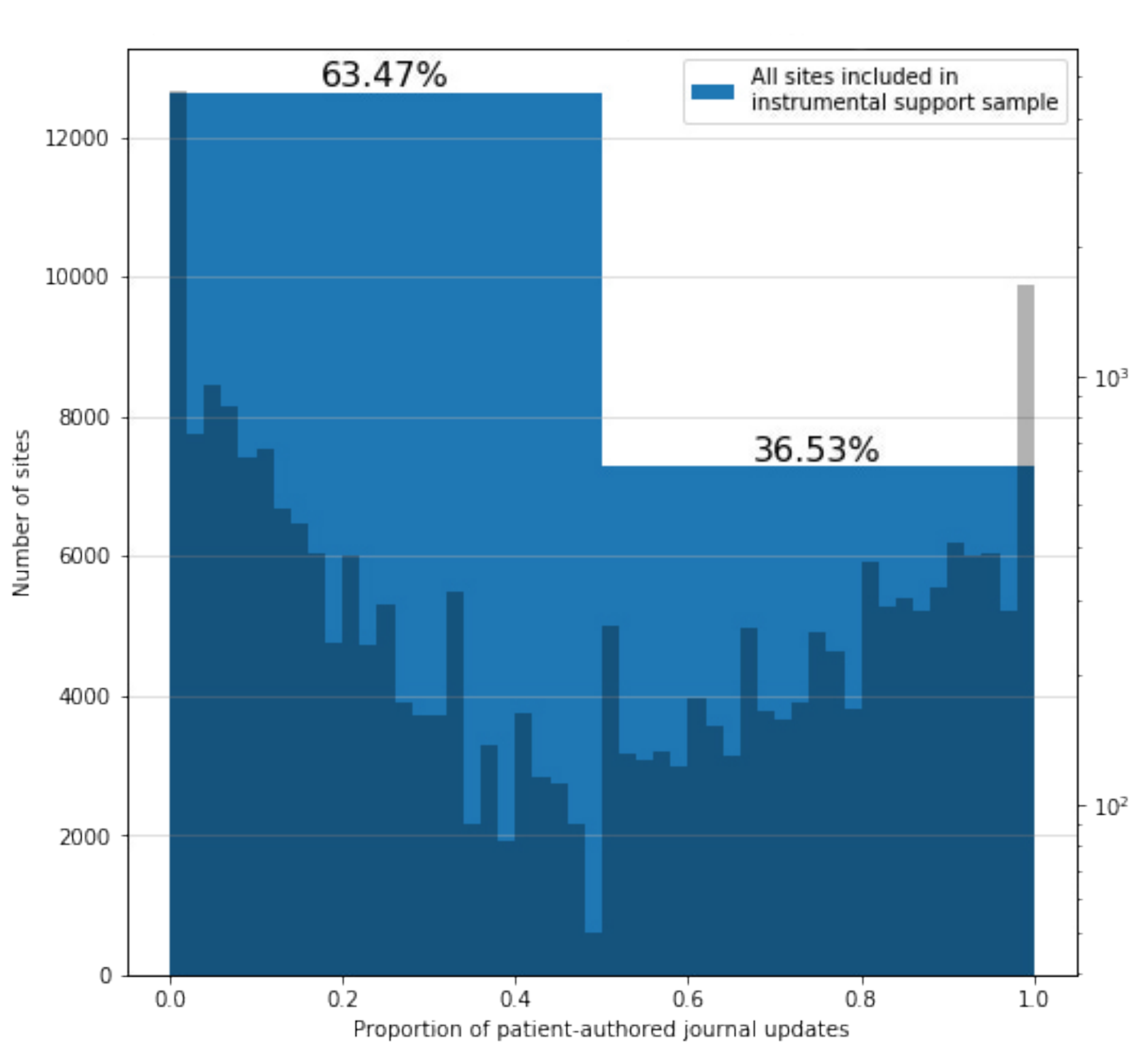}
\caption{Threshold where 63.5\% of sites are majority non-patient authored (i.e. >50\% of updates classified as non-patient authored).}
\label{fig:freq}
\end{subfigure}
\caption{Distribution of sites by proportion of patient-authored journal updates.}
\label{fig:ML_authorship}
\end{figure}
}

\newcommand{\highlevelsupport}{
\begin{table}
\centering
\begin{tabular}{|m{1.1in}|l m{0.5in}|l m{0.5in}|l|}
\hline
\textbf{Support Type}  & \multicolumn{2}{l|}{\textbf{P/CG} ($m = 576$)} & \multicolumn{2}{l|}{\textbf{FFA} ($n=415$)} & \textbf{Mann-Whitney $U$} \\ \hline
\begin{tabular}[c]{@{}l@{}} Prayer \\ Support \\ (PR) \end{tabular}  & \begin{tabular}[c]{@{}l@{}}$M=3.18$\\ $SD=1.14$ \end{tabular} & \includegraphics[width=0.5in]{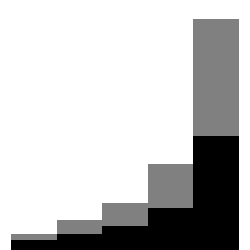} & \begin{tabular}[c]{@{}l@{}}$M=3.39$\\ $SD=1.01$\end{tabular} & \includegraphics[width=0.5in]{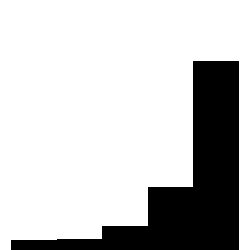} & \begin{tabular}[c]{@{}l@{}} $U$=130617 \\ $p=0.005$** \\ CLES=0.546 ($d$=0.16) \end{tabular} \\ \hline
\begin{tabular}[c]{@{}l@{}} Remote \\ Emotional Support \\ (EMO\textsubscript{R})\end{tabular}  & \begin{tabular}[c]{@{}l@{}}$M=2.87$\\ $SD=1.10$ \end{tabular} & \includegraphics[width=0.5in]{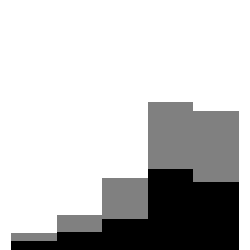} & \begin{tabular}[c]{@{}l@{}}$M=3.10$\\ $SD=0.94$\end{tabular} & \includegraphics[width=0.5in]{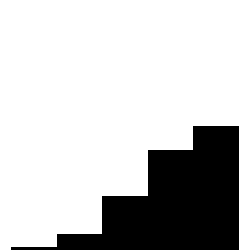} & \begin{tabular}[c]{@{}l@{}} $U$=132131.5 \\ $p=0.003$** \\ CLES=0.553 ($d$=0.19) \end{tabular} \\ \hline
\begin{tabular}[c]{@{}l@{}} Instrumental \\ Support \\ (INSTR) \end{tabular}  & \begin{tabular}[c]{@{}l@{}}$M=2.78$\\ $SD=1.19$ \end{tabular} & \includegraphics[width=0.5in]{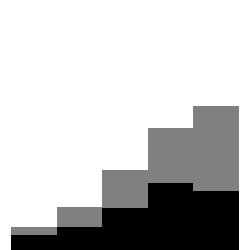} & \begin{tabular}[c]{@{}l@{}}$M=2.73$\\ $SD=1.24$\end{tabular} & \includegraphics[width=0.5in]{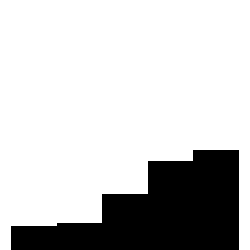} & \begin{tabular}[c]{@{}l@{}} $U$=121706.5 \\ $p=0.609$ \\ CLES=0.509 ($d$=0.03)\end{tabular} \\ \hline
\begin{tabular}[c]{@{}l@{}} Co-Located \\ Emotional Support \\(EMO\textsubscript{CO})\end{tabular}  & \begin{tabular}[c]{@{}l@{}}$M=2.51$\\ $SD=1.14$ \end{tabular} & \includegraphics[width=0.5in]{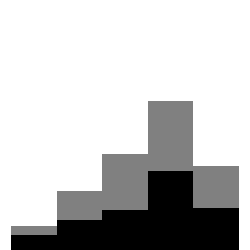} & \begin{tabular}[c]{@{}l@{}}$M=2.70$\\ $SD=1.15$\end{tabular} & \includegraphics[width=0.5in]{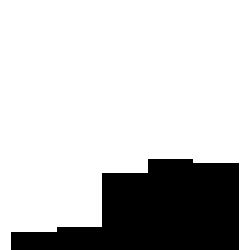} & \begin{tabular}[c]{@{}l@{}} $U$=130975 \\ $p=0.008$** \\ CLES=0.548 ($d$=0.17)\end{tabular} \\ \hline
\begin{tabular}[c]{@{}l@{}} Informational \\ Support \\(INFO) \end{tabular}  & \begin{tabular}[c]{@{}l@{}}$M=2.26$\\ $SD=1.23$ \end{tabular} & \includegraphics[width=0.5in]{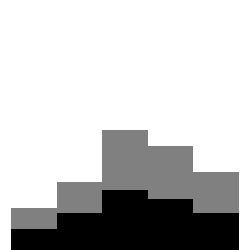} & \begin{tabular}[c]{@{}l@{}}$M=1.88$\\ $SD=1.37$\end{tabular} & \includegraphics[width=0.5in]{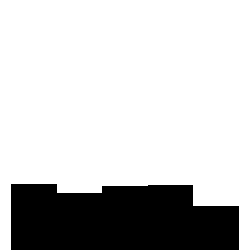} & \begin{tabular}[c]{@{}l@{}} $U$=138153.5 \\ $p<0.001$*** \\ CLES=0.578 ($d$=0.27)\end{tabular} \\ \hline
\end{tabular}
\caption{High-level Support Types. Respondents rated high-level support types on a 5-point unipolar Likert scale from 0: ``not at all'' to 4: ``extremely'' important. Patients/caregivers (P/CG) rated perceived importance to themselves, whereas FFA rated perceived importance of \textit{providing} help. Histograms show distributions of Likert scores. P/CG column: gray is CG, black is P. Mann-Whitney $U$ tests are shown with uncorrected $p$-values. Common Language Effect Size ($U/mn$) is shown with corresponding estimates of Cohen's $d$.}
\label{tab:support}
\end{table}
}

\newcommand{\demotable}{
\begin{table}[t]
\centering
\begin{tabular}{|m{0.9in}|m{1.15in}|m{0.8in}|m{1.7in}|}
\hline
\textbf{\begin{tabular}{l} Gender \end{tabular} }  & \begin{tabular}{m{0.75in} r}
Female & 841 \\ Male & 147 \\ Other & 3
\end{tabular} & \textbf{Education}  &  \begin{tabular}{m{1.25in} l}
High School (or Equiv) & 117\\ College (or Equiv) & 497 \\
Post Bachelor    & 377   
\end{tabular} \\

\hline
\textbf{\begin{tabular}{l} Race \end{tabular}} & \begin{tabular}{m{0.75in} r}
White & 852 \\ Non-white & 73 \\ Did Not Say & 66
\end{tabular} & \textbf{Employment} &\begin{tabular}{m{1.25in} l}
Employed & 576 \\ Retired & 271 \\ Other & 144
\end{tabular} \\
\hline
\textbf{\begin{tabular}{l} Country \end{tabular}} & \begin{tabular}{m{0.75in} r}
United States & 960 \\ Canada & 16 \\ Other & 15
\end{tabular} & \textbf{Household Income} & \begin{tabular}{m{0.8in} m{0.5in}} \rule{0pt}{0.5in}\includegraphics[width=1in,height=0.4in]{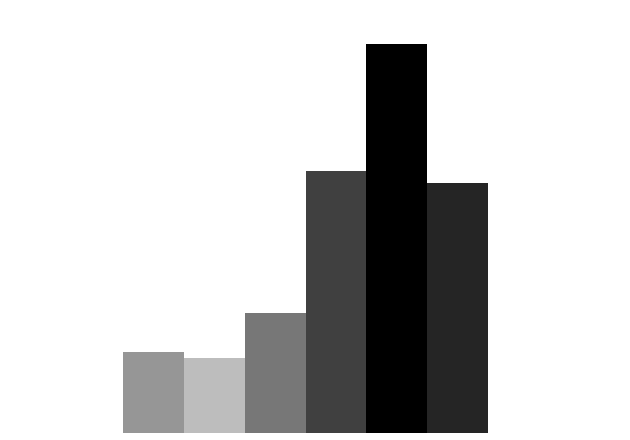} & \begin{tabular}{m{0.5in}}
Did Not Say:  247 \\
\end{tabular}\\
\end{tabular} \\
\hline
\textbf{\begin{tabular}{l} Population \\ Class \\ (county size) \end{tabular}} & \begin{tabular}{m{0.75in} l}
> 1 million & 331 \\ 50k to 1 mil & 338 \\ < 50k  & 322
\end{tabular}    & \textbf{Age} & \begin{tabular}{m{0.8in} m{0.5in}} \rule{0pt}{0.5in}\includegraphics[width=1in,height=0.4in]{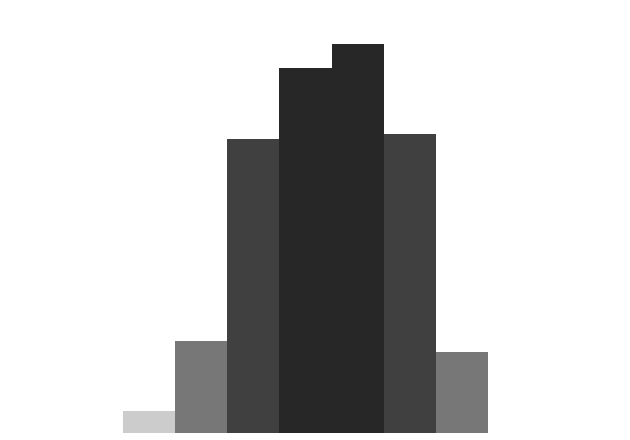} & \begin{tabular}{l}
Mode is: \\
55-64
\end{tabular}\\
\end{tabular} \\
\hline
\end{tabular}           
\caption{Demographics of 991 Survey Respondents. Income bins range from <\$25k to >\$150k. Age bins range from 18-24 to 75+ years old.}
\label{tab:demo}
\end{table}
} 

\newcommand{\challenges}{
\begin{table}[H]
\centering
\begin{tabular}{|c|c|c|c|}
\hline
\multicolumn{2}{|c|}{\textbf{P/CG (Binary Multiple Choices)}} & \multicolumn{2}{l|}{\textbf{FFA (Mean Likert Response)}} \\ \hline
I don't want to be a burden. & 80.2\% & Distance & -0.5 \\ \hline
I'm a very private person. & 42.9\% & Time & -0.31 \\ \hline
I'm embarrassed to ask. & 29.0\% & Monetary Cost & -0.17 \\ \hline
I don't know how to ask. & 22.9\% & Social Awkwardness & 0.17 \\ \hline
It's not fair for me to ask. & 20.5\% & Professional Boundaries & 0.41 \\ \hline
I don't think anyone is able to help. & 14.2\% & \multicolumn{2}{l|}{\multirow{3}{*}{\textit{\begin{tabular}[c]{@{}l@{}}(5-point bipolar mean Likert \\ responses from -2: "Extremely \\ difficult" to 2: "Extremely easy")\end{tabular}}}} \\ \cline{1-2}
I don't know whom to ask. & 7.6\% & \multicolumn{2}{l|}{} \\ \cline{1-2}
I lack trust in my community. & 4.2\% & \multicolumn{2}{l|}{} \\ \hline
\end{tabular}
\caption{Challenges related to asking for help (P/CG) versus offering help (FFA), questions ~\ref{ask_challenge}, ~\ref{ask_challenge_ffa}.}
\label{tab:challenges}
\end{table}
}

\newcommand{\usefulness}{
\begin{table}[H]
\centering 
\begin{tabular}{r|cc|cc|c|cl}
\hline
\multicolumn{1}{c|}{\textbf{INSTRUMENTAL}} & \multicolumn{2}{c|}{\textbf{PATIENTS}} & \multicolumn{2}{c|}{\textbf{CAREGIVERS}} & \textbf{Difference} & \multicolumn{2}{c}{\textbf{Mann-Whitney $U$}} \\
\multicolumn{1}{c|}{\textbf{USEFULNESS}} & $M$ & $SD$ & $M$ & $SD$ & \textbf{of Means}& $U$ & $p$-value  \\ \hline \hline
Chores (CH) & 2.67 & 1.24 & 2.98 & 1.07 & 0.31 $\pm$ 0.19\textsuperscript{$\dagger$} & 36067 & $p=0.005$** \\ 
Food (FO) & 2.53 & 1.24 & 2.77 & 1.13 & 0.24 $\pm$ 0.19\textsuperscript{$\dagger$} & 37319.5 & $p=0.031$* \\
Transportation (TR) & 2.53 & 1.29 & 2.05 & 1.43 & 0.48 $\pm$ 0.22\textsuperscript{$\dagger$} & 49350 & $p < 0.001$*** \\
Exercise (EX) & 2.08 & 1.24 & 1.79 & 1.27 & 0.29 $\pm$ 0.21\textsuperscript{$\dagger$} & 46830 & $p=0.006$** \\
Financial Assist. (FA) & 1.97 & 1.53 & 2.57 & 1.42 &  0.60 $\pm$ 0.24\textsuperscript{$\dagger$} & 32376 & $p < 0.001$***  \\
Personal Care (PC) & 1.91 & 1.29 & 2.07 & 1.27 & 0.16 $\pm$ 0.21 & 38596.5 & $p=0.141$ \\
Practical Item (PI) & 1.5 & 1.26 & 1.6 & 1.25 & 0.10 $\pm$ 0.21 & 39569.5 & $p=0.328$  \\
Familycare (FC) & 1.24 & 1.44 & 1.75 & 1.57 &  0.51 $\pm$ 0.25\textsuperscript{$\dagger$} & 34208.5 & $p < 0.001$*** \\
Shelter (SH) & 1.11 & 1.34 & 1.66 & 1.51 & 0.55 $\pm$ 0.23\textsuperscript{$\dagger$} & 32992 & $p < 0.001$*** \\ \hline
\end{tabular}
\caption{Usefulness of Instrumental Support Types to Patients and Caregivers. 5-point unipolar mean Likert responses on a scale of 0: "Not at all" to 4: "Extremely" useful, question ~\ref{useful_help}. The confidence interval (CI) for the difference of the means is given at the 95\% level. \textsuperscript{$\dagger$} indicates that CI does not contain 0.}
\label{tab:useful}
\end{table}
}

\newcommand{\channels}{
\begin{wraptable}{br}{5cm}
\centering
\begin{tabular}{rcc}
\hline
 & \textbf{P/CG} & \textbf{FFA} \\ \hline \hline
\textbf{Social Media} & -0.77 & -0.22 \\ 
\textbf{In-person} & -0.39 & -0.33 \\
\textbf{Email} & -0.10 & 0.47 \\ 
\textbf{Phone/text} & -0.05 & 0.49 \\ 
\textbf{CaringBridge} & -0.05 & 0.64 \\ \hline
\end{tabular}
\caption{Ease of Communication about Providing Instrumental Support. Mean of bipolar Likert response from "-2: Extremely difficult" to "2: Extremely easy." Questions ~\ref{ask_diff}, ~\ref{ask_diff_ffa}.}
\label{tab:channels}
\end{wraptable}
}

\newcommand{\anovaresults}{
\begin{table}[H]
\centering
\begin{tabular}{crlll}
\multicolumn{5}{c}{\textbf{ANOVA Model (Squared Multiple Correlation: In Person  $R^{2}=0.40$; Funding $R^{2}=0.21$)}} \\
\hline
\textbf{SQ} & \multicolumn{1}{c}{\textbf{Independent Variable}} & \textbf{In Person} & \textbf{Funding} & \multicolumn{1}{c}{\textbf{Trend (Greater Interest)}} \\ \hline \hline
~\ref{demo_age} & age & $p<0.001$*** & $p<0.001$*** & older age \\
~\ref{demo_income} & income & $p=0.043$* & $p=0.005$** & higher income \\
~\ref{MSA} & population class & $p=0.130$ & N/A & N/A \\
~\ref{relationship} & relationship & $p<0.001$*** & $p=0.008$** & closer relationship \\
~\ref{ask_challenge_ffa} & monetary challenge & N/A & $p<0.001$*** & lower monetary challenge \\
~\ref{ask_challenge_ffa} & time challenge & $p=0.003$** & N/A & less time challenge \\
~\ref{ask_challenge_ffa} & distance challenge & $p=0.005$** & N/A & lower distance challenge \\
~\ref{ask_diff_ffa} & difficulty phone/text & $p<0.001$*** & N/A & higher difficulty of phone/text \\
~\ref{ask_freq_ffa} & help needed & $p=0.002$** & N/A & higher perceived need \\
~\ref{ask_freq_ffa} & help asked for & $p=0.023$* & N/A & less help requested \\ \hline
\end{tabular}
\caption{Factors Affecting Interest in Providing Instrumental Support. Dependent variables are summative Likert scores of FFA interest in providing instrumental support (questions ~\ref{in_person_ffa}, ~\ref{funding_ffa}). SQ refers to survey question where independent variable was gathered. Trend shows direction of effect towards greater interest; e.g. older age predicts higher interest in providing instrumental support in person and by funding a service.}
\label{tab:anova}
\end{table}
}

\newcommand{\trustregression}{
\begin{table}[H]
\centering
\begin{tabular}{rrrll}
\hline
\multicolumn{5}{l}{\textbf{APP-BASED BUSINESSES}} \\
\hline
\textbf{SQ} & \multicolumn{1}{l}{} & \multicolumn{2}{c}{\textbf{Independent Variable}} & \multicolumn{1}{c}{\textbf{Trend}} \\ \hline \hline
~\ref{demo_age} & Financial Assist. (FA) & age & $p = 0.025$* & complex, no general trend \\
~\ref{demo_income} &  & income & $p = 0.048$* & lower income more likely to trust \\
~\ref{role} &  & role & $p = 0.003$** & caregiver more likely to trust\\
~\ref{role} & Transportation (TR) & role & $p = 0.002$** & patient more likely to trust\\
~\ref{app_used} &  & prior use & $p < 0.001$*** & prior use more likely to trust \\
~\ref{app_used} & Shelter (SH) & prior use & $p = 0.039$* & prior use more likely to trust \\
~\ref{demo_age} & Chores (CH) & age & $p = 0.016$* & complex, no general trend \\ \hline \hline 
\multicolumn{5}{l}{\textbf{TRADITIONAL BUSINESSES}} \\ \hline
\textbf{SQ} & \multicolumn{1}{l}{} & \multicolumn{2}{c}{\textbf{Independent Variable}} & \multicolumn{1}{c}{\textbf{Trend}} \\ \hline \hline
~\ref{demo_age} & Shelter (SH) & age & $p = 0.029$* & complex, no general trend \\
~\ref{demo_gender} & Personal Care (PC) & gender & $p = 0.001$*** & female more likely to trust \\ \hline
\end{tabular}
\caption{Factors Affecting Patient and Caregiver Trust in App-Based or Traditional Businesses. Only independent variables that remain statistically significant after Bonferroni correction are displayed.}
\label{tab:trustregression}
\end{table}
}

\newcommand{\trustmatrix}{
\begin{figure}
  \includegraphics[height=10.2cm]{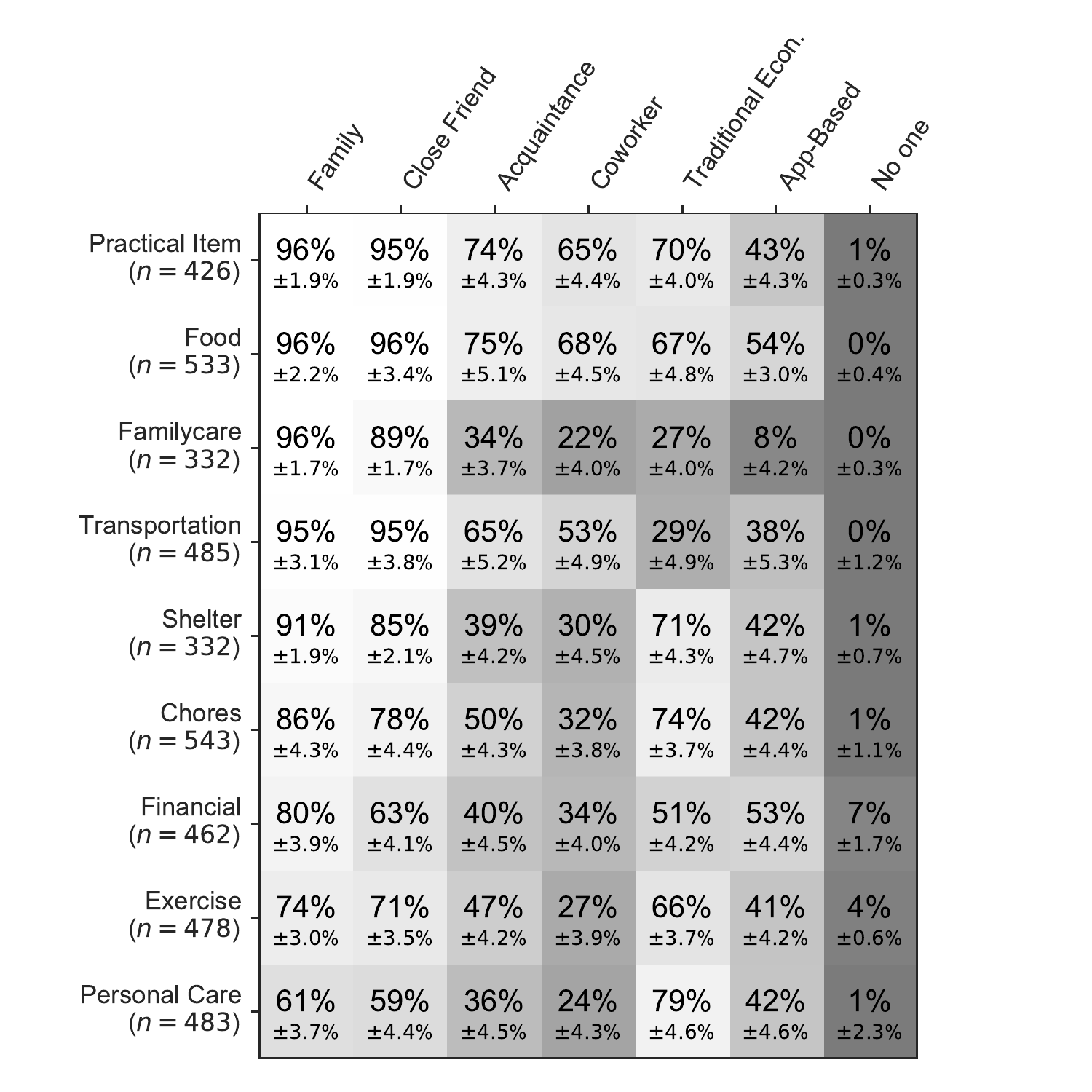}
  \caption{Patient and Caregiver Trust in People or Businesses to Provide Instrumental Support. Rows show each INSTR type; $n$ indicates number of patients/caregivers rating that INSTR type as at least ``slightly useful'' (question ~\ref{useful_help}). Columns show people or business-types that might provide help. Each cell shows the percentage of patients/caregivers (normalized by $n$) indicating trust in that person or business-type to provide that INSTR type, along with the 95\% confidence interval. Cell shading is solely a visual aid; the lighter the cell shading, the larger the proportion of patients/caregivers indicating trust in that supporter group.}~\label{fig:trust}
\end{figure}
}

\newcommand{\altffameans}{
\begin{table}[H]
\centering
\begin{tabular}{r|cc|c|cc|c}
\hline
\multicolumn{1}{c|}{\textbf{FFA INTEREST}} & \multicolumn{2}{c|}{\textbf{CO-LOCATED} ($M$)} & \textbf{Difference} & \multicolumn{2}{c|}{\textbf{REMOTE} ($M$)} & \textbf{Difference} \\ 
\multicolumn{1}{c|}{\textbf{IN PROVIDING}} & \textbf{In-P} & \textbf{Funding} & \textbf{of Means} & \textbf{In-P} & \textbf{Funding} & \textbf{of Means} \\
 \hline \hline
Food (FO) & 2.92 & 2.39 & 0.52$\pm$0.26\textsuperscript{$\dagger$} & 2.15 & 2.17 & 0.02$\pm$0.25 \\
Practical Item (PI) & 2.85 & 2.12 & 0.73$\pm$0.25\textsuperscript{$\dagger$} & 2.11 & 2.21 & 0.10$\pm$0.25 \\
Transportation (TR) & 2.7 & 1.97 & 0.73$\pm$0.25\textsuperscript{$\dagger$} & 1.85 & 1.69 & 0.15$\pm$0.25 \\
Chores (CH) & 2.47 & 2.0 & 0.47$\pm$0.28\textsuperscript{$\dagger$} & 1.9 & 1.76 & 0.14$\pm$0.26 \\
Familycare (FC) & 2.34 & 1.7 & 0.65$\pm$0.28\textsuperscript{$\dagger$} & 1.65 & 1.4 & 0.24$\pm$0.25 \\
Exercise (EX) & 2.16 & 1.49 & 0.67$\pm$0.28\textsuperscript{$\dagger$} & 1.54 & 1.32 & 0.22$\pm$0.25 \\
Shelter (SH) & 1.83 & 1.78 & 0.06$\pm$0.30 & 1.53 & 1.56 & 0.04$\pm$0.26 \\
Personal Care (PC) & 1.71 & 1.75 & 0.04$\pm$0.29 & 1.32 & 1.68 & 0.36$\pm$0.26\textsuperscript{$\dagger$} \\ \hline
\end{tabular}
\caption{FFA Interest in Providing Instrumental Support Types to Patients and Caregivers. ``In-P'' indicates ``in-person'' (question~\ref{in_person_ffa}) while ``Funding'' refers to interest in funding a paid service for that support type (question~\ref{funding_ffa}). 5-point unipolar mean Likert responses on a scale of 0:"Not at all" to 4:"Extremely" interested. The confidence interval (CI) for the difference of the means is given at the 95\% level.  \textsuperscript{$\dagger$} indicates that the CI does not contain 0. Financial Assistance (FA) is not included as a row in the table, since the distinction between In-P vs. Funding is not meaningful for FA; FFA mean interest in providing FA is $M\textsubscript{\textit{co-located}}=2.07$, $M\textsubscript{\textit{remote}}=2.03$, and the difference between these means is not significant.}
\label{tab:ffainterest1}
\end{table}
}

\newcommand{\completecodebook}{
\begin{table}[H]
\centering
\caption{Full Codebook for Author Codes. Examples are based on real anonymized user data; some examples fabricated for educational demonstration.}
\label{tab:authors}
\begin{tabular}{|m{1.2cm}|m{4.5cm}|m{6.8cm}|}
\hline
\textbf{Author} & \textbf{Definition} & \textbf{Example(s)} \\ \hline
patient (P) & The author appears to be personally going through a health journey and is writing about their own medical experiences. & \begin{tabular}[c]{m{6.6cm}}"I had my first round of chemo today."\\ \textbf{Special Case:} If an update is about a woman's pregnancy, then the woman is considered (P). If there are complications with a baby post-delivery, then the mother is (CG).\end{tabular} \\ \hline
caregiver (CG) & The author has some social connection to a patient and is posting about this person's health journey, or on this person's behalf. & \begin{tabular}[c]{m{6.6cm}}"Mom had her first round of chemo today."\\ \textbf{Special Case:} If parents write journal updates in the voice of their child(ren), updates may appear to be written by the patient, but if the writing was evidently completed by an adult, it should be coded (CG). I.e. "The doctor told mommy she should give me a special baby formula for my congenital hypothyroidism."\end{tabular} \\ \hline
\begin{tabular}[c]{@{}l@{}}both \\ patient \& \\ caregiver \\ (PC)\end{tabular} & The author appears to be writing details about their own medical condition, and about someone else's medical condition. & \begin{tabular}[c]{m{6.6cm}} "I had my first round of chemo today. ... Here's another update on Mom as well. She is still having symptoms from her surgery."\end{tabular} \\ \hline
unknown (U) & By default, mark journals (U) for which this is no explicit written evidence of medical symptoms/procedures/etc. & \begin{tabular}[c]{m{6.6cm}}"Click here to learn about chemo: <link>"\\ "We had time to go see a movie today. What a treat!"\end{tabular} \\ \hline
\end{tabular}
\end{table}

\begin{longtable}[H]{|c|m{6cm}|m{4.4cm}|}
\caption{Full Codebook for Help Types. Examples are based on real anonymized user data; some examples  fabricated for educational demonstration.} \\
\hline
\multicolumn{1}{|c|}{\textbf{Help Types}} & \multicolumn{1}{c|}{\textbf{Definition}} & \multicolumn{1}{c|}{\textbf{Example(s)}} \\ \hline \endhead
\begin{tabular}[c]{@{}l@{}}prayer \\ (PR)\end{tabular} & Prayers, positive karma, good ju-ju, warm thoughts, spiritual blessings, etc. (Generic religious or philosophical statements such as "Follow the way of the Lord for he is your savior" should NOT be coded in any category.) & \begin{tabular}[c]{m{4.1cm}}(PR) "Thanks for your thoughts and prayers"\\ (PR, GH) "Thank you for your warm thoughts and all the support" \end{tabular} \\ \hline
\begin{tabular}[c]{@{}l@{}}remote\\ emotional\\ support \\ (EMO\textsubscript{R})\end{tabular} & Emotional support through remote communication, art, and/or non-practical items. This includes posting on the CaringBridge site, sending messages, well wishes, giving cards, phone/skype/etc. calls, photos, photobooks, picture frames, paintings, drawings, and similar items, and does \textit{not} include physical in-person interactions. & \begin{tabular}[c]{m{4.1cm}}(EMO\textsubscript{R}, PR) "Thank you for your messages and prayers." \\ (EMO\textsubscript{R}) "Thank you for reading my journals."\end{tabular} \\ \hline
\begin{tabular}[c]{@{}l@{}}co-located \\ emotional \\ support \\ (EMO\textsubscript{CO})\end{tabular} & Support through socially interacting in-person. This includes hospital, hospice, or home visits, social events, and outings, or even simple physical interactions like hugs, high fives, taking photos together, spending time, or having fun together. This does not include phone calls, which should be coded (EMO\textsubscript{R}). EMO\textsubscript{CO} may possibly accompany other codes. & \begin{tabular}[c]{m{4.1cm}}(EMO\textsubscript{CO}, GD) "I had a great time with all of you at the race for the cure event yesterday!"\\ (EMO\textsubscript{CO}) "So glad I got to see Bill when he came to say hi yesterday."\end{tabular} \\ \hline
\begin{tabular}[c]{@{}l@{}}information \\ (INFO)\end{tabular} & Help finding information related to a health situation/medical condition from non-professionals. (If a dr. or nurse provides information or explanation about a medical situation, code (MED), not (INFO).) & (INFO) "The links you posted were really helpful. Thanks!" \\ \hline
\begin{tabular}[c]{@{}l@{}}food \\ (FO)\end{tabular} & Help in the form of food and drinks to be prepared or purchased. This includes meals, groceries, and restaurant food, as well as desserts, candy, etc. If the writing indicates that the food was consumed at a restaurant or cafeteria and/or with other people, it should be coded (FO, EMO\textsubscript{CO}). & (FO) "Well, I think I would have starved to death if it wasn't for Joann's tasty homecooked potroast. My tummy is happy for that!" \\ \hline
\begin{tabular}[c]{@{}l@{}}transportation \\ (TR)\end{tabular} & Help in the form of transportation for the patient or caregiver in-person, including rides to appointments or hospitals, flights, or other mechanisms for getting around (train, etc.). & \begin{tabular}[c]{m{4.1cm}}(TR, FO) "I'm thankful Jim gave me a ride to the grocery store," (since the Author was, themself, transported) \\ (FO) "I'm thankful Jim drove to the store to get me some groceries."\end{tabular} \\ \hline
\begin{tabular}[c]{@{}l@{}}chores \\ (CH)\end{tabular} & Help with household or current/future living area tasks like cleaning, heavy-lifting, laundry, errands, gardening, pet care, or moving. This does not include food related chores like cooking or grocery shopping. & (CH) "So grateful for the gardening help yesterday..." \\ \hline
\begin{tabular}[c]{@{}l@{}}shelter \\ (SH)\end{tabular} & Help in the form of a place to stay overnight for a temporary period. This does not include stays at the hospital, but it does include stays at other patient-oriented organizations, hotels, inns, airbnbs, friend's homes, etc. & (SH) "Thanks to Mindy for letting me stay with her when we had to go to Houston for the surgery." \\ \hline
\begin{tabular}[c]{@{}l@{}}exercise \\ (EX)\end{tabular} & Help with exercise, or motivation to exercise. This does not include physical therapy completed ONLY with a medical professional, which should be coded (MED). Instead, it refers to when exercise is completed with a social connection, or if a social connection's motivational presence or actions contributed to exercise. & (EX) "Joey came over to walk around the block with me, and I'm so happy he was able to do that. It really got me motivated to do a little activity." \\ \hline
\begin{tabular}[c]{@{}l@{}}personal care \\ (PC)\end{tabular} & Services or actions that provide comfort or assist with personal care, physical appearance. This includes doing hair, nails, massages, and activities related to well-being and taking care of the body, that do not explicitly state anything about exercise or food. This does not include medical needs. & (PC, EMO\textsubscript{CO}) "Lisa took me for a manicure and there was one little point of brightness in an otherwise difficult week. Really glad to have such a good time with a good friend." \\ \hline
\begin{tabular}[c]{@{}l@{}}financial \\ assistance \\ (FA)\end{tabular} & Help in the form of direct financial aid for the patient or their family. This includes cash, gift cards, bill payments, or donations through crowdfunding sites such as GoFundMe. This also includes personal benefit events. This does not include physical items or services purchased for a person; such physical items should be classified under a more specific category. & (FA) "Billy and Sarah helped to cover my medical expenses... I'm so grateful for their support." \\ \hline
\begin{tabular}[c]{@{}l@{}}practical \\ item\\ (PI)\end{tabular} & Help in the form of bringing over or transferring ownership of items that fulfill a specific need or improve quality of life. This can include items like flowers, decorations, toys, birthday or holiday gifts, games, music and magazines, clothes, jewelry, bedding, wigs, etc. It can also include item recommendations or digital sources of entertainment. However, if the entertainment is enjoyed as a part of a social group activity, it should be coded only (EMO\textsubscript{CO}). & (PI) "So thankful Sam dropped off a dvd for me to watch during chemo." \\ 
\hline
\begin{tabular}[c]{@{}l@{}}familycare \\ (FC)\end{tabular} & Help with watching or taking care of children or elderly. This can apply to either the focal patient, if the patient requires supervision, or to other people connected to the author. & (FC, MED) "So appreciative that Jeanna watched my son and helped him take his meds on time when we went out for a movie yesterday." \\ \hline
\begin{tabular}[c]{@{}l@{}}medical \\ (MED)\end{tabular} & Help in the form of medical assistance from specific person(s), including friends and family, doctors, nurses, therapists, organ donors, or experts/professionals. This does NOT include generic expressions of appreciation for good test results, recovery, progress, etc. & \begin{tabular}[c]{m{4.1cm}}(MED) "I just want to thank the whole team at St. Joes for their dedication to making dad's surgery such a big success."\\ Note the contrast to: (NH) "So thankful grandma's health is recovering after her fall."\end{tabular} \\ \hline
\begin{tabular}[c]{@{}l@{}}general \\ donation \\ (GD)\end{tabular} & Help in the form of donations or participation in a fundraising event, for which the proceeds do not specifically benefit the individual patient. This includes cancer walks and donations to research programs, or any other organization (such as CaringBridge). & (GD) "It means so much to me that you all came out for the Race for the Cure event." \\ \hline
\begin{tabular}[c]{@{}l@{}}caringbridge \\ (CB)\end{tabular} & Any help related to setting up, maintaining, or posting on a CaringBridge site. This includes co-authorship, sharing the CaringBridge site link, and setting up the account and preferences. & (CB) "Feeling blessed to have help from Sarah with keeping my journal up to date." \\ \hline
\begin{tabular}[c]{@{}l@{}}generic help \\ (GH)\end{tabular} & Help that does not fit in any specified category. This includes mentions of non-specified support, hospitality, nurturing, love, caring, kindness, etc. It also includes any specific description of help that does not match any codes above, including random help with things like coordinating schedules, paperwork, administrative, etc. & \begin{tabular}[c]{m{4.1cm}}(GH) "Thanks for everything! You're all wonderful!"\\ (GH) "The hospitality I have received has been wonderful and heartwarming."\end{tabular} \\ \hline
\begin{tabular}[c]{@{}l@{}}non-help \\ related \\  (NH)\end{tabular} & Appreciation is expressed for something that is not related to any form of help received. This includes expressions of appreciation for medical progress, healing, or good test results. & \begin{tabular}[c]{m{4.1cm}}(NH) "I appreciated the artistry of the movie we saw." \\ (NH) "Thank god the tests were negative."\end{tabular} \\ \hline
\end{longtable}
}

\newcommand{\classifiertable}{
\begin{table}[H]
\begin{tabular}{rcccc}
\hline
 & \textbf{Precision} & \textbf{Recall} & \textbf{F1-score} & \textbf{Support} \\
\hline \hline
\textbf{Non-Patient} & 0.95 & 0.93 & 0.94 & 191 \\
\textbf{Patient} & 0.96 & 0.98 & 0.97 & 384 \\
\textbf{Average/Total} & 0.96 & 0.96 & 0.96 & 575 \\ 
\hline
\end{tabular}
\caption{Analytical metrics for logistic regression classifier with L2 regularization of patient vs. non-patient authorship of individual journal updates.}
\label{author_classifier}
\end{table}
}

\newcommand{\completelex}{
\begin{table}[H]
\centering
\begin{tabular}{| lllll |}
\hline
\multicolumn{5}{| c |}{\textbf{Appreciation Lexicon Derived from CaringBridge.org Corpus}} \\ 
\hline
affection & blest & graciousness & kindess & suport \\
amazed & boundless & graditude & kindness & supprt \\
appeciate & closeness & graitude & kindnesses & sweetness \\
apppreciate & comforted & grateful & kudos & thak \\
apprciate & contentment & gratefull & love & thaks \\
appreaciate & deepest & gratefullness & loving & thanful \\
apprecate & deeply & gratefulness & loyalty & thank \\
apprecation & elation & gratful & offers & thankful \\
appreciate & embraced & gratified & outporing & thankfull \\
appreciated & emotion & gratitide & outpour & thankfullness \\
appreciates & empowered & gratitiude & outpourings & thankfully \\
appreciation & encouraged & gratitude & outreach & thankfulness \\
appreciative & encouragement & gratitute & overflowing & thanks \\
apprecitate & encourgement & grattitude & overjoyed & thanksful \\
apprecite & enouragement & greateful & overwhelemed & thankul \\
appriciate & evidently & happiness & overwhelmed & thanx \\
apprieciate & fortunate & heartened & payers & thoughtfullness \\
apreciate & fortunately & heartfelt & positivity & thoughtfulness \\
awed & friendship & helpfulness & prayerful & thx \\
awestruck & generocity & honoured & proud & touched \\
blesed & generosities & humbled & respect & truely \\
blessed & generousity & humbleness & selflessness & undying \\
blessing & gift & immeasurable & sincere & unworthy \\
blessings & glad & inspiration & sincerity & upheld \\
blesssed & godsend & joy & stunned & uplifted \\
blesssing & gracious & joyfulness & supoort & wonderful \\
\hline
\end{tabular}
\caption{Appreciation Lexicon. A vector space model (word2vec) was trained on a CaringBridge.org dataset of 588,210 sites with \textasciitilde15 million journal updates. Seed words identified from 250 randomly sampled journal updates were used to generate 200 candidate words. Final validated lexicon includes 130 words rated at least "weakly related" or better by a majority of crowdworkers.}
\end{table}
}

\newcommand{\recruitsection}{
\begin{table}[H]
\centering
\begin{tabular}{|c|m{11.5cm}|}
\hline
\textbf{Strategy} & \multicolumn{1}{c|}{\textbf{Advertisement Text}} \\ \hline
\begin{tabular}[c]{@{}c@{}}Banner 1\\ (April 2-3)\end{tabular} & Are you following or experiencing a cancer journey? \textbf{<<Take this online survey>>} and help us understand the types of help that matter most to cancer patients and their communities! \\ \hline
\begin{tabular}[c]{@{}c@{}}Banner 2\\ (April 4-9)\end{tabular} & Are you following a friend or loved one on a cancer journey? Would you be willing to take a short survey to help us gather your ideas on ways to help? Please consider taking this one-time online survey. \textbf{<<Learn More.>>} \\ \hline
\begin{tabular}[c]{@{}c@{}}Facebook \\ Post\\ (April 3)\end{tabular} & Have you or someone you know been diagnosed with cancer? If so, we would love to hear about what types of help matter most to cancer patients and their caregivers. Please consider taking this one-time survey. \textbf{<<Link>>} \\ \hline
\begin{tabular}[c]{@{}c@{}}Email\\ Text\\ (April 5)\end{tabular} &  \textit{[At the end of a longer email about several different topics]} Are you or your loved one on a cancer journey? Do you want to share your opinion about the types of help you want or don't want along the way? So that CaringBridge can help families on health journeys connect with the most meaningful support, we invite you to consider participating in an online survey sponsored by a research team at the {[}BLINDED{]}. The survey takes between 10 and 15 minutes. \textbf{<<Learn More.>>}\\ \hline
\end{tabular}
\caption{Recruitment Strategies facilitated by CaringBridge in April 2018. CaringBridge shared the same anonymous survey link in all locations. To protect anonymity, no strategies targeted individual users.}
\label{tab:recruit_strategies}
\end{table}
}

\newcommand{\usefulregression}{
\begin{table}[t]
\begin{tabular}{lrll}
\hline
\multicolumn{1}{c}{\textbf{Regression Model}} & \multicolumn{2}{c}{\textbf{Independent Variable}} & \multicolumn{1}{c}{\textbf{Trend}} \\ \hline \hline
1. Sum of all INSTR & gender & $p=0.024$* & females rate higher usefulness \\
 & age & $p<0.001$*** & complex, no general trend \\
 & income & $p=0.021$* & lower income rate higher usefulness \\
2. FA only & gender &  $p=4.775$ & females rate higher usefulness \\
 & age & $p<0.001$*** & complex, no general trend \\
 & income & $p<0.001$*** & lower income rate higher usefulness \\
3. Sum, excluding FA & gender & $p=0.015$* & females rate higher usefulness \\
 & age & $p=0.001$** & complex, no general trend \\ 
 & income & $p=0.142$ & lower income rate higher usefulness \\ \hline \hline
\end{tabular}
\caption{Effects of Demographics on Patient/Caregiver Ratings of Instrumental Support. 
For three regression models, we present the three demographic variables that are statistically significant in at least one model.
All $p$-values are Bonferroni-corrected and are for tests of overall influence, not for tests of difference between certain groups and categories.
Full model details in supplemental section~\ref{sec:fullFAregression}.
}
\label{tab:usefulregression}
\end{table}
}

\newcommand{\surveyquestions}{
\subsection{Role Selection}\label{RoleSelection}
\subsubsection{All Roles.} \label{all_roles} Please tell us about the role(s) you have played in recent cancer journey(s). A "family caregiver" is not a nurse or doctor, but is rather a spouse, partner, parent, child, sibling or close friend of a cancer patient who makes sure that the patient has their needs met. People in a "friend, family, or acquaintance" role occasionally help out, but they are not primarily responsible for caregiving. 

Select all the role(s) that have applied to you within the last three years: [Cancer patient (any type of cancer) | \\ Family caregiver of a cancer patient | Friend, family, or acquaintance of a cancer patient | None of the above ]

\subsubsection{Survey Track Selection.}\label{role} Now please select just one role that you are willing to answer questions about: 

[ (Track A) Cancer patient (any type of cancer) | (Track B) Family caregiver of a cancer patient | 

(Track C) Friend, family, or acquaintance of a cancer patient ]

\subsection{Additional Question If Track B or C: Relationship and Remoteness to Patient}\label{RelationshipToPatient}
\subsubsection{Relationship to Patient}\label{relationship}
You selected [Family Caregiver | Friend, Family, or Acquaintance]. As you complete the survey, think of just one specific person [for whom you are a caregiver | whose cancer journey you are following]. Please answer questions [considering only your perspective as a caregiver for this person | while keeping this one person, along with their non-professional family caregiver(s)] in mind. 

What is your relationship to the cancer patient you are thinking of? (If more than one response applies, select the response that feels most important to you.): [ Spouse or partner |
Other family member or relative |
Best or close friend |
Friend or acquaintance (non-professional connection) |
Co-worker, colleague, or other professional connection ]

\subsubsection{Remoteness to Patient.}\label{driving} Is this cancer patient within reasonable driving distance for you to help out with their day-to-day needs? [Yes | No]
\begin{itemize}
\item If Yes: Approximately how long would it take you to drive to where the cancer patient you know is located (in minutes)? If you currently reside in the same household as the patient, please enter the number 0: [Numerical Input] 
\end{itemize}

\subsection{Tracks A and B, Parts 1 and 2: "Cancer Patient" and "Caregiver"}\label{TracksAB_12}
\subsubsection{High-Level Help Types.}\label{support_types}
As you navigate this cancer journey, CaringBridge wants to help you and your loved one(s)feel supported in ways that you want to be supported. 

How important to you are each of the support types listed below? [5-point unipolar Likert: "Not at all" to "Extremely" important]
\begin{itemize}
\item Recommendations for informative links or resources about cancer or cancer treatments 
\item Cards, phone calls, CaringBridge messages, or other social media support 
\item In-person visits and/or social activities 
\item Prayer support and/or positive energy and thoughts 
\item Physical or in-person assistance such as rides to appointments, meal deliveries, help with chores, childcare,  financial support, etc. 
\end{itemize}

\subsubsection{Frequency of Asking.}\label{ask_freq} Thank you for telling us what types of support are most important to you!

For this survey "real world help" refers to any type of physical or in-person assistance, such as rides to appointments, meal deliveries, help with chores, childcare, financial support, etc. For the rest of the survey, we will focus on real world help specifically, so that we can understand how CaringBridge might consider new features for patient's sites.

During [your | this] cancer journey...
\begin{itemize}
\item ...how frequently do you feel like you need any type of real world help to provide care for yourself and your loved one(s)? 
\item ...how frequently do you ask for real world help? 
\end{itemize}
[Almost Never |
More than once per year |
More than once per month |
More than once per week |
More than once per day]

\subsubsection{Difficulty of Asking.} \label{ask_diff} How challenging is it for you to ask friends and family for real world help... [5-point bipolar Likert: "Extremely difficult" to "Extremely easy"]
\begin{itemize}
\item ...in person? 
\item ...over the phone or text message? 
\item ...over email? 
\item ...over social media (such as Facebook or Twitter)? 
\item ...through CaringBridge?
\end{itemize}

\subsubsection{Challenges Related to Asking}\label{ask_challenge} What can make it challenging to ask for help sometimes? (Select all that apply.) [Multiple Selection : Binary]
\begin{itemize}
\item I'm a very private person. 
\item I lack trust in my community. 
\item I don't want to be a burden. 
\item I'm embarrassed to ask. 
\item It's not fair for me to ask. 
\item I don't think anyone is able to help. 
\item I don't know how to ask. 
\item I don't know whom to ask. 
\item None of the above. 
\end{itemize}

\subsubsection{Other Challenges?}\label{other_challenge} Are there other reasons that make it challenging to ask for help sometimes? (optional) [Free Text Input]

\subsubsection{Instrumental Support Types.}\label{useful_help}
Now we're interested in learning more about specific types of real world help. During [your | your loved one's] cancer journey, how useful is it for [you | you and/or your loved one(s)] to receive each of the following types of help? [5-point unipolar Likert: "Not at all" to "Extremely" useful]
\begin{itemize}
\item Transportation (e.g. a ride to the hospital or a flight) 
\item Childcare, eldercare, or special needs care 
\item Food (e.g. a homemade meal, a takeout delivery, or groceries) 
\item Overnight accommodations (e.g. a place to stay temporarily or long term) 
\item Practical goods (e.g. toys, games, books, clothes, jewelry, bedding, wigs, entertainment items, etc.) 
\item Personal care (e.g. spa services for well-being) 
\item Exercise (e.g. someone to help with regular physical activity) 
\item Household chores (e.g. cleaning, lawn mowing, pet care, maintenance) 
\item Financial assistance (e.g. cash, gift cards, bill payments, or donations for medical or life expenses)
\end{itemize}

\subsubsection{Other Instrumental.}\label{useful_other} Please list any other types of help that would be useful to you that we did not ask about. (optional) [Free Text Input]

\subsubsection{Concerns and Ideas.}\label{concerns} Before you move on to the next section, here's a chance to share your concerns and ideas for how CaringBridge might facilitate real world help.
\begin{itemize}
\item What are your biggest concerns with asking for real world help through CaringBridge? (optional) [Free Text Input]
\item In your opinion, what is the best way CaringBridge could make it easier for you to ask for real world help? (optional) [Free Text Input]
\end{itemize}

\subsection{Track C, Parts 1 and 2: "Friend, Family, or Acquaintance"}\label{TrackC_12}
\subsubsection{High-Level Help Types.}\label{support_types_ffa}
CaringBridge wants to provide compassionate ways for you to support cancer patients. First, please help us understand how important you feel it is for you provide different types of support.

How important is it to you to provide each of the support types listed below to the cancer patient you are thinking of? [5-point unipolar Likert: "Not at all" to "Extremely" important]
\begin{itemize}
\item Recommendations for informative links or resources about cancer or cancer treatments 
\item Cards, phone calls, CaringBridge messages, or other social media support 
\item In-person visits and/or social activities 
\item Prayer support and/or positive energy and thoughts 
\item Physical or in-person assistance such as rides to appointments, meal deliveries, help with chores, childcare,  financial support, etc. 
\end{itemize}

\subsubsection{Frequency of Asking.}\label{ask_freq_ffa} Thank you for telling us what types of support are most important to you!

For this survey "real world help" refers to any type of physical or in-person assistance, such as rides to appointments, meal deliveries, help with chores, childcare, financial support, etc. For the rest of the survey, we will focus on real world help specifically, so that we can understand how CaringBridge might consider new features for patient's sites.

Think about the cancer journey you are following. What is your impression of...
\begin{itemize}
\item ...how frequently the patient and/or their caregiver(s) need real world help?
\item ...how frequently the patient and/or their caregiver(s) actually ask for the help that they need?  
\end{itemize}
[Almost Never |
More than once per year |
More than once per month |
More than once per week |
More than once per day]

\subsubsection{Difficulty of Asking.} \label{ask_diff_ffa} How challenging is it for you to communicate about providing specific types of real world help to this cancer patient and/or their caregiver(s)... [5-point bipolar Likert: "Extremely difficult" to "Extremely easy"]
\begin{itemize}
\item ...in person? 
\item ...over the phone or text message? 
\item ...over email? 
\item ...over social media (such as Facebook or Twitter)? 
\item ...through CaringBridge?
\end{itemize}

\subsubsection{Challenges Related to Asking}\label{ask_challenge_ffa} How challenging does each of the following considerations make it for you to offer real world help to this cancer patient and/or their caregiver(s)? [5-point bipolar Likert: "Extremely difficult" to "Extremely easy"]
\begin{itemize}
\item Monetary cost 
\item Time 
\item Distance 
\item Social awkwardness 
\item Professional boundaries 
\end{itemize}

\subsubsection{Other Challenges?}\label{other_challenge_ffa} Is there anything else that makes it challenging to offer real world help? (optional) [Free Text Input]

\subsubsection{Instrumental Support Types: In-Person.} \label{in_person_ffa}
Now we're interested in learning more about specific types of real world help. For each of these real world help types, how interested are you in helping the patient and/or their caregiver(s)in person, using your own time?
 [5-point unipolar Likert: "Not at all" to "Extremely" interested]
\begin{itemize}
\item Transportation (e.g. you provide a ride to the hospital) 
\item Childcare, eldercare, or special needs care (e.g. you help out by babysitting) 
\item Food (e.g. a homemade meal) 
\item Overnight Accommodations (e.g. a stay in your own home) 
\item Practical goods (e.g. you drop off toys, games, books, clothes, jewelry, bedding, wigs, entertainment items, etc.) 
\item Personal care (e.g. you give them a massage or hair cut) 
\item Exercise (e.g. you help out as an exercise buddy) 
\item Household chores (e.g. you clean the house, mow the lawn, care for pets, etc.) 
\end{itemize}

\subsubsection{Other Instrumental.}\label{in_person_ffa_other} Please list any other types of real world help that you would be interested in providing in-person that we did not mention. (optional) [Free Text Input]

\subsubsection{Instrumental Support Types: Funding.} \label{funding_ffa}
How interested are you in helping this patient and/or their caregiver(s) by funding the following types of services? [5-point unipolar Likert: "Not at all" to "Extremely" interested]
\begin{itemize}
\item Transportation (e.g. a ride to the hospital by a paid driver, or a flight) 
\item Childcare, eldercare, or special needs care (e.g. a paid company) 
\item Food (e.g. a takeout delivery, or grocery delivery service) 
\item Overnight accommodations (e.g. you pay for their hotel or Airbnb) 
\item Practical items (e.g. you pay for an Amazon order or instant delivery from a local courier) 
\item Personal care (e.g. spa services at a spa or mobile spa) 
\item Exercise (e.g. you pay for a personal trainer to assist with regular physical activity) 
\item Household chores (e.g. you pay for a cleaning, yardwork, or handyman service to complete some chores for them) 
\item Financial assistance (e.g. cash, gift cards, bill payments, or donations for medical or life expenses) 
\end{itemize}

\subsubsection{Concerns and Ideas.}\label{concerns_ffa} Before you move on to the next section, here's a chance to share your concerns and ideas for how CaringBridge might facilitate real world help.
\begin{itemize}
\item What are your biggest concerns with providing real world help through CaringBridge, either in person or by hiring a paid service? (optional) [Free Text Input]
\item In your opinion, what is the best way CaringBridge could make it easier for you to provide real world help to cancer patients? (optional) [Free Text Input]
\end{itemize}

\subsection{Tracks A, B, and C, Part 3: Sharing Economy Prior Experience}\label{TracksABC_3}
\subsubsection{Sharing Economy Familiarity.} \label{app_aware} Thanks for sharing what types of help you're interested in. Now, we'd like to find out who you would trust to provide that help.

For example, a new technology trend is enabling people to use smartphone apps and/or websites to connect customers with people who provide paid services such as overnight accommodations in their own homes (e.g. Airbnb), rides in their own cars (e.g. Uber), household tasks like assembling furniture or yardwork (e.g. TaskRabbit), or grocery shopping (e.g. Instacart).

Are you generally familiar with app-based services, even if you do not use them yourself? [Yes | No] 
\begin{itemize}
\item If No: App-based companies allow people with their own cars, homes, or tools to provide paid services like giving rides, providing overnight accommodations, or fixing common household problems. The company usually performs a background check on workers and provides customers with a way to rate the quality of workers. That way, future customers can decide if they want to hire a specific worker or not.\\

For example, Uber is a popular sharing economy company. If a person has their own car, they can sign up to be an Uber Driver. Uber performs a background check on the person's criminal history and Motor Vehicle Record. The Uber smartphone app lets passengers request a ride. Then, the app finds a verified Uber Driver nearby, who will pick up the passenger and drive them to their destination. The experience is similar to a Taxi, but unlike Taxi Drivers, Uber Drivers do not need a special additional license, and they drive their own vehicles.
\end{itemize}

\subsubsection{Sharing Economy Usage.} \label{app_used} Have you ever used an app-based service for any reason? [Yes | No]

\subsubsection{Sharing Economy Opinion.} \label{app_opinion} What is your general opinion of these app-based services? (optional) [Free Text Input]

\subsubsection{If ~\ref{app_used} is Yes: Sharing Economy Usage for Cancer Journey.} \label{app_used_cancer} Have you ever used an app-based service to help with circumstances related specifically to this cancer journey?
\begin{itemize}
\item If Yes: Please describe how you used an app-based service along this cancer journey. (Include the name of the service if you can remember it.) [Free Text Input]
\end{itemize}

\subsection{Tracks A and B, Part 4: Trusted People or Businesses} \label{TracksAB_4}
For each of the following questions, please select ALL of the people or services that you would trust to provide this type of help. For example, you might select "No one at all", or you might select one or more of the other options.

Whom would you trust to provide [ category ] for you and/or your loved one(s)? 

[Note: Conditional logic was employed to display each response below only if this category was rated \textit{\textbf{greater than}} "Not at all useful" in question ~\ref{useful_help}.]
\begin{itemize}
\item transportation
\item childcare, eldercare, or special needs care
\item food 
\item overnight accommodations
\item practical goods, like toys, games, books, clothes, jewelry, bedding, wigs, entertainment items, etc. 
\item personal care services, like a massage, hair cut, manicure, etc. 
\item exercise or regular physical activity
\item household chores like cleaning, yard work, pet care, home maintenance
\item financial assistance with medical or life expenses
\end{itemize}
[Multiple Selection (Binary per option) : No one at all | Family member | Close friend | Friend or acquaintance | Coworker or colleague | An app-based service | A traditional business]

\subsection{Track C, part 4: Trusted People or Businesses} \label{TrackC_4}
If you were to fund services for your friend, how much do you trust traditional businesses (such as hotels or taxis) versus app-based businesses (such as Airbnb or Uber) to perform these services? [5-point bipolar Likert: "Strongly distrust" to "Strongly trust"]
\begin{itemize}
\item I trust traditional businesses to perform services for cancer patients. 
\item I trust app-based services to perform services for cancer patients. 
\end{itemize}

\subsection{Demographics}\label{demosection}
You're almost to the finish line! Don't stop now! Last but not least, we just need to gather demographic information.

\subsubsection{Gender.}\label{demo_gender}
What is your gender?
\begin{itemize}
\item Male 
\item Female 
\item Other 
\end{itemize}

\subsubsection{Age.}\label{demo_age}
What is your age?
\begin{itemize}
\item 18-24 years old 
\item 25-34 years old 
\item 35-44 years old 
\item 45-54 years old
\item 55-64 years old 
\item 65-74 years old 
\item 75 years or older
\end{itemize}
 
\subsubsection{Country.}
What country do you live in?
\begin{itemize}
\item USA 
\item Canada 
\item Other 
\end{itemize}

\subsubsection{Population Class.}\label{MSA}
What type of county do you live in?
\begin{itemize}
\item Big City or Suburb (county population > 1million) 
\item Medium City (county population between 50k-1million) 
\item Small Town or Rural (county population < 50k) 
\end{itemize}

\subsubsection{Education.}
What is the highest degree or level of school you have completed? (If currently enrolled, choose the highest degree you have received.)
\begin{itemize}
\item Did not complete high school 
\item High school graduate or equivalent (e.g. GED) 
\item Trade/technical/vocational training or associate degree 
\item Bachelor's degree 
\item Post-Bachelor's degree (Master's, PhD, MD, JD, etc.) 
\end{itemize}

\subsubsection{Employment.}
What is your current employment status?
\begin{itemize}
\item Employed (by self or for wages/salary) 
\item Unemployed 
\item Homemaker 
\item Student 
\item Military 
\item Retired 
\end{itemize}

\subsubsection{Income.}\label{demo_income}
What was your total household income before taxes during the past 12 months?
\begin{itemize}
\item Less than \$25,000 
\item \$25,000 to \$34,999 
\item \$35,000 to \$49,999 
\item \$50,000 to \$74,999 
\item \$75,000 to \$149,999 
\item \$150,000 or more 
\item Prefer Not to Say 
\end{itemize}

\subsubsection{Race.}
What is your race?
\begin{itemize}
\item White 
\item Hispanic or Latino 
\item Black or African American 
\item Native American or American Indian 
\item Asian / Pacific Islander 
\item Other 
\item Prefer Not to Say
\end{itemize}
}

\newcommand{\newusefulanova}{
\begin{table}[H]
\centering
\begin{tabular}{r|cc|cc|cc}
\toprule
\multirow{3}{*}{\begin{tabular}[r]{@{}c@{}} \textbf{Independent} \\ \textbf{(Demographic)} \\  \textbf{Variables} \end{tabular} }& \multicolumn{6}{c}{\textbf{Dependent Variable}} \\
 \cline{2-7}
 & \multicolumn{2}{c|}{\textbf{1. Sum of all INSTR}} & \multicolumn{2}{c|}{\textbf{2. FA only}} & \multicolumn{2}{c}{\textbf{3. Sum, excluding FA}} \\
 & \multicolumn{2}{c|}{\textit{Ordinary Least Squares}} & \multicolumn{2}{c|}{\textit{Ordered Logistic}} & \multicolumn{2}{c}{\textit{Ordinary Least Squares}} \\
 & $F$-value & $p$-value & Chi Square & $p$-value &$F$-value & $p$-value \\
 \hline
 Gender&8.865&0.003**&1.032&0.597&9.783&0.002**\\
 Age&5.423&$<$0.001***&28.670&$<$0.001***&4.572&$<$0.001***\\
 Country&1.132&0.323&2.111&0.348&1.084&0.339\\
 Population Class &0.683&0.506&0.914&0.633&0.940&0.391\\
 Education&1.780&0.131&9.253&0.055&1.479&0.207\\
 Employment&0.374&0.867&4.917&0.426&0.561&0.730\\
 Income & 3.406 &0.003**&39.144&$<$0.001***&2.585&0.018*
 \\
 Race&1.824&0.092&15.231&0.019*&1.976&0.067
 \\
 \hline  \hline
 Observations & \multicolumn{2}{c|}{576} & \multicolumn{2}{c|}{576} & \multicolumn{2}{c}{576} \\ R$^{2}$ & \multicolumn{2}{c|}{0.168} & \multicolumn{2}{c|}{\textit{N/A}} & \multicolumn{2}{c}{0.145} \\ Adjusted R$^{2}$ & \multicolumn{2}{c|}{0.119} & \multicolumn{2}{c|}{\textit{N/A}} & \multicolumn{2}{c}{0.095} \\ \begin{tabular}[r]{@{}r@{}}Residual Std. Error \\ (df = 543) \end{tabular} & \multicolumn{2}{c|}{6.951} & \multicolumn{2}{c|}{\textit{N/A}} & \multicolumn{2}{c}{6.191} \\ \begin{tabular}[r]{@{}r@{}} F Statistic \\ (df = 32; 543) \end{tabular} & \multicolumn{2}{c|}{3.427$^{***}$} & \multicolumn{2}{c|}{\textit{N/A}} & \multicolumn{2}{c}{2.889$^{***}$} \\
\bottomrule
\end{tabular}
\caption{ANOVA Tables of Three Regression Models Predicting Effects of SES on Instrumental Support Results. Model 1 used Linear Regression to predict the summative score of all 9 instrumental support categories. Model 2 used Ordered Logistic Regression to predict the score of Financial Assistance (FA) only. Model 3 used Linear Regression to predict the summative score of 8 instrumental support categories, excluding FA. We report \textbf{raw} $p$-values here, not corrected for multiple comparisons as in Table~\ref{tab:usefulregression}. Summary statistics such as R$^{2}$ do not apply to Ordered Logistic Regression; Model 2 has Residual Deviance of 1669.741 and AIC of 1741.741.}\label{tab:newusefulanova}
\end{table}
}

\newcommand{\fullFAregressionA}{
\begin{table}[H]
\centering
  \caption{Regression Results of Effects of Demographics on Patient/Caregiver Ratings of Instrumental Support}
  \label{tab:fullregressionA}
\begin{tabular}{@{\extracolsep{5pt}}lccc}
\\[-1.8ex]\hline
\hline \\[-1.8ex]
 & \multicolumn{3}{c}{\textit{Dependent variable:}} \\
\cline{2-4}
\\[-1.8ex] & Sum of all INSTR & FA only & Sum, excluding FA \\
\\[-1.8ex] & \textit{OLS} & \textit{ordered} & \textit{OLS} \\
 & \textit{} & \textit{logistic} & \textit{} \\
\\[-1.8ex] & (1) & (2) & (3)\\
\hline \\[-1.8ex]
  Intercept & 30.192$^{***}$ &  & 25.858$^{***}$ \\
  & (2.364) &  & (2.106) \\
  Gender: Female (vs. male) & 2.482$^{***}$ & 0.226 & 2.323$^{***}$ \\
  & (0.834) & (0.222) & (0.743) \\
  Age (young to old)  & $-$4.185 & $-$0.936 & $-$3.345 \\
  & (3.011) & (0.776) & (2.682) \\
  Country: Canada (vs. USA) & 3.057 & 0.292 & 2.838 \\
  & (2.265) & (0.599) & (2.017) \\
  Country: Other (vs. USA) & 2.001 & 1.137 & 1.186 \\
  & (2.908) & (0.848) & (2.590) \\
  Population Class (suburb to rural) & $-$0.196 & 0.103 & $-$0.255 \\
  & (0.544) & (0.145) & (0.485) \\
  Education (low to high) & $-$1.810 & $-$8.250$^{***}$ & $-$1.390 \\
  & (4.573) & (0.164) & (4.074) \\
  Employment: unemployed (vs. employed) & 0.013 & 0.087 & $-$0.094 \\
  & (1.339) & (0.359) & (1.192) \\
  Employment: homemaker (vs. employed) & 0.954 & 0.132 & 0.813 \\
  & (1.056) & (0.278) & (0.941) \\
  Employment: student (vs. employed) & $-$2.984 & $-$0.164 & $-$3.005 \\
  & (3.874) & (1.158) & (3.451) \\
  Employment: military (vs. employed) & 0.792 & 1.136 & $-$0.194 \\
  & (5.134) & (1.140) & (4.573) \\
  Employment: retired (vs. employed) & 0.655 & $-$0.442$^{*}$ & 0.955 \\
  & (0.951) & (0.253) & (0.847) \\
  Income: \$25000-\$34999 (vs. $<$\$25000) & $-$3.871$^{*}$ & $-$1.168$^{*}$ & $-$3.206 \\
  & (2.245) & (0.628) & (2.000) \\
  Income: \$35000-\$49999 (vs. $<$\$25000) & $-$3.735$^{**}$ & $-$1.690$^{***}$ & $-$2.720 \\
  & (1.860) & (0.553) & (1.657) \\
  Income: \$50000-\$74999 (vs. $<$\$25000) & $-$3.773$^{**}$ & $-$1.169$^{**}$ & $-$3.124$^{**}$ \\
  & (1.699) & (0.517) & (1.514) \\
  Income: \$75000-\$149999 (vs. $<$\$25000) & $-$3.778$^{**}$ & $-$1.607$^{***}$ & $-$2.828$^{*}$ \\
  & (1.651) & (0.508) & (1.471) \\

  \end{tabular}
\end{table} 
}

\newcommand{\fullFAregressionB}{
\begin{table}[H]
\centering
\begin{tabular}{@{\extracolsep{5pt}}lccc}
\\[-1.8ex]\hline
\hline \\[-1.8ex]
 & \multicolumn{3}{c}{\textit{Dependent variable:}} \\
\cline{2-4}
\\[-1.8ex] & Sum of all INSTR & FA only & Sum, excluding FA \\
\\[-1.8ex] & \textit{OLS} & \textit{ordered} & \textit{OLS} \\
 & \textit{} & \textit{logistic} & \textit{} \\
\\[-1.8ex] & (1) & (2) & (3)\\
\hline \\[-1.8ex]
  Income: $\geq$\$150000 (vs. $<$\$25000) & $-$5.965$^{***}$ & $-$2.456$^{***}$ & $-$4.397$^{***}$ \\
  & (1.724) & (0.527) & (1.536) \\
  Income: prefer not to say (vs. $<$\$25000) & $-$5.964$^{***}$ & $-$1.973$^{***}$ & $-$4.732$^{***}$ \\
  & (1.670) & (0.512) & (1.488) \\ 
  Race: Hispanic or Latino (vs. White) & 1.181 & $-$0.472 & 1.678 \\
  & (2.167) & (0.561) & (1.930) \\
  Race: African American (vs. White) 
  & $-$0.394 & $-$0.255 & $-$0.239 \\
  & (2.899) & (0.781) & (2.582) \\
  Race: Native American (vs. White) 
  & 6.854 & 14.567$^{***}$ & 5.283 \\
  & (5.022) & (0.00000) & (4.473) \\
  Race: Asian/Pacific Islander (vs. White) & 2.718 & $-$1.137$^{*}$ & 3.410 \\
  & (2.496) & (0.649) & (2.223) \\
  Race: Other (vs. White) & 5.887$^{**}$ & 1.679$^{**}$ & 5.119$^{**}$ \\
  & (2.390) & (0.736) & (2.129) \\
  Race: prefer not to say (vs. White) & 1.985 & 0.289 & 1.848 \\
  & (1.365) & (0.372) & (1.216) \\
 
 \hline \\[-1.8ex]
Observations & 576 & 576 & 576 \\
R$^{2}$ & 0.168 &  & 0.145 \\
Adjusted R$^{2}$ & 0.119 &  & 0.095 \\
Residual Std. Error (df = 543) & 6.951 &  & 6.191 \\
F Statistic (df = 32; 543) & 3.427$^{***}$ &  & 2.889$^{***}$ \\
\hline
\hline \\[-1.8ex]
\textit{Note:}  & \multicolumn{3}{r}{$^{*}$p$<$0.1; $^{**}$p$<$0.05; $^{***}$p$<$0.01} \\
\end{tabular}
\end{table} 
}

%
%



\maketitle

\renewcommand{\shortauthors}{C.E. Smith et al.}

\input{body.tex}
\newpage
\input{appendix.tex}

\end{document}

%% file: body.tex
\section{Introduction}

\begin{quote}\textit{"I need help getting my kids from one event to the next. It was a challenge before, but now that I have multiple doctor's appointments every week, it has gotten much more tricky. I need help getting food in the house and keeping the pantry stocked. I need help remaining calm and respectful toward the hospital staff who tell me that the chemo they will inject could possibly give me a rare form of incurable cancer. ... So you can see I am struggling. I am not at all used to asking for help. I know that I cannot do all of this alone." \begin{flushright} \textasciitilde Anonymous Cancer Patient \end{flushright}} \end{quote} 

Written by a user of an online health community called CaringBridge (\url{www.caringbridge.org}), this passage reflects the challenging reality of day-to-day life with a cancer diagnosis. The sheer number of people facing such a reality is staggering: over 18 million new cancer cases were estimated to occur globally in 2018, with projected estimates continuing to rise~\cite{fund_worldwide_2012}. Life-threatening diagnoses disrupt people's ability to care not only for themselves, but also for their families and loved ones, especially when patients and family caregivers have differing value systems \cite{berry_how_2017}. Even for cancer survivors, symptoms like fatigue, depression, pain, or feelings of isolation can persist beyond the end of treatment, negatively impacting health-related quality of life, employment, and body image~\cite{wu_symptom_2015,jacobs_cancer_2016,eschler_im_2017}. As the cancer burden rises, so too does the importance of developing effective sociotechnical tools for helping cancer patients and their communities.

Technology offers great promise for supporting people affected by life-threatening health issues such as cancer. In contrast to medical support provided by healthcare professionals, social support refers to ``an exchange of resources between two individuals perceived by the provider or the recipient to be intended to enhance the well-being of the recipient'' \cite{shumaker_toward_1984}. A large body of work has explored how online health communities (OHCs) provide social support to people with a variety of health and wellness problems such as cancer, substance abuse disorders, or dementia (e.g. \cite{meier_how_2007,rubya_video-mediated_2017,glueckauf_online_2004}). 
Analyses of OHCs have usually considered three dimensions of social support, i.e. emotional, informational, and instrumental support \cite{cohen_social_2004,wills_supportive_1985}), and found evidence that OHC users primarily seek and provide informational and emotional support (e.g. \cite{meier_how_2007,yang_commitment_2017,ma_write_2017}). However, most OHCs are composed of strangers connecting with each other over the Internet.  As a journaling platform, CaringBridge provides a central online location for patients and family caregivers to share ongoing personal stories and health updates. Most CaringBridge users who follow a patient's site are acquainted with the patient through pre-existing offline groups, such as friends and family members, workplaces, or spiritual and religious organizations~\cite{anderson_uses_2011,ma_write_2017}; this study context affords new research opportunities to understand online social support during health crises in hybrid online/offline communities where peoples' priorities may collide in interesting ways.

\cbfig

In the context of CaringBridge, the present work aims to understand what OHC users acknowledge and prioritize in terms of social support as they navigate their own experiences of living---and sometimes dying---with a cancer diagnosis. We take a special focus on instrumental support---``the provision of material aid, for example, financial assistance or help with daily tasks''~\cite{cohen_social_2004}---in consideration of the reality that OHCs are now building or integrating features for instrumental support. For example, CaringBridge has developed an integration with GoFundMe (\url{www.gofundme.org}) in order to provide users with access to financial support through personal medical crowdfunding~\cite{caringbridge_caringbridge_nodate}. Some sites designed for patients and caregivers even exist \textit{primarily} for instrumental support. For example, Lotsahelpinghands (\url{lotsahelpinghands.com}) and Mealtrain (\url{www.mealtrain.com}) are online calendar tools that allow friends and family members to organize meal drop-offs or other instrumental forms of support for patients and caregivers. CaringBridge also offers a calendar-based ``Planner'' tool for authors to post instrumental help requests, e.g. for meals, rides, etc.  
Yet there remains a gap in our understanding of what types of instrumental support-related technological interventions or features that patients and caregivers might \textit{want} or \textit{prefer} during a cancer experience, as well as how patient and caregiver preferences compare to those of their friends, family, and acquaintances who are in a position to \textit{provide} the support.

To address this gap, we take an exploratory mixed methods approach in two phases. The first phase is a directed content analysis of over 600 CaringBridge journal updates written by cancer patients and family caregivers; its purpose is to assess whether categories used in prior literature suit the data, and to derive appropriate community-tailored language for building a survey instrument of high ecological validity. The second phase includes survey design, deployment, and analysis, based on categories from the first phase. Nearly 1,000 users participated in our \textasciitilde15-minute survey.

The results of our content analysis quantitatively illustrate what CaringBridge users appreciate receiving help with, as measured through their online writing. Our survey results, on the other hand, show how users rate the importance and usefulness of the different help types they write about in their journals. Our survey provides evidence that patients and caregivers differ from each other, as well as from their support networks, in terms of what types of instrumental support they find to be most useful vs. what their networks want to provide; these findings can inform the design of features intended to help users provide instrumental support to patients and caregivers. 

Another finding of our work is the unanticipated emergence and dominance of a ``prayer support'' category.  When we began, we did not seek to measure prayer support, as this category mostly did not appear in prior OHC literature (see \cite{flickinger_social_2017} for exception). However, we realized during codebook development that a prayer support category was necessary to accurately describe the data. Our content analysis shows that CaringBridge authors acknowledge prayer support more frequently than any other type of support. Furthermore, our survey shows that patients and caregivers both rate prayer support to be most important. Our discussion builds upon these empirical findings to suggest design implications and new directions for future work.

\section{CaringBridge Research Collaborative}
\label{collab}
This work proceeds from a research collaboration between CaringBridge and an interdisciplinary team from the College of Computer Science and Engineering, School of Nursing, and Earl E. Bakken Center for Spirituality \& Support at the University of Minnesota.\footnote{Note that the boilerplate text contained in this section may be identical, or nearly identical with small project-dependent adjustments, to text appearing in other works that result from this collaboration, such as \cite{ma_write_2017}.} CaringBridge is a 501(c)(3) non-profit organization that was established in 1997 with the objective of using ``compassionate technology'' to enhance social connectedness. CaringBridge brings people together in an online social network to help overcome the isolation often experienced with a health crisis. In 2019, CaringBridge served nearly 300,000 people daily and over 40 million unique users annually from 237 countries and territories around the world.\footnote{These figures were acquired through email correspondence with CaringBridge leadership.}

\subsection{Platform Description}\label{definitions}
In contrast to OHCs where patients primarily seek informational support from strangers~\cite{yang_commitment_2017}, CaringBridge primarily brings together users with existing social ties to a patient~\cite{anderson_uses_2011}. This paper adopts the same terminology used by Ma et al. in prior work; see~\cite{ma_write_2017} for detailed definitions. CaringBridge is differentiated from other OHCs by its primary tool, an online ``journal.'' Rather than incidental questions or comments to a forum or thread, the journal resembles a personal blog for a person who is either the person who is ill and/or their family caregiver. The journal contains a collection of self-narrated and successive health ``updates'' arranged in a timeline. Figure~\ref{fig:CB} demonstrates the appearance of a journal update in 2018. 

CaringBridge journals can be authored by patients, family caregivers, or both. CaringBridge authors often begin a site after receiving a diagnosis or having a serious accident, surgery, premature birth, etc. They can then invite friends, family, and acquaintances to follow the site and stay up-to-date on their health status. CaringBridge authors control their site's privacy level, and can choose to make the site visible \textit{only} to invited visitors, to any registered CaringBridge users, or to anyone online. Allowed visitors can offer support by posting comments on individual journal updates, or they can post well wishes and photos to the site. Visitors can choose to ``follow'' an individual's CaringBridge site in order to receive notifications whenever a new journal update is posted. Through this mechanism, CaringBridge sites often become the \textit{de facto} online location where members of the patient's professional, school, hobby, or spiritual community can go to find out how they are doing. Some CaringBridge sites may become more widely known, especially in the case of celebrities, or people of high visibility within some larger offline community---e.g. a pastor writing about their condition for the congregation. 

Because our work seeks to measure the needs of patients and family caregivers, we next provide context about authorship trends on the CaringBridge sites sampled in this paper. We trained a classifier to distinguish patient- vs. non-patient authored journal updates and achieved 96\% accuracy on a validation set (see supplemental section~\ref{author_ML}). We define a site to be ``independently authored'' when $\geq$90\% of updates are predicted as either patient or non-patient (i.e. a patient's caregiver or other social connection) but not both, and ``collaboratively edited'' when $<$90\% are independently authored. By this definition, 45.1\% of sites are collaboratively edited, 15.5\% are independently authored by patients, and 39.4\% are independently authored by non-patients. Thus, caregivers are more likely to independently author a site; on average, caregivers also post a greater proportion of journal updates. Although posting on CaringBridge may add some labor to general caregiving burdens, maintaining a site can also help to reduce stress, time, and chaos by reducing the effort required to update people in the patient's support network one person at a time.




\subsection{Data Description and Ethical Considerations}
 Conducted with permission of the CaringBridge leadership, this study was reviewed and deemed exempt from further IRB review by the University of Minnesota Institutional Review Board. The complete dataset used for this analysis includes de-identified information about 588,210 CaringBridge sites and 22,333,379 users between June 1, 2005 and June 3, 2016. 
We acknowledge the tension in HCI-related fields between open data dissemination~\cite{hornbaek_is_2014} and the ethical necessity to protect participants' rights and privacy~\cite{bruckman_cscw_2017} given the imperfection of de-identification techniques~\cite{narayanan_myths_2010}. Due to the latter two considerations, we cannot publicly release the dataset used for analysis in this paper. CaringBridge data are highly sensitive, and were acquired with the permission and collaboration of CaringBridge leadership in accordance with CaringBridge's Privacy Policy \& Terms of Use Agreement. In compromise between replicable science and the priorities of ethical protection of participants, we welcome inquiries about the dataset or the project by contacting the investigators who conducted this study, or CaringBridge directly. 

\section{Related Literature}
Prior work has frequently examined how OHCs can provide social support throughout many health crises, e.g., cancer~\cite{meier_how_2007}, substance use disorders~\cite{rubya_video-mediated_2017}, mood disorders~\cite{lamberg_online_2003}, 
and dementia~\cite{glueckauf_online_2004}. Communities for cancer patients are often forum-based; users create accounts and post questions or responses to forum threads, typically without any personal acquaintance to the vast majority of the other community members (e.g.~\cite{meier_how_2007,vlahovic2014support,yang_commitment_2017}). Recent work also explores well-being and mental health in non-health specific online communities like Tumblr~\cite{chancellor_recovery_2016}, Instagram~\cite{andalibi_sensitive_2017}, and Reddit~\cite{andalibi_social_2018} or sensitive disclosures (e.g. pregnancy loss~\cite{andalibi_social_2018}, sexual abuse~\cite{andalibi_responding_2018}, schizophrenia~\cite{ernala_linguistic_2017}, death~\cite{jiang_tending_2018,gach2020experiences}) on platforms like Facebook, Twitter, and MySpace. OHCs can offer benefits like ease of access without barriers related to time and location, social distance that allows more openness, and plentiful resources from others with similar experiences~\cite{walther2002attraction,wright2000perceptions}. Prior work on CaringBridge has explored factors related to expressive writing~\cite{ma_write_2017} and user behaviors related to cancer phases~\cite{levonianbridging}, however our work here characterizes the types of support that CaringBridge users appreciate receiving help for, focusing on instrumental and prayer support. Therefore, we next summarize prior work on social support and provide interdisciplinary context from the research on spirituality and healing. 


\subsection{Social Support in Online Health Communities}\label{sec:socialsupport}
Different online communities where people seek support for health-related conditions are perceived and used in different ways by different groups of people. For example, Facebook users may be impacted by ``positivity bias'' and view the platform as a place that is more suited to positive than negative news \cite{vitak_``you_2014}, which may make it more difficult and nuanced for people who need help to make sensitive disclosures and ask for support \cite{andalibi_sensitive_2017}. On the other hand, condition-specific Facebook groups (rather than Facebook's Newsfeed feature), or other OHCs, may offer better sources of support than even family or friends, especially for people afflicted by rare illnesses~\cite{macleod_be_2017}. To younger cancer patients (e.g., in their 20's and 30's), Facebook may be considered an accessible tool for sharing information, but only according to carefully considered boundaries that protect peoples' self image and personal information~\cite{eschler_im_2017}. For some younger users, forum-based OHCs or CaringBridge may even feel like foreign online spaces which are not easy to integrate into their lives~\cite{eschler_im_2017}. Furthermore, not all online communities are actually supportive to peoples' recovery from a health condition. For instance, some Tumblr and Instagram communities focused on eating disorders are oriented towards recovery and cessation of symptoms, whereas others continue to encourage symptomatic behaviors and views of the self~\cite{chancellor_recovery_2016, andalibi_sensitive_2017}. Most online communities that are specifically tailored towards health problems (e.g. OHCs for chronic conditions like cancer or diabetes) do not attract users who come to promote symptomatic behavior, but rather to help provide support towards healing \cite{chancellor_recovery_2016,sharma_mental_2018}.

Apart from these considerations, online communities may provide intrinsic benefits to users due to the therapeutic effects of expressive writing, which has been shown to have positive outcomes both offline (e.g. \cite{baikie_emotional_2005}) and online. For example, \cite{ernala_linguistic_2017} showed that following disclosure of a schizophrenia diagnosis on Twitter, measures of users' wellness improved in a manner consistent with expert psychiatric evaluation, while \cite{ma_write_2017} showed that increased expressive writing contributes to increased user engagement on CaringBridge. However, most prior work has focused on how users derive social support from \textit{others} in their online communities, often by directly or indirectly asking for it. Sensitive Interactions Systems Theory (SIST) describes ``direct'' help-seeking as explicit requests for help or specific descriptions of problems so that others can provide support, whereas ``indirect'' help-seeking involves complaints or hints that a problem exists without actually asking for help~\cite{barbee_experimental_1995}. 
Furthermore, receiving social support may be most effective for users when the type of support received is consistent with their actual needs \cite{mattson_health_2011}. Using these concepts, several recent works have examined the content and linguistic features of people's direct or indirect support-seeking behaviors and measured the different quantity or types of support that those behaviors receive in return, in the form of online comments or written responses \cite{vlahovic2014support, sharma_mental_2018, andalibi_social_2018}.

Different conceptual frameworks exist for understanding social support, such as the five categories presented by Cutrona and Suhr. Social support is considered across two ``action-facilitating'' help types---i.e. informational (resources, links, or knowledge about diseases, treatment options, and symptoms) and instrumental support (material aid and financial assistance)---and three ``nurturant'' help types---i.e. emotional support (encouraging affirmations and messages of care, concern, and empathy), network support (a sense of belonging in a particular network), and esteem support (regarding a person's intrinsic value and abilities) \cite{cutrona_controllability_1992}. Many works exploring social support use some or all of these categories. For example, in an analysis of Reddit, Andalibi et al. found that direct (rather than indirect) requests for informational, esteem, and instrumental support were more likely to receive comments including references to those types of support, and that users tended to provide emotional and network support in the comments, regardless of whether they were asked for~\cite{andalibi_social_2018}. Most prior work specific to OHCs (rather than general purpose social media) primarily examines emotional, informational, and less consistently, instrumental support ~\cite{cohen_social_2004,wills_supportive_1985,sharma_mental_2018,choi_toward_2017}. Therefore, we initially designed our methods to study these three categories. We acknowledge the limitation that our study does not examine all possible support categories (esp. network and esteem support), thus we cannot draw conclusions about their relative importance to CaringBridge users.

Many studies of social support in OHCs have used content analyses to show that a higher proportion of discussion or online writing relates to informational and emotional support. For example, Meier et al. found that emails sent to the Association of Cancer Online Resources primarily seek informational support~\cite{meier_how_2007}. Frost and Massagli found that users of the PatientsLikeMe OHC search for patients similar to them, primarily for informational support~\cite{frost_social_2008}. Civan and Pratt found informational, and next emotional, support were most prevalent in threads posted to three breast cancer OHCs, with a much smaller percent of threads discussing instrumental support~\cite{civan_threading_2007}. In HIV-affiliated support groups, Flickinger et al. found emotional (41\%), network (27\%), esteem (24\%), informational (18\%), and instrumental (2\%) support~\cite{flickinger_social_2017}, while Coursaris and Liu found informational (42\%), emotional (16\%), network (6.8\%), esteem (6.4\%), and instrumental (1\%) support~\cite{coursaris_analysis_2009}. Apart from OHCs, Jacobs et al. studied patients' information-sharing practices through a seven-month deployment study, concluding that more resources should be designed for ``personalized support systems for other health situations with complex information access models''~\cite{jacobs_mypath:_2018}. Examining top Reddit posts about cancer, Eschler et al. found that different self-identified illness phases are accompanied by different informational needs~\cite{eschler_self-characterized_2015}. Eschler and Pratt also found that young cancer patients are often overwhelmed in their efforts to find information, and other patients whose experiences are relevant to them~\cite{eschler_im_2017}. These studies suggest that dealing with information overload and getting emotional support are crucial parts of managing a health crisis, and are also popular topics of conversation in OHCs---but what about instrumental support?
 
Instrumental support has recently been studied apart from OHCs. For example, Rho et al. found that an online community for supporting low income students at elite universities (who often feel isolated and stigmatized) afforded users a useful way to anonymously request help getting resources like food and housing~\cite{rho_class_2017}. Wohn et al. studied the motivations of viewers of live streaming masspersonal platforms for contributing instrumental and financial support to streamers~\cite{wohn_explaining_2018}. Freeman and Wohn studied eSports communities and showed that in-game actions of instrumental support could not only lead to expressions of emotional and esteem support online, but could also evolve into in-person relationships, which is important because they also found that lonely people are more likely to provide intentions of instrumental support~\cite{freeman_social_2017}. 

In the OHC context however, where instrumental support is a much less prevalent topic of online conversation (possibly because users may experience shame or embarrassment to publicly ask for instrumental support when their identity is known~\cite{rho_class_2017,kim_not_2017}), many studies understandably do not aim to operationalize or identify instrumental support. For example, a survival analysis of the American Cancer Society's Cancer Support Network by Yang et al. showed that informational support-seeking occurs more frequently than emotional support-seeking~\cite{yang_commitment_2017}, yet instrumental support was not studied. However, most OHCs connect \textit{strangers} with similar medical experiences. On the other hand, CaringBridge connects patients with their existing social networks. Consistent with~\cite{wang_stay_2012,wang_social_2014}, Ma et al. conducted a survival analysis of CaringBridge and showed that receiving emotional support (i.e. visitors' comments) correlated strongly with long-term user engagement~\cite{ma_write_2017}. However, Ma et al. measured ``emotional support'' in terms of \textit{quantity} of supportive messages to and interactions with online journals, without looking at the \textit{content} of either authors' writings or the messages they received in response, and they did not examine any measures of informational or instrumental support. Therefore, we do not know what types of help CaringBridge users write about, or how those categories compare to other OHCs. RQ1 directly addresses this gap:

\begin{description}
\item [RQ1a:] \textit{What kinds of support do CaringBridge journal authors positively acknowledge in online writing about cancer experiences, and what is their relative prevalence?}
\item [RQ1b:] \textit{Do patient and caregiver authors differ in terms of the types of help they write about appreciating?}
\end{description}

To address RQ1, the first phase of our study is a directed content analysis that provides a codebook of help types and a measure of what support types users have written about receiving online. This contextualizes the rest of our study by situating instrumental support against other types of support. However, traces of online behavior have the intrinsic limitation that peoples' behaviors may or may not align with peoples' internal needs, priorities, or values. Therefore, further inquiry is required to understand how people prioritize different types of instrumental support.


\subsection{Instrumental Support for Patients and Caregivers}\label{sec:instrumental}
Even though instrumental support is \textit{discussed} less frequently online, a body of qualitative work in HCI provides evidence that instrumental support is crucial to cancer patients and their communities; we rely on these studies as a starting point for developing our phase one codebook of instrumental support types. For example, Jacobs et al. showed, through a yearlong mobile tablet deployment, that technology for cancer patients should be customizable to patients' goals and values, integrating many different types of medical, non-medical, and instrumental support resources~\cite{jacobs_lessons_2015}, and through focus groups, that experiences of cancer are marked by a wide range of differing needs, dependent on the stage of the illness~\cite{jacobs_cancer_2016}. Skeels et al. conducted participatory design sessions with breast cancer survivors, seeking to understand what types of help would have been useful during the timeframe of cancer treatment and recovery, providing specific examples of medical-related tasks, everyday chores, coordination tasks, and ``other'' tasks~\cite{skeels_catalyzing_2010}. 

Apart from specific tasks or material goods, financial support is also considered to be a form of instrumental support~\cite{cohen_social_2004}. Cancer is known to be financially devastating, especially for low-income patients or those without insurance~\cite{zafar_financial_2013}. Thus some prior work has looked at medical crowdfunding online---e.g. examining factors that affect the success of crowdfunding campaigns~\cite{rhue_emotional_2018}. Unfortunately, campaign beneficiaries may fear judgement for asking for money, and their friends may feel pressured to contribute funds they cannot truly afford~\cite{kim_not_2017}. For these reasons, Kim et al. suggest that medical crowdfunding campaigns would be improved by also allowing people to contribute support through \textit{non}-monetary ways offline, specifically pointing to CaringBridge as a promising place for coordinating instrumental support that is not necessarily financial~\cite{kim_not_2017}.

In the medical literature, some work has studied correlations between social support received and mortality rates, with mixed results (e.g.~\cite{thong_social_2007,cousson-gelie_anxiety_2007}). One study found that receiving social support had no effect on mortality, but that \textit{providing} instrumental support to others significantly reduced mortality~\cite{brown_providing_2003}. Another study of Alzheimer patients' satisfaction with social support reported that 96\% of spouse-caregivers received no financial support, 65\% received no instrumental support, and 30\% received it insufficiently~\cite{drentea_predictors_2006}. If patients and caregivers need instrumental support, and are receiving it insufficiently, how might technology help to close the gap? Whereas most studies focus on patient and caregiver perspectives, our study context allows us to capture and compare patient and caregiver perspectives against those of their friends, family, and acquaintances. This is especially salient, given that patients expect their friends and families---rather than strangers online---to assist with their instrumental needs~\cite{macleod_be_2017}. RQ2 helps us to evaluate the most promising directions for instrumental technology interventions:

\begin{description} 
\item [RQ2a:] \textit{What kinds of instrumental support are most useful to patients and caregivers on CaringBridge? Do patients and caregivers align in their perceptions of what is useful?}
\item[RQ2b:] \textit{What kinds of instrumental support are CaringBridge users who are friends, family, or acquaintances most interested in providing? How does this compare to patients and caregivers?}
\end{description}

\subsection{Possible Instrumental Technology Interventions for Patients and Caregivers}\label{sec:sharingecon}
One of the major benefits of CaringBridge is that it allows people who are either geographically nearby \textit{or} remote to the patient to maintain connections and provide meaningful social support. However, most forms of instrumental support in particular are challenging or impossible to facilitate purely online. Instrumental support can require material resources such as a car, physical goods, money, or potentially a larger time commitment than it takes to write online messages. These requirements introduce logistical challenges, and they impose a heavier burden on the people who are in a position to provide instrumental support---especially those who are closer to the patient geographically. Prior literature points at two possible strategies for using technology to mediate instrumental support: friendsourcing (which could help incentivize new or more people co-located with the patient to provide instrumental support) and/or providing new technical ways to hire paid services, e.g. through the ``gig'' or ``sharing economy.'' These solutions could offer remote supporters new ways to provide instrumental support, and possibly alleviate the heavy burdens placed on caregivers and those who are geographically closest.

Friendsourcing is a form of crowdsourcing within a socially-connected group of individuals that uses ``motivations and incentives over a user's social network to collect information or produce a desired outcome''~\cite{bernstein_personalization_2010}. For example, Bernstein et al. designed a game on Facebook that allows a person's friends to create fun and informative tags about them~\cite{bernstein_personalization_2010}. These tags can enhance peoples' profiles by adding new information about them that is of social interest, and players are rewarded with points for tagging new friends to join the game. The game was able to collect accurate personal information about active users~\cite{bernstein_personalization_2010}. Similarly, Martins et al. developed a friendsourcing application for patients that collects and verifies information for people with dementia~\cite{martins_friendsourcing_2014}. To be effective, such systems face challenges with recruiting a large enough support network~\cite{bernstein_personalization_2010,martins_friendsourcing_2014} without violating patients' privacy~\cite{martins_friendsourcing_2014}. In the context of cancer, friendsourcing technology for instrumental support could provide online mechanisms to organize and incentivize users to achieve instrumental rather than informational outcomes, such as providing patients and caregivers with meals, rides, childcare, etc. However, without specific affordances to bridge the online and offline worlds, this solution primarily applies to people who are geographically co-located with the patient. 

Paid services, on the other hand, can generally be provided by anyone with access to pay for them online or over the phone, regardless of geographic proximity to the patient. Sharing/gig economy, collaborative consumption, peer-to-peer exchange, and on-demand service technologies may be well-suited to provide instrumental support since they connect the offline and online worlds by matching people who own goods or offer services to others who need such goods or services~\cite{botsman_defining_2015,dillahunt_sharing_2017}. As advised by CaringBridge leadership, we decided to use the term ``app-based'' in the language of our survey because this term improves accessibility of the language to CaringBridge users, and it can also be used to refer to the full breadth of services in this space. We will also use the term ``app-based'' in the paper as well, in order to maintain consistent language throughout. 

Some app-based services are already being used to provide patients and caregivers with instrumental support. For example, Lyft.com (a ride-sharing company that allows drivers to accept passengers in their own cars) has developed partnerships with healthcare organizations to help patients overcome transportation obstacles for medical appointments~\cite{larock_lyft_2009,japsen_lyft_nodate}. Airbnb.com (a travel company that allows property owners to hosts guests in their properties) has developed programs for reduced cost stays for patients and caregivers during medical treatments~\cite{airbnb_airbnb_2019}. Yet these technologies also face problems related to trust and safety~\cite{mclachlan_you_2016,dillahunt_promise_2015}, geographical disparities in availability or willingness to participate~\cite{thebault-spieker_avoiding_2015,thebault-spieker_toward_2017}, and other issues like achieving critical mass and establishing a belief in the commons~\cite{dillahunt_promise_2015,botsman_whats_2011}. A systematic literature review on the sharing economy suggested that most of the work in HCI has provided qualitative insights related to sociotechnical design, motivations, and social relationships and community, but that little work has explored special contexts of use, and that quantitative evidence is lacking to understand how qualitative themes play out at scale~\cite{dillahunt_sharing_2017}. Thus, little is known about peoples' attitudes or preferences towards the use of app-based services in a health context. 

Patients and caregivers already have elevated concerns and risks related to privacy, safety, personal information sharing, and misleading health information online~\cite{hartzler_sharing_2011,newman_its_2011,rahman_prisn:_2014,jacobs_comparing_2015,skeels_sharing_2010,mankoff_competing_2011}. Trust is also a factor that affects sensitive disclosures and use of technology such as social media and OHCs during health crises (e.g.~\cite{joinson_self-disclosure_2009,balani_detecting_2015}). Trust is an overloaded term that can mean a great variety of things, and the academic community does not have a single definition that can be used in every context~\cite{mcknight_meanings_1996}. In HCI, the term has been used in terms of trust in platforms or networks~\cite{fogel_internet_2009,leskovec_signed_2010,gach2020experiences}, predicting strength of social ties~\cite{gilbert_predicting_2009} or the trustworthiness of people or information~\cite{kittur_can_2008}. Rather than these types of trust, this work aligns most closely with Hardin in considering trust as a three-part relation in which ``A trusts B to do X''~\cite{hardin2001conceptions}, and the product of the trust is a specific action. In our context, A are patients or caregivers, B are members of their support network, and X are a variety of instrumental needs. Peoples' willingness to share sensitive or private information depends on their differential trust of people at differing levels of social closeness~\cite{olson_study_2005}, but there is a gap in our knowledge of peoples' trust in others to provide them with instrumental support. This work therefore provides a benchmark measure of OHC users' trust in a variety of people and businesses that might provide them with instrumental support. This is important because it can help designers understand which features might be most acceptable to OHC users.

\begin{description} 
\item [RQ3:] \textit{Whom do CaringBridge users trust to provide cancer patients and caregivers with instrumental support?} \end{description}

\subsection{Spiritual Care and Prayer Support}\label{sec:spiritualsupport}

Our initial study design was intended to address RQ1-3. However, the overwhelming prevalence of prayer in CaringBridge journals led us to add a post-hoc RQ4 in order to better communicate our most prominent result. Therefore, we conclude with background information from the literature on spirituality and healing. 

During serious or terminal illnesses, patients often desire spiritual care in addition to physical and psychosocial care~\cite{puchalski_improving_2009,xing_are_2018}. Furthermore, spiritual interventions may improve quality of life, and reduce depression, anxiety and hopelessness~\cite{xing_are_2018}. Thus it is considered the responsibility of health care professionals to assess and respond to patients' spiritual care needs~\cite{institute_of_medicine_dying_2015,ferrell_national_2018}. Moreover, the Joint Commission (which accredits US healthcare organizations) has established standards for spiritual assessment which mandate that every patient's spirituality and spiritual needs be assessed, accommodated and attended to~\cite{joint_commission_spiritual_2019}. However, many peoples' expectations for spiritual support are not currently perceived to be met by medical professionals~\cite{kalish_evidence-based_2012,phelps_addressing_2012,siler_interprofessional_2019}.
  
``Spiritual support'' is a broad term that means many different things to many different people. In this study of CaringBridge, we specifically observed a high prevalence of prayer, which should be considered as a sub-category of the broader concept of spiritual support. CaringBridge authors frequently include ``prayer requests'' (e.g. \textit{``Please pray for me.''}) and expressions of appreciation for prayer (e.g. \textit{``Thanks for all the prayers and warm thoughts.''}) in their journal updates. Our codebook defines prayer support as ``prayers, spiritual blessings, positive karma, good juju, warm thoughts.'' While Flickinger et. al~\cite{flickinger_social_2017} categorize ``prayers'' as a sub-category of emotional support, our survey result that CaringBridge users rate it differently from emotional support provides evidence that users experience and perceive prayer support to be distinct from emotional support. Under the typology of Cutrona and Suhr~\cite{cutrona_controllability_1992}, we suggest that prayer support be categorized as a distinct form of nurturant support, since it is intended to improve the wellbeing or spiritual/energetic state of the recipient, but does not have a material outcome as do action-facilitating forms of support. Rather, prayer support appeals to peoples' religious or spiritual beliefs to create a sense of connection, and prior work suggests that many cancer patients and their families depend on spiritual resources such as self- or group-prayer to cope with their cancer~\cite{taylor_spiritual_2005,hsiao_role_2008}. 

Although we do not know of other work in HCI that examines prayer support in OHCs as its own category, Wyche et al. characterized the use of technology across a variety of Christian denominational churchs, and found that pastors often issue prayer requests via email to ``ask members of the congregation to pray for someone else in the congregation, or a family member'' \cite{wyche_technology_2006}. Additionally, Shaw et al. showed that breast cancer patients who used more religious or prayer-related words in online support group messages experienced lower negative emotions and better well-being, possibly because religious coping mechanisms allowed them to experience less fear of death, place trust in God, and view their cancer experiences with more positivity~\cite{shaw2007effects}. Our work builds upon this concept of online prayer support by demonstrating that CaringBridge authors seek and receive prayers through their online community via journals and comments. Our content analysis provides evidence that users \textit{write} about prayer more often than any other category of support, yet we do not know how to contextualize the \textit{importance} of this type of support, relative to other types of support that can be facilitated through online mechanisms:

\begin{description} 
\item [RQ4:] \textit{How do users perceive the importance of prayer support, relative to the perceived importance of other types of social support? Do patients and caregivers differ from their support networks in their evaluations of prayer and other high level support categories?} 
\end{description}

\section{Overview of Dual Synergistic Phases}
We performed a two-phase study, using the first primarily to derive support categories from the data, and the second to directly ask users for their input on those categories. Phase 1 addresses RQ1 through a directed content analysis of journal updates. We developed a codebook suitable for our user-generated content and introduced a measure for quantifying social support received, which we term an ``Expression of Appreciation'' (EOA). EOAs are statements written by a user which positively acknowledge that they have received a specific type of support, including expressions of thanks, gratitude, blessing, or happiness to have received something. We use the EOA concept as a proxy for measuring what users have valued receiving. Following our content analysis, we conducted Phase 2, a large-scale survey deployed to CaringBridge users. In the survey design process, our codebook provided the benefit of helping us to structure both the language and categories used in our survey, so that we could most effectively address RQ2, RQ3, and post-hoc RQ4.

\section{Phase 1: Directed Content Analysis of Journal Updates}
\label{qual}
\begin{figure}
  \frame{\includegraphics[width=\textwidth]{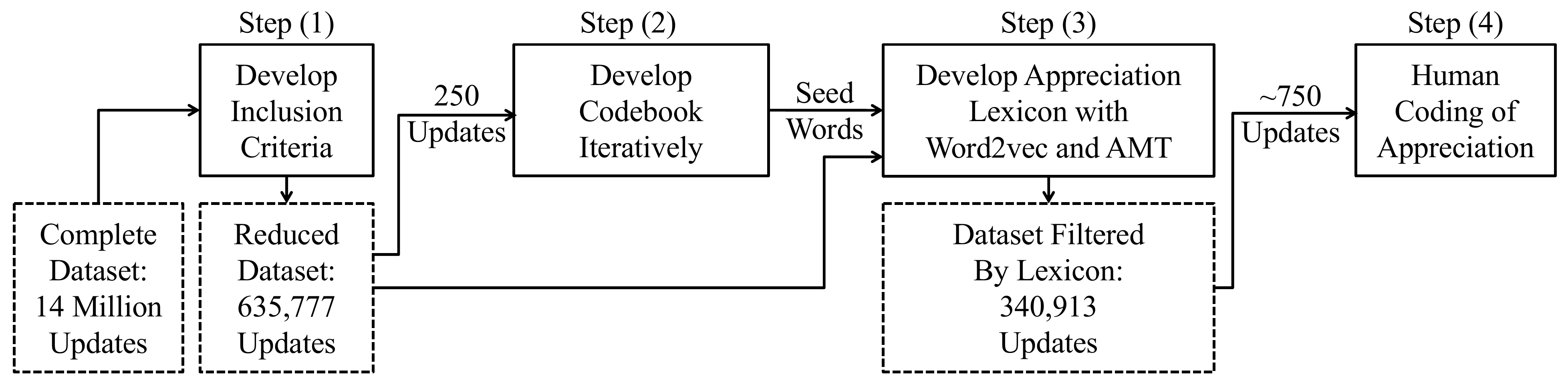}}
  \caption{Summary Visualization of Directed Content Analysis. Boxes with solid lines indicate actions taken by our research team. Boxes with dotted lines indicate the size of eligible dataset included at each step.}~\label{fig:P1method}
\end{figure}

We completed a directed content analysis~\cite{hsieh_three_2005} in four steps (see Fig.~\ref{fig:P1method}): (1) development of data inclusion criteria; (2) iterative codebook development; (3) lexicon development and filtering for journal updates containing EOAs; and (4) human coding of EOAs in filtered journal updates.

\subsection{Inclusion Criteria}
\label{inclusion}
First, we limited our analysis to self-identified cancer sites. Cancer (of any type) represents the largest group of all self-identified conditions on CaringBridge \cite{ma_write_2017}, and is a chronic condition for which instrumental support might potentially be useful to patients and caregivers over a lengthy timespan. Second, many users who create a site do not return to the community after one or two visits. Specifically, in our entire dataset prior to June 3, 2016, 42.2\% of created sites have 0, 1, or 2 updates (27.1\% have no updates, 11.6\% have 1 update, 3.5\% have 2), while the remainder (57.8\%) have 3 or more updates. Similar to~\cite{ma_write_2017}, we examined sites with 3 or more journal updates; these sites represent more engaged community members rather than one-time sign-ups. Finally, although our dataset contains sites dating from 2005, the CaringBridge platform has evolved over time and expanded its userbase; we included sites created within the last three years of the dataset (June 3, 2013 - June 3, 2016) to capture a more recent slice of the data. Applying these criteria, we retained 19,535 unique sites with 635,777 total journal updates (average of 32.5 journal updates/site). Additionally, \textasciitilde800,000 unique visitors left comments on these sites (average of \textasciitilde41 visitors/site). 

\subsection{Iterative Codebook Development for Categorical Support Types} 
\label{codebook} 
\subsubsection{Preliminary Codebook Development} Directed content analysis utilizes both prior work and emergent codes from the data for codebook development~\cite{hsieh_three_2005}. Two researchers developed a preliminary codebook draft by discussing and clustering the specific help tasks described by~\cite{skeels_catalyzing_2010} into help type categories under the overarching categories of informational, emotional, instrumental, or other support. We decided to subdivide emotional support into co-located (EMO\textsubscript{CO}) vs. remote emotional support (EMO\textsubscript{R}) because as we read journal updates, we noticed that authors' descriptions of having received supportive messages (either online or on handwritten cards) were easily distinguishable from in-person interactions. With an eye to RQ2 and RQ3, we found it helpful to distinguish between support types that can be facilitated remotely (e.g. online) vs. requiring physical interaction (e.g. in-person, offline), since some CaringBridge site visitors are geographically co-located with the patient. We completed an exploratory reading of 50 randomly selected journal updates that met inclusion criteria (see~\ref{inclusion}) and refined our preliminary codebook by writing code definitions to closely describe how CaringBridge authors write about each category, along with examples from the data.

\afterpage{
\irrtable
\prelimcodes
\newpage
}

In our exploratory reading, we observed that CaringBridge authors infrequently write direct requests for help from members of their support community, but more often use journal updates to thank community members for help they have already received. Based on this observation and the SIST definition of direct help-seeking~\cite{barbee_experimental_1995}, we initially sought to code the following help-related expressions: (1) explicit acknowledgements of when help had been received and (2) ``direct'' requests for help.\footnote{Note that we also initially attempted to code ``'indirect'' help-seeking. Prior work suggests that indirect help seeking is subtle and less informative~\cite{barbee_experimental_1995}, and less effective at eliciting helpful responses online~\cite{andalibi_responding_2018}. Our preliminary IRR indicated that indirect requests were too difficult to identify reliably, thus we omit our analysis of indirect help-seeking.} We coded 17 types of help derived from prior literature \cite{skeels_catalyzing_2010} and our reading of the data, e.g. ``transportation,'' ``food,'' or ``shelter''. Finally, similar to Reddit posts about cancer~\cite{eschler_self-characterized_2015}, most updates provide clear textual evidence that an author is either a ``patient'' or ``caregiver.'' However, a small number of authors are ``both'' a patient \textit{and} a caregiver to someone else, or if the writing provides no textual evidence about their role, ``unknown,'' therefore we coded four author types. Table ~\ref{tab:codebook} provides abbreviated descriptions of help type codes; complete verbatim codebook is available in the supplementary materials (section ~\ref{sec:complete}).

\subsubsection{Codebook Iteration using Interrater Reliability Scores} Three research assistants\footnote{Undergraduates at the University of Minnesota who received training on appropriate conduct regarding sensitive data} joined to revise and refine the codebook. We used Interrater Reliability scores (Krippendorff's alpha) to evaluate and refine codes over four rounds, with each researcher independently coding 50 new updates at each round (see Table ~\ref{tab:IRR}). In this preliminary sample of 200 randomly selected updates, 67\% included acknowledgements of received help, whereas only 34\% included direct requests. Moreover, there was less variety in what people asked for help with; the majority of direct requests sought prayer support (PR), continued comments on the CaringBridge site (EMO\textsubscript{R}), and in-person social activity (EMO\textsubscript{CO}) (See Figure ~\ref{fig:prelimcodes}). Some categories had low agreement due to very few cases of that category appearing in the data. Additionally, coders were able to achieve higher reliability on acknowledgements of help received, especially because most of them were accompanied by words that signal \textit{appreciation}. 

\subsubsection{Using Expressions of Appreciation as a Measurement Tool} While prior work has often explored support exchanges, esp. of emotional and informational support (e.g.~\cite{vlahovic2014support}), Coursaris and Liu reveal three additional categories of positive group interactions, including (1) sharing personal experiences, (2) offering congratulations, and (3) expressions of gratitude~\cite{coursaris_analysis_2009}. Prior work has not quantified these interaction types. Conceptually similar to expressions of gratitude, we introduce a measurement concept we term an ``Expression of Appreciation'' (EOA). Broader than gratitude alone, appreciation and may include gratitude, a positive emotional connection to, and/or an acknowledgment of help received from others~\cite{adler_appreciation:_2005}. EOAs have the limitation that they do not indicate what types of support users seek or need, however they do provide a useful measure of help that has been both received and positively acknowledged---allowing us to identify both the act of appreciation itself, as well as the specific categories of help that have been the target of appreciation. Furthermore, we anecdotally observed that most expressions about help received \textit{did} include appreciation words, since CaringBridge authors are very often writing to an audience of people they know, and whom they would like to publicly thank. However, we note the limitation that EOAs do not reflect counts of help types that may \textit{not} have been accompanied by words indicating positive acknowledgement; thus the actual total of all counts of help received may be slightly higher. To assist coders, we sought to filter out journal updates that did not contain any EOAs. However, lacking a standardized ``appreciation'' lexicon (e.g. within the LIWC dictionary), our next task was to develop one.

\codebook

\subsection{Development of an Appreciation Lexicon \& Journal Filtering}
\label{lex_dev}
During codebook development (section ~\ref{codebook}), we collected seed words related to appreciation.\footnote{Seed words include: ``appreciate,'' ``blessed,'' ``blessing,'' ``gratitude,'' ``grateful,'' ``humbled,'' ``support,'' ``thank," ``thankful,'' ``thankfully,'' ``thanks''} We then followed the Empath approach to lexicon development~\cite{fast_empath:_2016}. We used our seed words to retrieve the 200 nearest neighbors in a word2vec model trained on the entire CaringBridge journal corpus~\cite{mikolov_efficient_2013,mikolov_distributed_2013}.
Using the exact instructions and pay as Empath~\cite{fast_empath:_2016}, we hired Amazon Mechanical Turk crowdworkers to evaluate whether each candidate word was related to the concept of appreciation, and narrowed down to 130 final words, including common misspellings. (See supplementary materials, section ~\ref{lexicon} for complete lexicon.) Finally, we filtered for journal updates including at least one lexicon word. Of 635,777 journal updates that met inclusion criteria, 340,913 (\textasciitilde53.6\%) journal updates contained at least one appreciation word. 

\subsection{Application of codebook to CaringBridge journal updates}\label{coding}
After codebook development and appreciation filtering, four new research assistants received training on how to apply the codebook (see supplementary material section ~\ref{protocol} for coding protocol). Researchers were required to tag a test set of 20 journal updates with >95\% accuracy before moving on to code unknown data; all researchers achieved >95\% accuracy within 2-3 attempts to code the test set. Finally, we randomly sampled CaringBridge journal updates from data that met inclusion criteria and contained at least one appreciation word. Figure ~\ref{fig:CB} shows an example of anonymized user data, with codes applied according to the described protocol. Note that codes were applied in a binary manner to each journal update (i.e., a code was either present or not present). 

\firstresults

\subsection{Directed Content Analysis Findings}

\subsubsection{\textbf{RQ1a: \textit{What kinds of support do CaringBridge journal authors positively acknowledge in online writing about cancer experiences, and what is their relative prevalence?}}}

Five coders tagged 763 randomly selected updates from the 340,913 updates that contained at least one word from our appreciation lexicon (see section~\ref{lex_dev}); 427 (56\%) authors were identified as ``caregivers'', 292 (38.2\%) as ``patients'', 13 (1.7\%) as ``both'' patients and caregivers, and 31 (4.1\%) as ``unknown''. Of 732 journals with identifiable authorship, 89.3\% contained an expression of appreciation (EOA). The 44 updates by either ``both'' or ``unknown'' authors, as well as the 78 updates that did not contain an EOA, were excluded from subsequent analysis. Thus the remainder of our analysis considers 641 journal updates, which cumulatively contained 1405 EOAs.  Figure~\ref{fig:counts_of_help} shows the counts of each help type code, and we discuss the most prominent categories below.   


Wyche et al. observed that Christian communities use communication technologies such as email to issue prayer requests \cite{wyche_technology_2006}. In our content analysis, we found that such prayer requests also take place via CaringBridge journals, and that CaringBridge authors frequently acknowledge receiving prayer support in their journal updates. Other than generic or non-help related EOAs, prayer support was the most common of the high-level support categories, with authors acknowledging it in 42\% of updates (269 EOAs). We note that \cite{flickinger_social_2017} observed ``prayer,'' but included it as a subcategory of emotional support. We considered this strategy, but as our most prevalent code, prayer would have subsumed most of the emotional support category. \textbf{Thus, our first major finding is that prayer support should exist as its own distinct high-level category of social support.} Our survey in phase two provides further evidence that users rate prayer and emotional support differently, suggesting that users may experience, prioritize, or value them as distinct forms of support.

We also find that EMO\textsubscript{CO} (173 EOAs) and EMO\textsubscript{R} (142 EOAs) were individually prominent; when combined, emotional support was acknowledged in 41.5\% of updates. Low-level instrumental support categories were individually present in lower counts, however adding together all types of instrumental support shows that as a high-level category, instrumental support is acknowledged in 24.5\% of updates and is thus a prevalent target of appreciation. In descending order of EOA counts, Figure~\ref{fig:counts_of_help} shows that patients and caregivers positively acknowledged receiving help with nine different types of instrumental support: food, practical items, chores, transportation, financial assistance, family care, personal care, shelter, and exercise.

Informational support was rarely acknowledged (11 EOAs, or 1.7\% of updates). Rather than interpreting this finding about informational support as a signal that it is ``unimportant'' to CaringBridge users, we suggest that this finding may be specific to the unique context of CaringBridge. In particular, CaringBridge is an OHC designed to connect patients and caregivers with family and friends, who may not be in a good position to offer informational support. Other OHCs often connect patients who are strangers, but who share the same or similar disease conditions, and are more likely to have experiences, resources, or information specifically relevant to their disease~\cite{meier_how_2007,frost_social_2008,civan_threading_2007}. In a study of patients with rare diseases, MacLeod et al. found that participants did not expect their needs for informational support to be filled by family and friends, but rather from ``Dr. Google'' or other patients online~\cite{macleod_be_2017}---our results may be consistent with this. \textbf{Thus, our second major finding is that EOAs for prayer support, emotional support, and instrumental support are relatively common (in descending order of frequency), whereas EOAs for informational support are uncommon on CaringBridge.} 

\subsubsection{\textbf{ [RQ1b:] \textit{Do patient and caregiver authors differ in terms of the types of help they write about appreciating?}}}


Figure \ref{fig:freq_of_help} shows the relative frequency of each help type code applied. A Chi Squared test indicates a significant difference between patient and caregiver proportions of the \textit{high-level} support types ($\chi^2=12.27$, d.o.f.=4, $p<0.05$). A Fisher Exact test confirms that the significant difference between the high-level support types exists at the low-level EOAs, as well ($p<0.01$). No significant difference existed in the EOA diversity of journal updates between patient and caregiver authors ($p>0.05$). \textbf{Thus, we find that patients and caregivers express appreciation for different types of help in their online writing.} Our Phase 2 survey explores these differences in more depth. 

\subsubsection{Phase 1: Summary of Content Analysis Results to Address RQ1} Our content analysis has shown that prayer support is especially prevalent in CaringBridge journals, thus we suggest it should be considered as its own independent analytical category. Furthermore, CaringBridge authors acknowledge prayer, emotional, and instrumental support, but rarely acknowledge informational support. Examining instrumental support specifically, we find that patients and caregivers express appreciation for food, practical items, chores, transportation, financial assistance, family care, personal care, shelter, and exercise. However, patients and caregivers do not write about receiving these types of help at the same frequencies; rather, there exist suggestive differences between the two author groups.

While content analysis allows us to draw conclusions about users' online writing, it does not enable us to understand their internal priorities or unstated needs. Furthermore, it does not provide insights into the motivations of CaringBridge \textit{visitors}, i.e. friends, family, or acquaintances who follow patients' health journeys. Phase 2 helps us explore these issues.

\section{Phase 2: Quantitative Survey Analysis}
\label{quant}
We used support categories derived from Phase 1 to design a survey instrument that quantitatively addresses RQs 2-4. While prior work has deployed surveys to measure variables like satisfaction with social network structures~\cite{stokes_predicting_1983} or the amount of social support received as a predictor of survival~\cite{thong_social_2007, sonderen_het_1993}, our research questions required the design of a survey to understand users' underlying attitudes towards and interests in different types of instrumental support.

\subsection{Survey Design}
We followed guidelines suggested by Mueller et al. \cite{mueller_survey_2014} to create a survey instrument on Qualtrics~\cite{qualtrics_qualtrics_2018}. We first developed a preliminary survey based on our codebook. Internal CaringBridge employees then helped us to iterate on the language and flow of survey questions to ensure an accessible experience for CaringBridge users. We piloted survey questions extensively throughout development.

Figure ~\ref{fig:survey} depicts an overview of the survey structure, while supplemental section~\ref{survey} contains the complete verbatim survey. The survey had three separate tracks for (A) cancer patients, (B) family caregivers, or (C) friends, family and acquaintances (FFA). Tracks for patients and caregivers were almost identical: minor word differences reflected ``your'' or ``this'' cancer journey. Track FFA inverted perspective on most of the same questions, with a few omissions and additions. Most questions forced responses on a 5-point unipolar or bipolar Likert scale. We also included a small number of optional questions with textual free response fields. Each track consisted of four similar question blocks and ended with demographic questions:

\begin{figure}
  \frame{\includegraphics[width=\textwidth]{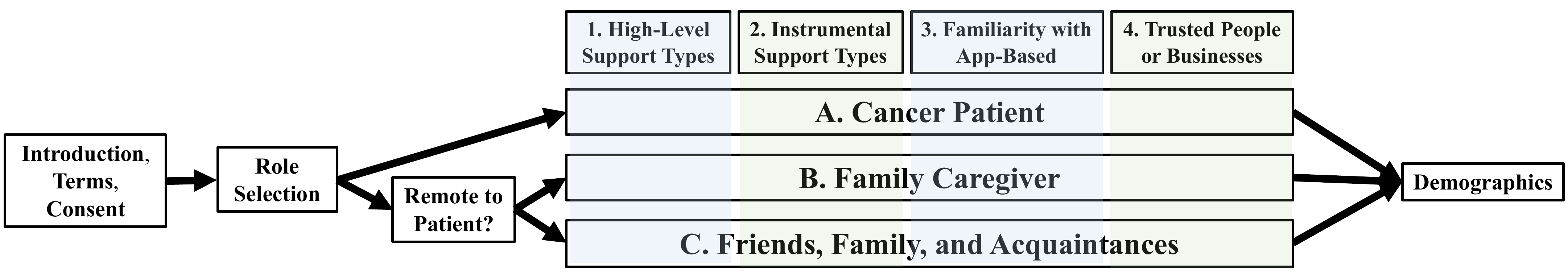}}
  \caption{Survey Structure. Participants selected one of three tracks based on their role in a cancer experience. Each track had four question blocks; question text was verbatim between tracks where possible, or adjusted to account for different perspectives. Complete survey available in supplementary materials, section ~\ref{survey}. }~\label{fig:survey}
\end{figure}

\begin{enumerate}
\item \textit{High-Level Support Types.} All tracks rated the importance of high-level support types. 
\item \textit{Instrumental Support Types.} Respondents were asked to rate how frequently instrumental support was needed vs. asked for, and how challenging it was to communicate about providing help. Tracks for patients and caregivers were asked to rate the usefulness of nine instrumental support types. Track FFA was asked to rate how interested they were in providing these nine support types, either in-person or by funding a service.
\item \textit{App-Based Business Familiarity.} All tracks were asked identical questions about whether they were familiar with app-based services and whether they had ever used them.
\item \textit{Trusted People or Businesses.} 
For each instrumental support type rated higher than ``Not at all useful,'' patients and caregivers were asked to select all the people/businesses they trust to provide services. FFA were asked to rate their trust in traditional vs. app-based services.
\end{enumerate}

\subsection{Survey Sampling Frame and Deployment}
We approximated our survey sampling frame to our content analysis inclusion criteria. Eligible respondents must have been involved in a cancer experience at any point in the last three years\footnote{Note that the timeframe of the ``past three years'' (i.e. roughly 2015-2018) only partially aligns with the last three years of our Phase 1 dataset of CaringBridge journal updates (i.e. June 3, 2013 - June 3, 2016).}\label{timespan} as either patients, caregivers or FFA of a cancer patient. 19,535 unique sites with \textasciitilde800,000 unique visitors met eligibility criteria for the content analysis. To represent these populations with 95\% (+/- 5\%) confidence, we calculated that we required sample sizes of at least 377 patients and caregivers and 384 FFA \cite{mueller_survey_2014}. 

Participation was anonymous and CaringBridge featured a survey link through a banner message on their website, a Facebook post, and an email (uncompensated). 1565 responses were initiated and 1094 completed (70\% completion rate), with 288 patients, 288 family caregivers, and 415 FFA; the remaining 103 selected responses indicating ineligibility due to lack of recent involvement in a cancer experience. As a validity check, we measured time spent on key survey questions, looking for evidence of respondents who may have answered questions too rapidly. We found no outliers below the first quartile (near \textasciitilde15 seconds). Similarly, the overall survey response time IQR was between 8-15 minutes, median of 11 minutes, with outliers only on the higher end. Therefore, we included all 991 respondents in our analysis. 

\demotable

\subsection{Demographics and Contextual Factors Related to Survey Respondents}\label{demographics}
Table~\ref{tab:demo} summarizes respondent demographics. We observe that respondents are predominantly white, female, middle-aged Americans, trending towards higher socio-economic status (SES). While CaringBridge usage is prevalent in a broad spectrum of health conditions and user types, CaringBridge confirms that this sample is representative of their user base of authors who specify a cancer condition (except that the average age of survey respondents is approximately 10 years older than the average age of all CaringBridge users). Although these sample demographics do not mirror the general public, prior work suggests a similar trend in other OHCs. For example, surveyed OHC users in a mental health context are more likely to be female, white, and college-educated~\cite{deandrea_online_2013,rains_social_2016}. These trends raise concerns about uneven knowledge about and use of OHCs and the benefits they stand to provide~\cite{anderson_uses_2011}. For more context, we next summarize respondents' geographic and social relationships, their perceived need for instrumental support, and potential effects of SES on our results. 

\subsubsection{Geographic and social relationships}
Geographic location is an important social context that can affect dispositions towards support seeking online and app-based services~\cite{thebault-spieker_avoiding_2015}. For example, \cite{helft2005use} found that only 10\% of disadvantaged rural cancer patients use the Internet to find information. Because of this, we approximated Metropolitan Statistical Area by asking participants about their county size (question ~\ref{MSA}). Respondents are evenly split across rural and urban areas. Finally, we asked caregivers and FFA about their relationship to the cancer patient (question ~\ref{relationship}), and whether they live within driving distance (question ~\ref{driving}). Most caregivers are the partner (44.8\%) or family member (54.2\%) of a cancer patient, with only 3 close friend caregivers (1.0\%). FFA are more diversely represented: 0.7\% are partners; 30.8\% are family members; 16.1\% are close friends; 41\% are non-professional friends or acquaintances; 11.3\% are coworkers or colleagues. 181 FFA (43.6\%) are co-located (i.e. within driving distance, $M=21$ minutes, $SD=22.8$). By comparison, 255 CG (88.5\%) are co-located ($M=7.3$ minutes, $SD=25.7$).

\subsubsection{Respondents' Perceived Need for Instrumental Support}
Prior literature asserts the importance of instrumental support to cancer patients and caregivers~\cite{jacobs_lessons_2015,jacobs_cancer_2016,skeels_catalyzing_2010}. However, we also wanted to validate the focus of our study by finding out if respondents actually felt that they need instrumental support, and whether or not their needs were being addressed. Therefore, we asked respondents about how frequently they \textit{needed} instrumental support, and how frequently they \textit{asked} for it (~\ref{ask_freq}, ~\ref{ask_freq_ffa} ). Responses were recorded on a 5-point scale from ``0: almost never'' to more than once per ``1: year,'' ``2: month,'' ``3: week,'' or ``4: day''. 

Only 14.2\% of patients and 10.4\% of caregivers reported almost never needing or asking for instrumental support. Moreover, 45.5\% of patients and 58.7\% of caregivers expressed needing instrumental support more frequently than they ask for it, while 21.5\% of patients and 31.3\% of caregivers almost never ask for help, even if they expressed needing it sometimes. Examining responses within subjects, we do find that some patients/caregivers need and ask for instrumental support at the same rate, and no respondents ask more often than they need help. However, on average, patients/caregivers felt they needed help (2.22) more frequently than they asked for help (1.39). FFA reported an even wider gap, perceiving a more frequent need for help (2.77) and an even lower frequency of asking for help (1.32). Taken together, these responses clearly indicate that: (1) many CaringBridge users feel a need for instrumental support, (2) this need may be even greater for caregivers than patients, and (3) a gap exists in their needs being met.




\usefulregression

\subsubsection{Potential Effects of SES on Instrumental Support Results}\label{SES}
A cancer diagnosis can result in a financially toxic state for patients who do not have adequate insurance or income to cover high medical costs~\cite{zafar_financial_2013}. Thus, Financial Assistance (FA) is a critical need for many cancer patients. Because our sample has relatively high SES, we wanted to understand how participants' SES relates to their perceived usefulness of instrumental support categories, especially FA. Therefore, we built three regression models to check which demographic features significantly predicted ratings of instrumental support, as we describe next.


We included demographic responses (Survey Section~\ref{demosection}) as independent variables. For dependent variables, we used the usefulness ratings of different instrumental support types (Question~\ref{useful_help}). In Table~\ref{tab:usefulregression}, we report $p$-values only for the three variables---gender, age, and income---that retained significance at the 95\% significance level in at least one of three models: 
Model 1 used Linear Regression to predict the summative score of all 9 instrumental support categories, and showed that income appeared to be a significant predictor of usefulness (R$^{2}$=0.168, Adjusted R$^{2}$=0.119, F=3.427***). Model 2 used Ordered Logistic Regression to predict the score of \textit{FA only}, because the response variable (usefulness of FA) was now an ordered factor; unsurprisingly, lower income respondents rated FA to be of higher usefulness than higher income respondents in this model (Residual Deviance=1669.741, AIC=1741.741). However, Model 3 used Linear Regression to predict the summative score of 8 instrumental support categories \textit{excluding FA}. In Model 3, income was no longer a significant predictor of usefulness, suggesting that regardless of SES, non-financial forms of instrumental support are perceived as useful (R$^{2}$=0.145, Adjusted R$^{2}$=0.095, F=2.889***). Additionally, Figure ~\ref{fig:useful} shows a distribution of responses to usefulness ratings of FA which suggests that FA is a polarizing category. In none of the models was Education Level significant. (For reference, supplemental section~\ref{sec:fullFAregression} includes complete details for all three models.) 

In summary, our sample trend towards high SES may have only limited impacts on our main findings. However, given the split in usefulness ratings by high and low SES respondents, future work should especially examine financial support (e.g. medical crowdfunding) in low SES contexts. Moreover, future work should also more closely examine demographic trends across OHC platforms and barriers to OHC access, such as a lack of exposure to their existence~\cite{anderson_uses_2011} or membership in stigmatized or marginalized groups (e.g.~\cite{rho_class_2017}).

\subsection{Survey Limitations}
While the survey allowed us to gather a large sample of responses regarding peoples' attitudes and preferences about instrumental support, this method has some inherent limitations. First, the total number of impressions of the survey invitation is unknown. Self-selection bias may have impacted who took the survey. Second, surveys measure only self-reported data. Our content analysis provides one measure of users' online writing, however future work might examine or compare user behaviors related to instrumental help-seeking or help-receiving on other platforms (e.g. lotsahelpinghands.com, posthope.org, or facebook.com). Third, we did not recruit via any other OHCs or platforms because we wanted to maintain the ecological validity of our work within one OHC, thus limiting our scope to CaringBridge users. Fourth, it is possible that people may overvalue (i.e. rate more highly) those forms of support that are currently accessible for them to receive or provide through already existing social practices and technical systems; they may not have a clear understanding of how other forms of support that aren't as accessible right now may be useful to them in the future. This limitation can be addressed through repeating surveys over time to compare how ratings change, and by engaging with people in more depth through qualitative methods, as we will suggest in our discussion.

Finally, we cannot infer respondents' \textit{actual} roles on CaringBridge, however we \textit{approximate} ``authors'' with the self-selected roles of patient or caregiver, and ``visitors'' with the role FFA. With 576 patients or caregivers and 415 FFA, we achieve a large enough sample to represent our sampling frame with 95\% (+/- 5\%) confidence. However, since some respondents may have selected roles that do not align with their corresponding author or visitor role on CaringBridge, and since the three-year timeframe of data inclusion does not exactly align with the past three years of the survey, our sampling frame is approximate. Future work could recruit users via email addresses associated with their user account, so that their role in the OHC would be known. We considered this approach, however opted against it for two reasons. First, by agreeing to CaringBridge's terms of service, users consent to their de-identified data being used for analytic purposes, however they haven't self-selected to be contacted for research. Second, as a small non-profit with limited resources, we wanted to minimize strain required internally of CaringBridge. 

\afterpage{

\usefulplot
\pcguseful
\pagebreak
}

\subsection{Survey Findings}
In this section, we present responses to relevant survey questions, organized according to research question. For RQ2 and RQ4 we asked respondents to rate support categories from our content analysis on a 5-point unipolar Likert scale. We present summaries of the mean Likert responses in order to demonstrate which categories are, on average, most useful or important to patients and caregivers, and which categories FFA are most interested in providing. We also calculate confidence intervals on the differences of the means, in order to see where differences are significant, and to assess if different respondent groups align or mis-align in their ratings.

For RQ3, we asked patients/caregivers about which possible supporters (e.g. family, close friend, acquaintance, coworker, app-based business, or traditional business) they trust to provide each type of instrumental support. To summarize the responses, we present the percentages of respondents who indicated trusting each category of supporter, and conduct T-tests to assess whether some supporters are trusted more than others. We also asked FFA about whether they generally trust app-based or traditional businesses to provide patients with instrumental support, and evaluated the difference of means.

Supplemental section~\ref{supplemental_regressions} contains additional explanatory models related to RQ2 and RQ3. These models provide information about factors that affect FFA interest in providing instrumental support (section~\ref{fund_vs_inperson}), and patient/caregiver trust in app-based or traditional businesses to provide them with instrumental support (section~\ref{trust_regression}). We omit these models from the main paper because they do not directly address our stated research questions, however we wanted to make this information available to readers, should they be interested to see how demographic factors affect the responses.

\subsubsection{\textbf{RQ2a: What kinds of instrumental support are most useful to patients and caregivers on CaringBridge? Do patients and caregivers align in their perceptions of what is useful?}} We asked patients and caregivers to rate the nine different instrumental support types from our content analysis from ``not at all'' to ``extremely'' useful. Figure~\ref{fig:usefulplot} summarizes responses according to mean values of the ratings, and Figure~\ref{fig:useful} shows the distributions of the ratings.

In descending order, patients ranked chores, food and transportation (equal means), exercise, and financial assistance, as the top five most useful categories, whereas caregivers ranked chores, food, financial assistance, personal care, and transportation as their top five categories. Importantly, patients and caregivers align in rating chores and food as the two of the most useful types of instrumental support, suggesting that these categories may be excellent foci for future study and innovation. We draw particular attention to the financial assistance category, for which the mean rating value between patients and caregivers differs most substantially of all categories. Figure~\ref{fig:useful} shows that there appears to be a split in the ratings of financial assistance from patients and caregivers; both distributions have peaks at ``not at all'' and ``extremely'' useful, with more caregivers skewing heavily towards rating financial assistance as ``extremely'' useful. This suggests that caregivers, who must manage not only care for the patient, but also entire households, may feel a greater degree of financial worry and strain. 

Practical items, familycare, and shelter were, on average, rated as least useful; both familycare and shelter have noticeable peaks of ``not at all'' useful ratings, suggesting that there is broadly less need, or perhaps that people simply do not especially want help with these two categories. 

For all instrumental support types other than personal care and practical items, Mann-Whitney $U$ tests indicate that the distributions of responses differ significantly between patients and caregivers (see Table~\ref{tab:useful} in appendix for $U$ values.) Similarly, we performed two-sided t-tests to compare the means of all distributions, and found that the means for all categories except personal care and practical items differ significantly. \textbf{From these results, we conclude that patients and caregivers systematically diverge in their perceptions of what types of instrumental support are most useful to them during a cancer experience.}

\afterpage{
\interestplot
\ffadistributions
\pagebreak
}

\subsubsection{\textbf{RQ2b: What kinds of instrumental support are CaringBridge users who are friends, family, or acquaintances most interested in providing? How does this compare to patients and caregivers?}} 

We asked FFA to rate their interest in providing different types of instrumental support in person using their own time and resources, or by funding a service. In our sample, 43.6\% of FFA are co-located (i.e. within driving distance) to the patient. Because co-located FFA are capable of providing many types of instrumental support in person, whereas those who are remote cannot do so, we calculated average Likert responses separately for FFA who are co-located vs. remote. Figure~\ref{fig:interestplot} shows mean Likert responses, and Figure~\ref{fig:interest} shows the distributions of ratings for FFA interest in providing instrumental support types in-person. 
Finally, we discuss the financial assistance category separately from the other categories because there is not a meaningful distinction between providing support in person vs. funding for this category.

In descending order, co-located FFA rated their interest in providing food, practical items, transportation, chores, familycare, exercise, shelter, and finally personal care in person. For all categories other than the two of least interest (shelter and personal care), we find that co-located FFA are significantly more interested in providing instrumental support in person, rather than funding a service. Given their geographic proximity, co-located FFA may feel it is an important show of support to be physically present.

For remote FFA, most of the mean differences between providing support in person or by funding services are not statistically significant; only personal care retains a significant difference of means. Figure~\ref{fig:interest} shows that for every category except financial assistance, there exist considerable peaks of ``not at all'' interested ratings. (Co-located ratings also have peaks of ``not at all'' interested ratings only for exercise, shelter, and personal care.) This heightened prevalence of ``not at all'' interested ratings lowers mean ratings across the board by remote FFA. However, excluding the ``not at at all'' interested ratings, the shape of the distributions are visually similar to co-located FFA ratings. We interpret this as an indication that there may exist two distinct attitudes held by remote FFA towards providing instrumental support---a considerable proportion may have little to no interest in providing instrumental support at all, whereas others may be similarly interested to co-located FFA, even if they do not have any practical way to provide instrumental support.

Regardless of proximity to the patient, FFA are, on average, ``moderately'' interested ($M$\textsubscript{co-located}=2.07,  $M$\textsubscript{remote}=2.03) in providing financial assistance, with no significant difference of means between the two groups. For co-located FFA, their interest in providing most support types in person is higher than their interest in providing financial assistance. However, other than food and practical items, their interest in providing generic financial assistance is higher than their interest in funding all other categories. For remote FFA, interest in providing financial assistance is ranked third (behind practical items and food), and is higher than their interest in providing support in person or funding all other categories. Mean FFA ratings of financial assistance are between the mean ratings of patients ($M$\textsubscript{patients}=1.97) and caregivers ($M$\textsubscript{caregivers}=2.57).

\textbf{Comparing mean values across instrumental support types, we find that FFA systematically diverge from patients/caregivers in terms of the types of help they are most interested in providing.} To illustrate the trend, we will use mean ratings for in-person support by co-located FFA. In patients/caregivers' rankings, both rate chores as most useful ($M$\textsubscript{patients}=2.67, $M$\textsubscript{caregivers}=2.98), whereas FFA rate interest in providing help with chores as fourth ($M$\textsubscript{FFA}=2.47). FFA rated interest in providing food highest ($M$\textsubscript{FFA}=2.92), whereas patients/caregivers rated it second highest ($M$\textsubscript{patients}=2.53, $M$\textsubscript{caregivers}=2.77). This general misalignment extends across the categories we measured.

Interestingly, FFA ratings for interest in providing support types roughly align with our Phase~1 results. That is, comparing Figures~\ref{fig:counts_of_help} and~\ref{fig:interestplot} shows that patients/caregivers generally express appreciation for the types of support that FFA are most interested in providing. In our discussion, we will discuss ways to address the divergent preferences of patients/caregivers and FFA.

\subsubsection{\textbf{RQ3: Whom do CaringBridge users trust to provide cancer patients and caregivers with instrumental support?}}

For each instrumental support type that they rated at least ``slightly useful'', patients/caregivers were asked to select all people or businesses they trust to provide that type of help from among: no one at all, family member, close friend, friend or acquaintance, coworker or colleague, an app-based business, or a traditional business (multiple choice, binary selection, Question~\ref{TracksAB_4}). Figure~\ref{fig:trust} depicts a matrix of responses, showing the proportion of trusted individuals or businesses, including 95\% confidence intervals. 

Prior work has shown that relationship closeness is one factor (among others) affecting support provision by social media users to people in need of social support (e.g.~\cite{andalibi_responding_2018,macleod_be_2017}). Figure~\ref{fig:trust} demonstrates the clear trend that larger proportions of patients and caregivers trust people who are socially closer to them to provide most forms of help. For example, in the food category (a commonly acknowledged form of support in CaringBridge journals, as well as highly rated in our survey by both both patients/caregivers and FFA), 96\% trust family and close friends, 75\% trust friends/acquaintances, and 68\% trust coworkers/colleagues. The proportions trusting traditional businesses are somewhere between those trusting close friends and friends or acquaintances, while the proportions trusting app-based businesses are generally lower, and more difficult to identify clear trends. In the case of food, 67\% trust traditional businesses, while 54\% trust app-based businesses.

\trustmatrix

In order to statistically evaluate how social closeness affects patients' and caregivers' trust in possible supporters in our data, we conducted Bonferroni-corrected independent two-sample Z-tests for each pair of the four personal relations for each instrumental support type. We found that for all instrumental support types, the proportion of patients/caregivers trusting family or close friends is greater than the proportion trusting acquaintances or coworkers (all $p<0.001$). Patients/caregivers trust family at a significantly higher rate than close friends only for familycare ($p=0.028$) and financial assistance ($p<0.001$). Patients/caregivers trust acquaintances at a significantly higher rate than coworkers for transportation ($p=0.008$), familycare ($p=0.04$), personal care ($p=0.001$), exercise ($p<0.001$), and chores ($p<0.001$). \textbf{Thus, we conclude that patients/caregivers generally have a higher degree of trust in possible supporters to provide them with instrumental support when they share a closer social connection.} 


We did not ask FFA to rate their trust in each of the 9 instrumental support categories, however FFA were asked to \textit{generally} rate their trust in traditional businesses or app-based businesses to provide instrumental support for patients (question ~\ref{TrackC_4}) on a 5-point bipolar Likert scale from -2: ``Strongly distrust'' to 2: ``Strongly trust.'' On average, FFA rated traditional businesses at $M=0.58$, $SD=1.03$, while app-based businesses were rated at $M=0.21$, $SD=1.01$. A paired T-test shows that this difference is significant ($T = 7.324$, $p < 0.001$). Furthermore, for all but two categories, we observe that a greater proportion of patients/caregivers trust traditional businesses than app-based businesses. \textbf{Therefore we conclude that for both patients/caregivers and FFA, we find that trust is \textit{generally} higher in traditional businesses than in app-based services.} This may be because app-based businesses are relatively new, and prior research has shown that people have concerns about safety, willingness to participate, and disparities in where app-based services are geographically available~\cite{mclachlan_you_2016,dillahunt_promise_2015,thebault-spieker_avoiding_2015,thebault-spieker_toward_2017}. However, we were intrigued to observe that more patients/caregivers trust app-based services for financial assistance and transportation (e.g., crowdfunding such as GoFundMe, or Lyft/Uber) than traditional services (e.g., bank loans, or taxis), likely due to greater availability of these particular services than others which are less ubiquitous or well known (e.g., TaskRabbit). Regression results in Table ~\ref{tab:trustregression} in the appendix also show that prior use of app-based businesses for transportation and shelter predict higher trust.

\highlevelsupport

\subsubsection{\textbf{RQ4: \textit{How do users perceive the importance of prayer support, relative to the perceived importance of other types of social support? Do patients and caregivers differ from their support networks in their evaluations of prayer and other high level support categories?}}}

We asked respondents to rate the importance of all high-level support types (questions ~\ref{support_types} and ~\ref{support_types_ffa}); Table ~\ref{tab:support} shows the distributions of responses. \textbf{Importantly, prayer support is more commonly rated ``Extremely important'' by both patients/caregivers and FFA than any other category.} This may be because people become increasingly concerned with spiritual matters as they navigate life-threatening and end-of-life concerns~\cite{puchalski2011making,brubaker_workshop}, while prayer and other spiritual practices increase quality of life and well being in ways that other forms of support do not~\cite{vallurupalli_role_2012}.

In order to understand whether patients/caregivers align in their evaluations of each high-level support type, we conducted two-sample independent t-tests with Bonferroni correction (d.o.f.=989) to compare the mean response between patients/caregivers and FFA. On average, patients/caregivers rate informational support ($p < 0.001$) as more important than FFA rate it, whereas FFA rate prayer ($p=0.019$) and EMO\textsubscript{R} ($p=0.003$) more highly than patients/caregivers rate them.  At the 95\% confidence level, we fail to reject the null hypothesis of equal means between patients/caregivers and FFA for instrumental support ($p=2.43$) and EMO\textsubscript{CO} ($p=0.053$). With the exception of instrumental support, Mann-Whitney tests show that support type importance distributions are significantly different between patients/caregivers and FFA. We used the Common Language Effect Size (CLES) to describe the degree of separation, finding these differences to be small~\cite{mcgraw_common_1992}. \textbf{Thus we conclude that patients/caregivers differ from FFA in their importance ratings for prayer, informational, and remote emotional support, however their ratings for instrumental and co-located emotional support do not differ significantly.}

\subsubsection{Phase 2: Summary of Survey Results to Address RQ2-4} 

\quad \\
\phantom{--} RQ2: Our survey provides evidence that patients and caregivers diverge from each other, as well as from FFA, in their ratings of instrumental support categories. Caregivers rate most forms of instrumental support as more useful than patients do, except for transportation and exercise (with no significant differences between personal care and practical items). Patients and caregivers align in ranking chores and food as their top two priorities, however the ranking order diverges for other categories. On the other hand, co-located and remote FFA rate food and practical items as their top two priorities, with chores in fourth place. Co-located FFA are more interested in providing instrumental support than remote FFA, and they are more interested in providing it in person with their own time and resources rather than hiring paid services. 

RQ3: Our survey demonstrates the trend that patients and caregivers generally trust people are who are socially closer to them to provide them with support. Furthermore, patients/caregivers and FFA generally trust traditional businesses more than app-based businesses to provide instrumental support, although respondents also indicated higher trust in two of the more well-established app-based services, i.e. ride-sharing and crowdfunding services rather than traditional taxi and banking services, to provide them with transportation and financial support. 

RQ4: Finally, in considering high-level support categories, patients/caregivers and FFA align in ranking prayer support as the most important form of social support to them. However, patients/caregivers rate informational support more highly than FFA, and FFA rate prayer support and remote emotional support more highly than patients/caregivers. On the other hand, patients/caregivers and FFA's mean ratings of instrumental and co-located emotional support do not differ with statistical significance.

\section{Discussion}
CaringBridge offers compelling opportunities for socio-technical innovation to help cancer patients and caregivers get social support from their pre-existing networks of friends, family, and acquaintances (FFA). The primary functionality of CaringBridge has historically been its journal feature, where patients and caregivers can author updates about their health, ask for help, and receive comments from FFA. This paper quantitatively measures the online writing of CaringBridge users through a content analysis of journal updates. Building on categories derived from the content analysis, we deploy a survey to \textasciitilde1000 users to explore their attitudes towards different types of social support.

CaringBridge has shared with our research team their goal to become a ``hub of help and healing'', and to break down barriers to asking for and providing help. In service of their vision---\textit{``a world where no one goes through a health journey alone''}---CaringBridge began offering new features intended to streamline support seeking within patients' sites in an area called ``Ways to Help'' in the past few years. This area contains tools like a narrative section allowing families to articulate what is helpful to them, a Planner tool for coordinated help and care, an integration with the personal fundraising tool GoFundMe, and the option to disclose what care facilities they are in and whether they want in-person visits. It also offers a ``Support Links'' space where users can post links to other articles or sites. Throughout the rest of our discussion, we will refer to some of these features and share how our results can inform the ongoing development of tools such as these, that are intended to provide patients/caregivers and FFA with ways to exchange social support.  

Our content analysis provides us with the categories of social support that users write about, and shows that users positively acknowledge prayer, emotional, and instrumental support in their journals, but rarely acknowledge informational support. We focus our discussion primarily on our survey results, and synthesize these results to provide implications for designing OHCs that are better equipped to facilitate instrumental support for people experiencing cancer. Finally, we discuss the need for future research to investigate ways to empower caregivers and to support people spiritually during health crises.

\subsection{Accommodate Divergent Preferences for Social Support} 
We find that disconnects exist between the types of help that patients and caregivers find useful or important, and the types of support that FFA want to provide. To accommodate divergent preferences for social support means to design mechanisms that suitably address the needs of patients/caregivers while also allowing their support networks to provide the types of help that they are prepared to provide. On average, Table~\ref{tab:support} shows that patients/caregivers rate the high-level category of informational support as more important to them than FFA. While mean ratings of instrumental support as a high level category did not differ significantly between patients/caregivers and FFA, we found that they do diverge in terms of \textit{what types} of instrumental support are of interest. Furthermore, FFA rate that providing prayer and emotional support are more important to them than to patients/caregivers.\footnote{Note that the difference of means was statistically significant only for remote emotional support; for co-located emotional support, $p=0.053$.} We do not imply that any forms of support are unimportant; patients/caregivers rated all forms of social support we measured to be at least ``moderately important'', while prior work suggests that technologies for patients/caregivers should provide flexible, personalized experiences that meet peoples' unique needs at various stages of their cancer experiences~\cite{jacobs_cancer_2016,jacobs_mypath:_2018,choi_toward_2017,skeels_catalyzing_2010}. However, these data suggest the trend that on CaringBridge, patients/caregivers may perceive \textit{action-facilitating}~\cite{cutrona_controllability_1992} types of social support to be more important than FFA (and they also write less frequently about receiving action-facilitating support). On the other hand, FFA find it more important to provide \textit{nurturant}~\cite{cutrona_controllability_1992} forms of social support (and patients/caregivers write more often about receiving nurturant support). Thus, the socio-technical characteristics of OHCs comprised primarily of FFA may result in more nurturant social support and less action-facilitating support being given to patients/caregivers. 
While prior literature suggests that OHCs should provide mechanisms for patients/caregivers to find and connect with similar strangers who are better equipped than FFA to provide informational support~\cite{eschler_im_2017,simoni_peer_2011,frost_social_2008}, our study has taken a sharper focus on instrumental support. Thus we next focus our discussion on accommodating divergent preferences for instrumental support. 

\subsubsection{Divergent Instrumental Support Preferences of Patients vs. Caregivers}
Our survey shows that patients and caregivers do not differ significantly in their mean ratings of the importance of instrumental support as a high level support category, however they do diverge in how useful they find specific types of instrumental support. This quantitative finding resonates with broader qualitative themes from work by Berry et al. who found that challenging asymmetries can exist in patient/caregiver values, that caregiver prerogatives can conflict with patient autonomy, and that responsibilities in patient/caregiver dyads often need to shift over the course of health crises~\cite{berry_how_2017}. Berry et al. suggest that technology design may help to navigate these differences---and our results point to specific ways that the instrumental support divergences uncovered in our survey can impact design. 

For example, an OHC like CaringBridge could implement a feature into their journaling or planning tools that provides ways for users to view and select from different types of instrumental support tasks or activities---we will refer to this generically as an ``instrumental support recommender'' throughout this section. An instrumental support recommender could use UI prompts, drop-down menus, or pre-composed text suggestions to reduce the burden on authors of thinking of what might be helpful to them when they are creating tasks in the Planner or composing a narrative description of what is useful to them. One way that our findings can specifically inform design is by impacting the default settings of an instrumental support recommender. For example, while patients/caregivers both rate chores and food to be two of the most useful forms of instrumental support, patients next rate transportation and exercise, whereas caregivers next rate financial assistance and personal care. When an author starts a site and there is little information known about them, prompts could be delivered in a different sequence. Both patients and caregivers could receive an initial prompt to consider asking for chores and food, whereas in subsequent prompts, patients might receive messaging around transportation and exercise, whereas caregivers might receive prompts for help organizing a GoFundMe campaign (i.e. financial assistance) or a trip to the spa (i.e. personal care). Recent work has also begun exploring how to predict the types of support that users need, based on features such as demographic information, reading or posting behavior, and self-perceived roles~\cite{choi_toward_2017}. As authors use the site more frequently, suggestions could be automatically refined based on their behavior and use of the system.


\subsubsection{Divergent Instrumental Support Preferences of Patients and Caregivers vs. Friends, Family, and Acquaintances}
Furthermore, FFA also diverge from patients/caregivers in terms of what types of instrumental support they are interested in providing. For example, patients/caregivers weigh help with chores (e.g. pet care, cleaning, errands, etc.) as the most useful form of instrumental support, whereas FFA rate interest in providing chores as fourth.

One way to bridge these differences is to make FFA more aware of what types of help are most needed; perhaps if FFA knew what patients/caregivers really needed, their interest in providing certain types of help could be shifted. Although our quantitative data do not point to specific ways to achieve this through design, we will elaborate on the ``instrumental support recommender'' mentioned above, as a conceptual exploration. If patients/caregivers had an instrumental support recommender, this feature could include explanations such as: \textit{``Let visitors know what you would most like help with,''} and allow patients/caregivers to select help types that would be most useful to them. In response, UI prompts to FFA while they are writing comments on Journal updates could suggest that FFA offer to provide the types of help that are most needed. For example, a pop-up window could say: \textit{``[Patient's name] could really use help with cleaning the house. Do you want to offer this type of help in your message?''} In this way, CaringBridge could nudge patients/caregivers to communicate their needs through the system rather than expending the emotional energy required to directly ask for specific things. Some work already points to the utility of system-directed support for OHC users. For example, Peng et al. built a bot that provides writing support to users posting supportive comments in a simulated OHC for people with depression~\cite{pengexploring}. They found that the bot helped to increase the amount of emotional and informational support in the comments, and it improved users' confidence in their comments, however some users expressed concerns about authenticity and sincerity, if the bot were to be deployed in a real community~\cite{pengexploring}. Thus, future research should carefully examine not only the effectiveness, but also the acceptability of using technology to provide nudges intended to align the needs of patients and caregivers with the interests of FFA. 

Another possibility to bridge these differences would be an interface that allows FFA to transform cash flow directly into the types of help that are most needed. Our results suggest that for both co-located and remote FFA, interest in providing generic financial assistance is higher than their interest in funding all instrumental support types other than food and practical items, whereas for some patients and caregivers, help with chores or transportation might be more useful. For example, in the Planner, if patients/caregivers create a calendar task related to chores, FFA could be offered an option to either sign up themselves, or else pay for services such as house cleaning, dog walking, or handy work. Such a tool would require some degree of configuration ahead of time, to ensure that appropriate and trusted service providers were available, as we discuss next.

\subsection{Expand the Instrumental Support Network}\label{expand}
To expand the instrumental support network is to develop new mechanisms or systems that enable a broader set of people and/or businesses to contribute to meeting the instrumental needs of patients and caregivers. Figure ~\ref{fig:trust} reveals the unsurprising trend that the closer a person's relationship is to the patient, the more they are trusted by patients/caregivers to provide instrumental support. Table ~\ref{tab:anova} (in appendix) also shows that FFA are more interested in providing help when they have a closer relationship and geographical proximity to the patient, and when they perceive greater support needs, but a lower frequency of asking for help. However, in our survey sample, over 52\% of FFA respondents are less socially close (i.e. acquaintances or coworkers) and over 56\% indicated that they are not located within driving distance of the patient they know. These results suggest that a greater burden falls on those who are closest to the patient, both socially and geographically---a concept supported by prior work. For example, more socially distant FFA may initially provide some support, but their support tends to taper off over time, leaving most of the burden on only a small number of very close social ties~\cite{martins_friendsourcing_2014}. Thus, we join others (e.g.~\cite{martins_friendsourcing_2014,macleod_be_2017,skeels_catalyzing_2010}) in emphasizing that expanding instrumental support networks is an important challenge for technology designers.

\subsubsection{Coordinating Trusted Social Connections}
Given that people already trust their closer social connections to provide them with instrumental support, new tools on CaringBridge could make it easier to offload caregiving tasks to coordinate and facilitate FFA support activities. For instance, ``friendsourcing'' has been explored in the context of information gathering to improve the information available to or about users of social networks, in both health~\cite{martins_friendsourcing_2014} and non-health specific online networks~\cite{bernstein_personalization_2010}. Friendsourced applications ``recruit a user's motivated friends rather than require the user to do work on their own behalf'', and they do so using social incentives and mechanisms to retain existing users and spread to new users~\cite{bernstein_personalization_2010}. Bernstein et al. designed an online game on Facebook that allows people to create tags with new information about users that does not currently exist on their profiles, and rewards points for tagging new friends. Such friendsourcing mechanisms are able to gather unique, accurate and useful information about people~\cite{bernstein_personalization_2010,martins_friendsourcing_2014}, and might be conceptually expanded to allow FFA to collaborate to provide instrumental support. For example, CaringBridge could create affordances that allow patients/caregivers to indicate which visitors to their site are close social connections who are trusted to provide instrumental support. Rather than relying exclusively on patients and caregivers, who are already overburdened, these trusted FFA could have the ability to suggest tasks in the Planner tool, and tag others in the network who are trusted to help, or who patients/caregivers should consider providing with trusted status, thus allowing the support network to grow organically over time. Skeels et al. suggest that online tools for providing support should allow patients/caregivers to politely reject unwanted help~\cite{skeels_catalyzing_2010}; such a functionality should always allow patients/caregivers to approve or decline help offers.


\subsubsection{Integrating App-Based Services}
Alongside traditional brick-and-mortar businesses, an ever-increasing array of app-based services could be deployed to help patients and caregivers, such as the efforts now being made by companies like Lyft~\cite{larock_lyft_2009} and Airbnb~\cite{airbnb_airbnb_2019}. Our survey shows that patients/caregivers trust their social connections more than businesses, and that they generally trust traditional businesses more than app-based businesses. However, Table ~\ref{tab:trustregression} shows that prior use of app-based businesses for transportation (e.g. Uber, Lyft) and shelter (e.g. Airbnb) predicts trust in app-based businesses to provide these services for cancer patients. These platforms are some of the first, most well-established, and most available; interestingly, some app-based businesses appear to earn more trust from patients/caregivers than traditional businesses that offer the same services. Figure ~\ref{fig:trust} shows that 37\% of patients/caregivers trust app-based transportation services while only 28\% trust traditional services, like taxis. Similarly, 52\% trust app-based services for financial support (e.g. GoFundMe), while 50\% trust traditional businesses (e.g. banks). 

For some instrumental support types, trust in acquaintances and colleagues is comparable to trust in traditional or app-based businesses. For example, chores was rated the most useful instrumental support type by both patients and caregivers. Coworkers are less trusted than traditional or app-based businesses to provide chores, but perhaps a coworker might fund a service through an integration between an app-based company and an OHC, if it were simple and convenient to do so. While trust and safety are especially challenging for adoption of app-based businesses in disadvantaged communities~\cite{dillahunt_promise_2015}, McLaclan et al. suggest that trust may be built over time through a Minimal Sharing Paradigm, where initial small amounts of trust can eventually accrue to a larger amount of trust~\cite{mclachlan_you_2016}; our results support this concept. Furthermore, although more rural areas may have less access to or trust in app-based services~\cite{thebault-spieker_toward_2017}, our data indicate that the degree to which users reside in rural or urban areas does not affect trust or interest in providing support to cancer patients and caregivers. To summarize, encouraging and enabling more distant FFA to provide the most needed types of instrumental support could alleviate the burden on closer social connections, yet future research will need to untangle \textit{how} to design systems that patients/caregivers and FFA trust. While some work has begun investigating peoples' motivations to provide emotional and instrumental support outside of OHCs~\cite{wohn_explaining_2018}, we join others (e.g.~\cite{macleod_be_2017}) in suggesting that more in-depth work engage with FFA to understand how to elicit more and better social support for patients and caregivers dealing with many different types of challenging health conditions.

\subsection{Future Work to Empower Family Caregivers}
In our survey, 45.5\% of patients and 58.7\% of family caregivers ask for help less frequently than they need it. Furthermore, 21.5\% of patients and 31.3\% of family caregivers \textit{almost never} ask for help, even if they need it sometimes. While these numbers suggest that both patients and caregivers are not receiving all the help they need, caregivers are less likely than patients to seek it. Much work in HCI has focused on enhancing individual health management, yet caregivers are often overlooked despite being critical contributors to the health of patients~\cite{schorch_designing_2016,chen_caring_2013}. Caregiving can be extremely demanding, with dramatic physical, emotional, and financial impacts~\cite{chen_caring_2013,chih_communicating_2013,umberson_social_2010}. Older caregivers may even have an increased risk of mortality~\cite{schulz_caregiving_1999}. In addition to our nine categories of instrumental support, Skeels et al. also describe the ``meta'' task of coordination~\cite{skeels_catalyzing_2010}, which constitutes another form of real work that is often taken on by caregivers. Previous technology designed for caregivers has focused on communication and coordination, yet these systems can be problematic if they make it simpler to unintentionally add \textit{even more} burdensome work to caregivers~\cite{chen_caring_2013,macleod_be_2017}.

We suggest that to empower family caregivers is to recognize these caregivers as essential stakeholders alongside patients, and to design for their challenges and needs as users, which may differ from those of patients. This divergence between patient and caregiver needs is especially salient on CaringBridge, since authorship trends suggest that caregivers are writing more journal updates than patients. In general, CaringBridge journals are useful for sharing ongoing updates about a health crisis in one central location, often saving time over other alternatives such as calling or emailing people individually, yet communication and coordination burdens are nonetheless falling more heavily on caregivers. The emotional toll of caregiving may be tricky or impractical to express publicly to FFA, since caregivers who discuss their own needs and challenges on patients' sites may run the risk of being perceived as blaming the patient or making the patient feel like a burden. DuBenske et al. have demonstrated that providing caregivers of advanced-stage cancer patients with online tools\footnote{See the Comprehensive Health Enhancement Support System (CHESS) system, https://center.chess.wisc.edu/research-projects/view/alcohol-and-other-drugs-help-for-parents-and-partners-of-abusers.} for communicating \textit{their own needs} (alongside those of patients) to \textit{clinicians} can reduce their emotional distress, improve understanding and coping skills and ease their burden~\cite{dubenske_chess_2014}. Could a separate online space or separate tools on CaringBridge designed specifically for caregivers only ease the burdens on them and help them get the support they need for themselves? As is the case with rare diseases~\cite{macleod_be_2017}, a more effective strategy may be for caregivers to elicit support from other caregivers in similar situations~\cite{nicholas_evaluation_2012}, and to have a more explicit focus on taking care of their own needs and concerns. Future work should explore ways to help caregivers care for themselves, even as they must navigate the challenges of caring for patients, as well.


\subsection{Future Work to Support Spirituality}
In her 2010 CHI keynote, Genevieve Bell asks, ``How do we study things that are deeply important to human beings'' including religion and spirituality, which are embraced by over 80\% of the world's population?~\cite{bell_chi_2010} Religious organizations and spiritual belief systems---as well as the people who belong to or adopt them---must now grapple with the relationship between technology and spiritual practice in daily life. As one interesting example, the Vatican recently launched an eRosary wearable product, intending to engage more youth in the Catholic faith in prayer activities~\cite{england_vatican_nodate}. Furthermore, hybrid online/offline practices such as ``prayer requests'' occur over email~\cite{wyche_technology_2006} and in OHCs like CaringBridge~\cite{anderson_uses_2011}. (Anecdotally, we observe that this phenomenon occurs on many social media sites (e.g., Facebook, Twitter, Reddit, etc.), although we have not specifically characterized it on those platforms.) Because prior literature on social support in OHCs has not included ``prayer support'', we did not originally seek to measure it. In contrast with prior work characterizing social support in OHCs, results from our content analysis and survey align, indicating that prayer support is not only the most frequently acknowledged support type in CaringBridge journals, but also, it is (on average) perceived as most important by both patients/caregivers and FFA. This result raises a number of critical considerations.

First, the demographics of OHC users do not necessarily align with the general public. CaringBridge user demographics skew white, female, upper class, and Christian~\cite{anderson_uses_2011}. Thus, it is possible that CaringBridge users have unique or different priorities related to prayer than the public at large. However, as a coping mechanism, patients with advanced cancers often rely on spiritual beliefs, while spiritual care also contributes to higher quality of life~\cite{vallurupalli_role_2012}. More than 39-50\% of ethnically diverse patients reported needs for overcoming fear, finding hope, meaning, peace of mind, and locating spiritual resources~\cite{moadel_seeking_1999}. Minority racial groups are even more likely than white people to report these needs~\cite{moadel_seeking_1999}. Furthermore, as people approach the end of their lives, most people experience an increase in concerns about the meaning of their lives, or their spiritual or religious beliefs~\cite{puchalski2011making}. Ma et al. show that \textasciitilde37\% of CaringBridge cancer sites end due to the death of the patient~\cite{ma_write_2017}---it may be the case that prayer is of elevated importance to these users because of proximity to death~\cite{brubaker_workshop}. Gender may also influence peoples' priorities. For instance, men are more likely to seek informational support, while women are likely to seek both emotional and informational support~\cite{gray_breast_1996}. To understand the importance of prayer and spirituality beyond CaringBridge, future work should evaluate this topic on other OHCs and general social media. 

Second, we suggest that our definition of prayer support (``prayers, spiritual blessings, positive karma, good juju, warm thoughts'') is useful for future content analyses of user-generated text in online communities (not necessarily health-only). However, given the importance of prayer to people, we suggest that a focus on spirituality more generally will open broader avenues of research in HCI. Along these lines, Wyche et al. present \textit{extraordinary computing} as a framework for ``systems that recognize, support, and honor meaningful aspects in users' domestic lives''~\cite{wyche_extraordinary_2009}; peoples' spirituality or religion certainly falls under this umbrella. Although our methods in this work are unable to indicate specific technological implementations that would serve users' senses of spirituality, our work points to the need and importance of future work in this area. 

Therefore, to address the limitations of our methodology, future work should engage in participatory design methods~\cite{muller_participatory_1993} with people who hold spiritual and religious values in order to: (1) determine a broader working definition of ``spiritual support'' for the HCI community than the definition of prayer support presented here, and (2) discern design principles for expressing and meeting peoples' needs for spiritual support. To support these aims and guide future work in extraordinary computing, we turn to the literature on spirituality and healing. While there are many definitions of spirituality~\cite{astrow_religion_2001}, a widely accepted definition is: \textit{``Spirituality is the aspect of humanity that refers to the way individuals seek and express meaning and purpose and the way they experience their connectedness to the moment, to self, to others, to nature, and to the significant or sacred''}~\cite{puchalski_improving_2009}. This definition has been used across a variety of research and medical contexts, and should be used to inform future efforts in HCI. Finally, we suggest that to support spirituality means to design and implement tools or interaction modalities that communicate, visualize, or embody the experience of requesting or receiving spiritual support, including but not limited to prayer support.

\subsection{Conclusion}
In this paper, we contribute a quantitative understanding of the sociotechnical design space for mediating instrumental support via OHCs. We introduce ``prayer support'' as a new category for study in online social support, and suggest that future work should further explore the relationship between technology and prayer/spirituality during health crises. We also contribute a data-derived lexicon of appreciation-related words and a codebook of instrumental support types, both of which support future work in this space. We offer the design implications that OHCs should accommodate divergent preferences for social support and expand the instrumental support network. Finally, we discuss the need for future work to empower the caregiver and to support spirituality. Participatory design may address the limitations of our methodologies and provide rich insights into designing compassionate technology for facilitating social support for patients with life-threatening illness. 

\begin{acks}
We gratefully acknowledge Wenqi Luo, Haonan Tian, Stephanie Herbers, Anna Meyer, and Mikhaila Friske for their assistance with our directed content analysis. We thank our colleagues Haiyi Zhu and Jacob Thebault-Spieker for their useful methodological suggestions, and Mo Houtti, in addition to our anonymous reviewers, for their thoughtful critiques which have helped us to improve this manuscript over time. Finally, we are deeply grateful to all members of the CaringBridge team who contributed to the design and deployment of our survey, and who provided feedback on the paper; this work could not have been completed without their help. The first author of this work was supported by the Graduate Assistance in Areas of National Need (GAANN) Fellowship.
\end{acks}

\begin{center}
\textasciitilde \\
\textit{This paper is dedicated in loving memory to Christine Elizabeth (Ringle) Smith, who died of a rare endometrial sarcoma on June 9, 2015. Her dearest friends maintained a CaringBridge site for her. At the time, her daughter (first author of this work) was considering PhD programs...and the rest is history.}
\end{center}

\bibliographystyle{ACM-Reference-Format}
\bibliography{acmsmall}

%% file: appendix.tex
\appendix

\MLauthors
\classifiertable

\section{Details of Authorship Classification}\label{author_ML}
Vowpal Wabbit \cite{langford_vowpal_2007} was used to train a binary logistic regression classifier with L2 regularization on human-annotated data. Humans completed annotations of 7154 journal updates, of which 6167 were patient-authored updates and 957 were non-patient-authored. These numbers include sites that were annotated for the directed content analysis reported in this paper. Additionally, these numbers include journal updates non-randomly selected for a different qualitative exploration of the data that is not presented in this work. Features used were the unigrams, bigrams, and trigrams in the training data. After downsampling patient-authored posts, the data was randomly split into 2296 training updates and 575 validation updates. As reported in Table~\ref{author_classifier}, average F1 score on the validation set was 96\%.

To analyze sites in the sample reported in this paper, we applied the trained author type classifier to journal updates with $>$50 characters. We used the predicted update author types to compute the proportion of updates on a site that were patient-authored. 36.53\% of sites (7276) were majority patient-authored (>50\% of updates were classified as patient-authored), versus 63.47\% of sites (12640) that were majority authored by non-patients.

\section{Coding Protocol and Complete Codebook}
\subsection{Coding Protocol and Instructions}\label{protocol}

CaringBridge journal updates can be very long, describing challenging topics related to illness, dying, and death. It can be cognitively and emotionally taxing to require researchers to code without any "hints." Thus, we designed our coding protocol to: (1) reduce reading of irrelevant updates, and (2) visually emphasize expressions of appreciation. To satisfy (1), eligible journals were required to contain at least one word from the appreciation lexicon (section~\ref{lex_dev}). To satisfy (2), all appreciation words were UPPERCASED, while all others were lowercased. Research assistants were provided with a comprehensive and detailed set of instructions. We briefly summarize the procedure here:

\begin{enumerate}
\item Read the first few sentences closely; skim the rest to get a sense of what is happening.
\item First, identify the Author. or skip the update if the Author is "unknown."
\item Next, identify whether the Author wrote any expressions of appreciation (EOA). Locate UPPERCASED appreciation-related word(s), and use your best judgment to decide if appreciation is actually being expressed. Code any additional EOAs you notice, even if they are not accompanied by all-caps words.
\item If there is an EOA, closely examine the immediate vicinity of UPPERCASED appreciation words (both a few sentences before and after) to locate the topic being appreciated. Use the codebook to decide which help type code to apply.
\item Indicate if you feel like you need a second opinion to ensure accurate coding.
\end{enumerate}
The first author revisited all codes where a second opinion was indicated and made final discernments of edge cases, in order to ensure the highest possible consistency of code application.
\newpage 

\subsection{Complete Codebook for Author and Help Type Codes} \label{sec:complete}
\completecodebook

\section{Appreciation Lexicon}\label{lexicon}
\completelex
\newpage

\section{Survey Questions}\label{survey}
\surveyquestions
\newpage

\section{Instrumental Support Ratings Tables}
\usefulness
\altffameans
\newpage

\section{Supplemental Regression Models}\label{supplemental_regressions}

\subsection{Complete Details of Regression Models Predicting Potential Effects of SES on Instrumental Support Results}\label{sec:fullFAregression}
This section contains fully expanded versions of Table~\ref{tab:usefulregression}, which is presented and described in Section~\ref{SES} of the main body of the paper. Table~\ref{tab:newusefulanova} contains ANOVA summary tables. Table~\ref{tab:fullregressionA} contains complete regression details and was generated using Stargazer~\cite{stargazer_package}. 
\newusefulanova
\newpage

\fullFAregressionA
\fullFAregressionB

\newpage

\subsection{FFA Interest in Funding Services vs. Providing Help in Person}\label{fund_vs_inperson}
We fit two multiple linear regression models to untangle the relationship between potential predictors of FFA interest in providing in person help versus funding a service. Because of a high internal consistency among providing INSTR types in person (Cronbach's alpha=0.91) and among funding services for INSTR (Cronbach's alpha=0.95), we use the summative score of FFA interest in providing different types of help in-person, and another summative score of FFA interest in funding a service, for dependent variables. For independent variables, we include answers to all multiple choice questions from sections ~\ref{demosection} (Demographics), ~\ref{RelationshipToPatient} (Relationship and Remoteness to Patient), and ~\ref{TrackC_12} (FFA Track), except question ~\ref{support_types_ffa} (High-Level Help Types), because FFA's perceived importance of providing INSTR is correlated with summative scores of FFA's interest in providing help in-person (corr=0.47) and funding a service (corr=0.30) both semantically and statistically.

We use three stepwise algorithms to distinguish important (active) from unimportant (inactive) predictors based on the Akaike Information Criterion (AIC), all of which separately lead to the same final model: (1) forward selection from null model with intercept only, (2) backward selection from full model with all predictors, and (3) forward-backward selection. For in person interest, the stepwise-selected model keeps age, population class, income, relationship to patient, time challenge, distance challenge, difficulty of providing help over phone/text, and the perceived frequencies of help needed and asked for as active regressors. For funding interest, the stepwise-selected model keeps age, income, relationship to patient, and monetary challenge as active regressors. Remoteness to patient highly correlates with distance challenge (|corr|=0.76); the selection algorithm retains distance challenge (question ~\ref{ask_challenge_ffa}) but drops remoteness (question ~\ref{driving}). Full regression results specify values for each level of unordered factors and each contrast of ordered factors, however we summarize the trend of regression outputs and present the statistical significance of each factor based on analysis of variance (ANOVA) outputs in Table ~\ref{tab:anova}. 
\anovaresults
\newpage

\subsection{Patients and Caregivers' Trust in App-Based vs. Traditional Businesses}\label{trust_regression}
To understand predictors of of patients/caregivers' trust in app-based versus traditional businesses, we fit two logistic regression models per each of the 9 instrumental support types. For independent variables, we include answers from sections ~\ref{role} (Survey Track Selection), ~\ref{demosection} (Demographics), and ~\ref{TracksABC_3} (Familiarity with App-Based Services). Table ~\ref{tab:trustregression} presents Bonferroni-corrected statistical significance of factors based on ANOVA. 
\trustregression

%% file: acmsmall.bbl

\begin{thebibliography}{131}


\ifx \showCODEN    \undefined \def \showCODEN     #1{\unskip}     \fi
\ifx \showDOI      \undefined \def \showDOI       #1{#1}\fi
\ifx \showISBNx    \undefined \def \showISBNx     #1{\unskip}     \fi
\ifx \showISBNxiii \undefined \def \showISBNxiii  #1{\unskip}     \fi
\ifx \showISSN     \undefined \def \showISSN      #1{\unskip}     \fi
\ifx \showLCCN     \undefined \def \showLCCN      #1{\unskip}     \fi
\ifx \shownote     \undefined \def \shownote      #1{#1}          \fi
\ifx \showarticletitle \undefined \def \showarticletitle #1{#1}   \fi
\ifx \showURL      \undefined \def \showURL       {\relax}        \fi
\providecommand\bibfield[2]{#2}
\providecommand\bibinfo[2]{#2}
\providecommand\natexlab[1]{#1}
\providecommand\showeprint[2][]{arXiv:#2}

\bibitem[\protect\citeauthoryear{Adler}{Adler}{2005}]%
        {adler_appreciation:_2005}
\bibfield{author}{\bibinfo{person}{Mitchel~G. Adler}.}
  \bibinfo{year}{2005}\natexlab{}.
\newblock \showarticletitle{Appreciation: {Individual} {Differences} in
  {Finding} {Value} and {Meaning} as a {Unique} {Predictor} of {Subjective}
  {Well}-{Being}.}
\newblock \bibinfo{journal}{\emph{Journal of Personality}}
  \bibinfo{volume}{73}, \bibinfo{number}{1} (\bibinfo{date}{Feb.}
  \bibinfo{year}{2005}), \bibinfo{pages}{79--114}.
\newblock
\showISSN{00223506}


\bibitem[\protect\citeauthoryear{Airbnb}{Airbnb}{2019}]%
        {airbnb_airbnb_2019}
\bibfield{author}{\bibinfo{person}{Airbnb}.} \bibinfo{year}{2019}\natexlab{}.
\newblock \bibinfo{title}{Airbnb {Expands} {Medical} {Stays} {Program} with
  \$1.2M in {Funding} to {New} {Partners}}.
\newblock
\newblock
\urldef\tempurl%
\url{https://press.airbnb.com/airbnb-expands-medical-stays-program-with-1-2m-in-funding-to-new-partners/}
\showURL{%
\tempurl}


\bibitem[\protect\citeauthoryear{Andalibi and Forte}{Andalibi and
  Forte}{2018}]%
        {andalibi_responding_2018}
\bibfield{author}{\bibinfo{person}{Nazanin Andalibi} {and}
  \bibinfo{person}{Andrea Forte}.} \bibinfo{year}{2018}\natexlab{}.
\newblock \showarticletitle{Responding to {Sensitive} {Disclosures} on {Social}
  {Media}: {A} {Decision}-{Making} {Framework}}.
\newblock \bibinfo{journal}{\emph{ACM Trans. Comput.-Hum. Interact.}}
  \bibinfo{volume}{25}, \bibinfo{number}{6} (\bibinfo{date}{Dec.}
  \bibinfo{year}{2018}), \bibinfo{pages}{31:1--31:29}.
\newblock
\showISSN{1073-0516}
\urldef\tempurl%
\url{https://doi.org/10.1145/3241044}
\showDOI{\tempurl}


\bibitem[\protect\citeauthoryear{Andalibi, Haimson, Choudhury, and
  Forte}{Andalibi et~al\mbox{.}}{2018}]%
        {andalibi_social_2018}
\bibfield{author}{\bibinfo{person}{Nazanin Andalibi},
  \bibinfo{person}{Oliver~L. Haimson}, \bibinfo{person}{Munmun~De Choudhury},
  {and} \bibinfo{person}{Andrea Forte}.} \bibinfo{year}{2018}\natexlab{}.
\newblock \showarticletitle{Social {Support}, {Reciprocity}, and {Anonymity} in
  {Responses} to {Sexual} {Abuse} {Disclosures} on {Social} {Media}}.
\newblock \bibinfo{journal}{\emph{ACM Trans. Comput.-Hum. Interact.}}
  \bibinfo{volume}{25}, \bibinfo{number}{5} (\bibinfo{date}{Oct.}
  \bibinfo{year}{2018}), \bibinfo{pages}{28:1--28:35}.
\newblock
\showISSN{1073-0516}
\urldef\tempurl%
\url{https://doi.org/10.1145/3234942}
\showDOI{\tempurl}


\bibitem[\protect\citeauthoryear{Andalibi, Ozturk, and Forte}{Andalibi
  et~al\mbox{.}}{2017}]%
        {andalibi_sensitive_2017}
\bibfield{author}{\bibinfo{person}{Nazanin Andalibi}, \bibinfo{person}{Pinar
  Ozturk}, {and} \bibinfo{person}{Andrea Forte}.}
  \bibinfo{year}{2017}\natexlab{}.
\newblock \showarticletitle{Sensitive {Self}-disclosures, {Responses}, and
  {Social} {Support} on {Instagram}: {The} {Case} of \#{Depression}}. In
  \bibinfo{booktitle}{\emph{Proceedings of the 2017 {ACM} {Conference} on
  {Computer} {Supported} {Cooperative} {Work} and {Social} {Computing}}}
  \emph{(\bibinfo{series}{{CSCW} '17})}. \bibinfo{publisher}{ACM},
  \bibinfo{address}{New York, NY, USA}, \bibinfo{pages}{1485--1500}.
\newblock
\showISBNx{978-1-4503-4335-0}
\urldef\tempurl%
\url{https://doi.org/10.1145/2998181.2998243}
\showDOI{\tempurl}
\newblock
\shownote{event-place: Portland, Oregon, USA.}


\bibitem[\protect\citeauthoryear{Anderson}{Anderson}{2011}]%
        {anderson_uses_2011}
\bibfield{author}{\bibinfo{person}{Isolde~K. Anderson}.}
  \bibinfo{year}{2011}\natexlab{}.
\newblock \showarticletitle{The {Uses} and {Gratifications} of {Online} {Care}
  {Pages}: {A} {Study} of {CaringBridge}}.
\newblock \bibinfo{journal}{\emph{Health Communication}} \bibinfo{volume}{26},
  \bibinfo{number}{6} (\bibinfo{date}{Sept.} \bibinfo{year}{2011}),
  \bibinfo{pages}{546--559}.
\newblock
\showISSN{1041-0236}
\urldef\tempurl%
\url{https://doi.org/10.1080/10410236.2011.558335}
\showDOI{\tempurl}


\bibitem[\protect\citeauthoryear{Astrow, Puchalski, and Sulmasy}{Astrow
  et~al\mbox{.}}{2001}]%
        {astrow_religion_2001}
\bibfield{author}{\bibinfo{person}{A.~B. Astrow}, \bibinfo{person}{C.~M.
  Puchalski}, {and} \bibinfo{person}{D.~P. Sulmasy}.}
  \bibinfo{year}{2001}\natexlab{}.
\newblock \showarticletitle{Religion, spirituality, and health care: social,
  ethical, and practical considerations.}
\newblock \bibinfo{journal}{\emph{The American journal of medicine}}
  \bibinfo{volume}{110}, \bibinfo{number}{4} (\bibinfo{date}{March}
  \bibinfo{year}{2001}), \bibinfo{pages}{283--287}.
\newblock
\showISSN{0002-9343}
\urldef\tempurl%
\url{https://doi.org/10.1016/s0002-9343(00)00708-7}
\showDOI{\tempurl}


\bibitem[\protect\citeauthoryear{Baikie and Wilhelm}{Baikie and
  Wilhelm}{2005}]%
        {baikie_emotional_2005}
\bibfield{author}{\bibinfo{person}{Karen~A. Baikie} {and} \bibinfo{person}{Kay
  Wilhelm}.} \bibinfo{year}{2005}\natexlab{}.
\newblock \showarticletitle{Emotional and physical health benefits of
  expressive writing}.
\newblock \bibinfo{journal}{\emph{Advances in Psychiatric Treatment}}
  \bibinfo{volume}{11}, \bibinfo{number}{5} (\bibinfo{date}{Sept.}
  \bibinfo{year}{2005}), \bibinfo{pages}{338--346}.
\newblock
\showISSN{1355-5146, 1472-1481}
\urldef\tempurl%
\url{https://doi.org/10.1192/apt.11.5.338}
\showDOI{\tempurl}


\bibitem[\protect\citeauthoryear{Balani and De~Choudhury}{Balani and
  De~Choudhury}{2015}]%
        {balani_detecting_2015}
\bibfield{author}{\bibinfo{person}{Sairam Balani} {and} \bibinfo{person}{Munmun
  De~Choudhury}.} \bibinfo{year}{2015}\natexlab{}.
\newblock \showarticletitle{Detecting and {Characterizing} {Mental} {Health}
  {Related} {Self}-{Disclosure} in {Social} {Media}}. In
  \bibinfo{booktitle}{\emph{Proceedings of the 33rd {Annual} {ACM} {Conference}
  {Extended} {Abstracts} on {Human} {Factors} in {Computing} {Systems}}}
  \emph{(\bibinfo{series}{{CHI} {EA} '15})}. \bibinfo{publisher}{ACM},
  \bibinfo{address}{New York, NY, USA}, \bibinfo{pages}{1373--1378}.
\newblock
\showISBNx{978-1-4503-3146-3}
\urldef\tempurl%
\url{https://doi.org/10.1145/2702613.2732733}
\showDOI{\tempurl}
\newblock
\shownote{event-place: Seoul, Republic of Korea.}


\bibitem[\protect\citeauthoryear{Barbee and Cunningham}{Barbee and
  Cunningham}{1995}]%
        {barbee_experimental_1995}
\bibfield{author}{\bibinfo{person}{Anita~P. Barbee} {and}
  \bibinfo{person}{Michael~R. Cunningham}.} \bibinfo{year}{1995}\natexlab{}.
\newblock \showarticletitle{An {Experimental} {Approach} to {Social} {Support}
  {Communications}: {Interactive} {Coping} in {Close} {Relationships}}.
\newblock \bibinfo{journal}{\emph{Annals of the International Communication
  Association}} \bibinfo{volume}{18}, \bibinfo{number}{1} (\bibinfo{date}{Jan.}
  \bibinfo{year}{1995}), \bibinfo{pages}{381--413}.
\newblock
\showISSN{2380-8985, 2380-8977}
\urldef\tempurl%
\url{https://doi.org/10.1080/23808985.1995.11678921}
\showDOI{\tempurl}


\bibitem[\protect\citeauthoryear{Bell}{Bell}{2010}]%
        {bell_chi_2010}
\bibfield{author}{\bibinfo{person}{Genevieve Bell}.}
  \bibinfo{year}{2010}\natexlab{}.
\newblock \bibinfo{title}{{CHI} 2010 {Opening} {Plenary}: {Messy} {Futures}:
  {Culture} (see clip at 32:40)}.
\newblock
\newblock
\urldef\tempurl%
\url{https://www.youtube.com/watch?v=-a2gj2clzTk}
\showURL{%
\tempurl}


\bibitem[\protect\citeauthoryear{Bernstein, Tan, Smith, Czerwinski, and
  Horvitz}{Bernstein et~al\mbox{.}}{2010}]%
        {bernstein_personalization_2010}
\bibfield{author}{\bibinfo{person}{Michael~S Bernstein},
  \bibinfo{person}{Desney Tan}, \bibinfo{person}{Greg Smith},
  \bibinfo{person}{Mary Czerwinski}, {and} \bibinfo{person}{Eric Horvitz}.}
  \bibinfo{year}{2010}\natexlab{}.
\newblock \showarticletitle{Personalization via friendsourcing}.
\newblock \bibinfo{journal}{\emph{ACM Transactions on Computer-Human
  Interaction (TOCHI)}} \bibinfo{volume}{17}, \bibinfo{number}{2}
  (\bibinfo{year}{2010}), \bibinfo{pages}{6}.
\newblock


\bibitem[\protect\citeauthoryear{Berry, Lim, Hartzler, Hirsch, Wagner, Ludman,
  and Ralston}{Berry et~al\mbox{.}}{2017}]%
        {berry_how_2017}
\bibfield{author}{\bibinfo{person}{Andrew B.~L. Berry},
  \bibinfo{person}{Catherine Lim}, \bibinfo{person}{Andrea~L. Hartzler},
  \bibinfo{person}{Tad Hirsch}, \bibinfo{person}{Edward~H. Wagner},
  \bibinfo{person}{Evette Ludman}, {and} \bibinfo{person}{James~D. Ralston}.}
  \bibinfo{year}{2017}\natexlab{}.
\newblock \showarticletitle{How {Values} {Shape} {Collaboration} {Between}
  {Patients} with {Multiple} {Chronic} {Conditions} and {Spousal}
  {Caregivers}}. In \bibinfo{booktitle}{\emph{Proceedings of the 2017 {CHI}
  {Conference} on {Human} {Factors} in {Computing} {Systems}}}
  \emph{(\bibinfo{series}{{CHI} '17})}. \bibinfo{publisher}{ACM},
  \bibinfo{address}{New York, NY, USA}, \bibinfo{pages}{5257--5270}.
\newblock
\showISBNx{978-1-4503-4655-9}


\bibitem[\protect\citeauthoryear{Botsman}{Botsman}{2015}]%
        {botsman_defining_2015}
\bibfield{author}{\bibinfo{person}{Rachel Botsman}.}
  \bibinfo{year}{2015}\natexlab{}.
\newblock \showarticletitle{Defining the sharing economy: what is collaborative
  consumption–and what isn't}.
\newblock \bibinfo{journal}{\emph{Fast Company}}  \bibinfo{volume}{27}
  (\bibinfo{year}{2015}).
\newblock


\bibitem[\protect\citeauthoryear{Botsman and Rogers}{Botsman and
  Rogers}{2011}]%
        {botsman_whats_2011}
\bibfield{author}{\bibinfo{person}{Rachel Botsman} {and} \bibinfo{person}{Roo
  Rogers}.} \bibinfo{year}{2011}\natexlab{}.
\newblock \showarticletitle{What's mine is yours: how collaborative consumption
  is changing the way we live}.
\newblock  (\bibinfo{year}{2011}).
\newblock


\bibitem[\protect\citeauthoryear{Brown, Nesse, Vinokur, and Smith}{Brown
  et~al\mbox{.}}{2003}]%
        {brown_providing_2003}
\bibfield{author}{\bibinfo{person}{Stephanie~L. Brown},
  \bibinfo{person}{Randolph~M. Nesse}, \bibinfo{person}{Amiram~D. Vinokur},
  {and} \bibinfo{person}{Dylan~M. Smith}.} \bibinfo{year}{2003}\natexlab{}.
\newblock \showarticletitle{Providing {Social} {Support} {May} {Be} {More}
  {Beneficial} {Than} {Receiving} {It}: {Results} {From} a {Prospective}
  {Study} of {Mortality}}.
\newblock \bibinfo{journal}{\emph{Psychological Science}} \bibinfo{volume}{14},
  \bibinfo{number}{4} (\bibinfo{date}{July} \bibinfo{year}{2003}),
  \bibinfo{pages}{320--327}.
\newblock
\showISSN{0956-7976}


\bibitem[\protect\citeauthoryear{Brubaker and Vertesi}{Brubaker and
  Vertesi}{2010}]%
        {brubaker_workshop}
\bibfield{author}{\bibinfo{person}{Jed~R. Brubaker} {and}
  \bibinfo{person}{Janet Vertesi}.} \bibinfo{year}{2010}\natexlab{}.
\newblock \showarticletitle{{Death and the Social Network}}. In
  \bibinfo{booktitle}{\emph{Workshop on HCI at the End of Life: Understanding
  Death, Dying, and the Digital at the 28th ACM Conference on Human Factors in
  Computing Systems - CHI'10}}.
\newblock


\bibitem[\protect\citeauthoryear{Bruckman, Fiesler, Hancock, and
  Munteanu}{Bruckman et~al\mbox{.}}{2017}]%
        {bruckman_cscw_2017}
\bibfield{author}{\bibinfo{person}{Amy~S. Bruckman}, \bibinfo{person}{Casey
  Fiesler}, \bibinfo{person}{Jeff Hancock}, {and} \bibinfo{person}{Cosmin
  Munteanu}.} \bibinfo{year}{2017}\natexlab{}.
\newblock \showarticletitle{{CSCW} {Research} {Ethics} {Town} {Hall}: {Working}
  {Towards} {Community} {Norms}}. In \bibinfo{booktitle}{\emph{Companion of the
  2017 {ACM} {Conference} on {Computer} {Supported} {Cooperative} {Work} and
  {Social} {Computing}}} \emph{(\bibinfo{series}{{CSCW} '17 {Companion}})}.
  \bibinfo{publisher}{ACM}, \bibinfo{address}{New York, NY, USA},
  \bibinfo{pages}{113--115}.
\newblock
\showISBNx{978-1-4503-4688-7}


\bibitem[\protect\citeauthoryear{CaringBridge}{CaringBridge}{2018}]%
        {caringbridge_caringbridge_nodate}
\bibfield{author}{\bibinfo{person}{CaringBridge}.}
  \bibinfo{year}{2018}\natexlab{}.
\newblock \bibinfo{title}{How CaringBridge Supports Personal Fundraisers
  Through GoFundMe}.
\newblock
\newblock
\urldef\tempurl%
\url{https://www.caringbridge.org/resources/gofundme/}
\showURL{%
\tempurl}


\bibitem[\protect\citeauthoryear{Chancellor, Mitra, and
  De~Choudhury}{Chancellor et~al\mbox{.}}{2016}]%
        {chancellor_recovery_2016}
\bibfield{author}{\bibinfo{person}{Stevie Chancellor},
  \bibinfo{person}{Tanushree Mitra}, {and} \bibinfo{person}{Munmun
  De~Choudhury}.} \bibinfo{year}{2016}\natexlab{}.
\newblock \showarticletitle{Recovery {Amid} {Pro}-{Anorexia}: {Analysis} of
  {Recovery} in {Social} {Media}}.
\newblock \bibinfo{journal}{\emph{Proceedings of the SIGCHI conference on human
  factors in computing systems . CHI Conference}}  \bibinfo{volume}{2016}
  (\bibinfo{date}{May} \bibinfo{year}{2016}), \bibinfo{pages}{2111--2123}.
\newblock
\urldef\tempurl%
\url{https://doi.org/10.1145/2858036.2858246}
\showDOI{\tempurl}


\bibitem[\protect\citeauthoryear{Chen, Ngo, and Park}{Chen
  et~al\mbox{.}}{2013}]%
        {chen_caring_2013}
\bibfield{author}{\bibinfo{person}{Yunan Chen}, \bibinfo{person}{Victor Ngo},
  {and} \bibinfo{person}{Sun~Young Park}.} \bibinfo{year}{2013}\natexlab{}.
\newblock \showarticletitle{Caring for {Caregivers}: {Designing} for
  {Integrality}}. In \bibinfo{booktitle}{\emph{Proceedings of the 2013
  {Conference} on {Computer} {Supported} {Cooperative} {Work}}}
  \emph{(\bibinfo{series}{{CSCW} '13})}. \bibinfo{publisher}{ACM},
  \bibinfo{address}{New York, NY, USA}, \bibinfo{pages}{91--102}.
\newblock
\showISBNx{978-1-4503-1331-5}


\bibitem[\protect\citeauthoryear{Chih, DuBenske, Hawkins, Brown, Dinauer,
  Cleary, and Gustafson}{Chih et~al\mbox{.}}{2013}]%
        {chih_communicating_2013}
\bibfield{author}{\bibinfo{person}{Ming-Yuan Chih}, \bibinfo{person}{Lori~L
  DuBenske}, \bibinfo{person}{Robert~P Hawkins}, \bibinfo{person}{Roger~L
  Brown}, \bibinfo{person}{Susan~K Dinauer}, \bibinfo{person}{James~F Cleary},
  {and} \bibinfo{person}{David~H Gustafson}.} \bibinfo{year}{2013}\natexlab{}.
\newblock \showarticletitle{Communicating advanced cancer patients' symptoms
  via the {Internet}: a pooled analysis of two randomized trials examining
  caregiver preparedness, physical burden, and negative mood}.
\newblock \bibinfo{journal}{\emph{Palliative medicine}} \bibinfo{volume}{27},
  \bibinfo{number}{6} (\bibinfo{year}{2013}), \bibinfo{pages}{533--543}.
\newblock


\bibitem[\protect\citeauthoryear{Choi, Kim, Lee, Kwon, Yi, Choo, and Huh}{Choi
  et~al\mbox{.}}{2017}]%
        {choi_toward_2017}
\bibfield{author}{\bibinfo{person}{Min-Je Choi}, \bibinfo{person}{Sung-Hee
  Kim}, \bibinfo{person}{Sukwon Lee}, \bibinfo{person}{Bum~Chul Kwon},
  \bibinfo{person}{Ji~Soo Yi}, \bibinfo{person}{Jaegul Choo}, {and}
  \bibinfo{person}{Jina Huh}.} \bibinfo{year}{2017}\natexlab{}.
\newblock \showarticletitle{Toward {Predicting} {Social} {Support} {Needs} in
  {Online} {Health} {Social} {Networks}}.
\newblock \bibinfo{journal}{\emph{Journal of Medical Internet Research}}
  \bibinfo{volume}{19}, \bibinfo{number}{8} (\bibinfo{year}{2017}),
  \bibinfo{pages}{e272}.
\newblock
\urldef\tempurl%
\url{https://doi.org/10.2196/jmir.7660}
\showDOI{\tempurl}


\bibitem[\protect\citeauthoryear{Civan and Pratt}{Civan and Pratt}{2007}]%
        {civan_threading_2007}
\bibfield{author}{\bibinfo{person}{Andrea Civan} {and} \bibinfo{person}{Wanda
  Pratt}.} \bibinfo{year}{2007}\natexlab{}.
\newblock \showarticletitle{Threading together patient expertise}. In
  \bibinfo{booktitle}{\emph{{AMIA} annual symposium proceedings}},
  Vol.~\bibinfo{volume}{2007}. \bibinfo{publisher}{American Medical Informatics
  Association}, \bibinfo{pages}{140}.
\newblock


\bibitem[\protect\citeauthoryear{Cohen}{Cohen}{2004}]%
        {cohen_social_2004}
\bibfield{author}{\bibinfo{person}{Sheldon Cohen}.}
  \bibinfo{year}{2004}\natexlab{}.
\newblock \showarticletitle{Social relationships and health.}
\newblock \bibinfo{journal}{\emph{American psychologist}} \bibinfo{volume}{59},
  \bibinfo{number}{8} (\bibinfo{year}{2004}), \bibinfo{pages}{676}.
\newblock


\bibitem[\protect\citeauthoryear{Commission}{Commission}{2019}]%
        {joint_commission_spiritual_2019}
\bibfield{author}{\bibinfo{person}{Joint Commission}.}
  \bibinfo{year}{2019}\natexlab{}.
\newblock \bibinfo{title}{Spiritual {Assessment}}.
\newblock
\newblock
\urldef\tempurl%
\url{http://www.jointcommission.org/standards_information/jcfaqdetails.aspx}
\showURL{%
\tempurl}


\bibitem[\protect\citeauthoryear{Coursaris and Liu}{Coursaris and Liu}{2009}]%
        {coursaris_analysis_2009}
\bibfield{author}{\bibinfo{person}{Constantinos~K. Coursaris} {and}
  \bibinfo{person}{Ming Liu}.} \bibinfo{year}{2009}\natexlab{}.
\newblock \showarticletitle{An analysis of social support exchanges in online
  {HIV}/{AIDS} self-help groups}.
\newblock \bibinfo{journal}{\emph{Computers in Human Behavior}}
  \bibinfo{volume}{25}, \bibinfo{number}{4} (\bibinfo{date}{July}
  \bibinfo{year}{2009}), \bibinfo{pages}{911--918}.
\newblock
\showISSN{0747-5632}
\urldef\tempurl%
\url{https://doi.org/10.1016/j.chb.2009.03.006}
\showDOI{\tempurl}


\bibitem[\protect\citeauthoryear{Cousson-Gelie, Bruchon-Schweitzer, Dilhuydy,
  and Jutand}{Cousson-Gelie et~al\mbox{.}}{2007}]%
        {cousson-gelie_anxiety_2007}
\bibfield{author}{\bibinfo{person}{Florence Cousson-Gelie},
  \bibinfo{person}{Marilou Bruchon-Schweitzer}, \bibinfo{person}{Jean~Marie
  Dilhuydy}, {and} \bibinfo{person}{Marthe-Aline Jutand}.}
  \bibinfo{year}{2007}\natexlab{}.
\newblock \showarticletitle{Do {Anxiety}, {Body} {Image}, {Social} {Support}
  and {Coping} {Strategies} {Predict} {Survival} in {Breast} {Cancer}? {A}
  {Ten}-{Year} {Follow}-{Up} {Study}}.
\newblock \bibinfo{journal}{\emph{Psychosomatics}} \bibinfo{volume}{48},
  \bibinfo{number}{3} (\bibinfo{date}{May} \bibinfo{year}{2007}),
  \bibinfo{pages}{211--216}.
\newblock
\showISSN{0033-3182, 1545-7206}
\urldef\tempurl%
\url{https://doi.org/10.1176/appi.psy.48.3.211}
\showDOI{\tempurl}


\bibitem[\protect\citeauthoryear{Cutrona and Suhr}{Cutrona and Suhr}{1992}]%
        {cutrona_controllability_1992}
\bibfield{author}{\bibinfo{person}{Carolyn~E. Cutrona} {and}
  \bibinfo{person}{Julie~A. Suhr}.} \bibinfo{year}{1992}\natexlab{}.
\newblock \showarticletitle{Controllability of {Stressful} {Events} and
  {Satisfaction} {With} {Spouse} {Support} {Behaviors}}.
\newblock \bibinfo{journal}{\emph{Communication Research}}
  \bibinfo{volume}{19}, \bibinfo{number}{2} (\bibinfo{date}{April}
  \bibinfo{year}{1992}), \bibinfo{pages}{154--174}.
\newblock
\showISSN{0093-6502}
\urldef\tempurl%
\url{https://doi.org/10.1177/009365092019002002}
\showDOI{\tempurl}


\bibitem[\protect\citeauthoryear{DeAndrea and Anthony}{DeAndrea and
  Anthony}{2013}]%
        {deandrea_online_2013}
\bibfield{author}{\bibinfo{person}{D.~C. DeAndrea} {and} \bibinfo{person}{J.~C.
  Anthony}.} \bibinfo{year}{2013}\natexlab{}.
\newblock \showarticletitle{Online peer support for mental health problems in
  the {United} {States}: 2004-2010}.
\newblock \bibinfo{journal}{\emph{Psychological Medicine}}
  \bibinfo{volume}{43}, \bibinfo{number}{11} (\bibinfo{date}{Nov.}
  \bibinfo{year}{2013}), \bibinfo{pages}{2277--2288}.
\newblock
\showISSN{0033-2917, 1469-8978}
\urldef\tempurl%
\url{https://doi.org/10.1017/S0033291713000172}
\showDOI{\tempurl}


\bibitem[\protect\citeauthoryear{Dillahunt and Malone}{Dillahunt and
  Malone}{2015}]%
        {dillahunt_promise_2015}
\bibfield{author}{\bibinfo{person}{Tawanna~R Dillahunt} {and}
  \bibinfo{person}{Amelia~R Malone}.} \bibinfo{year}{2015}\natexlab{}.
\newblock \showarticletitle{The promise of the sharing economy among
  disadvantaged communities}. In \bibinfo{booktitle}{\emph{Proceedings of the
  33rd {Annual} {ACM} {Conference} on {Human} {Factors} in {Computing}
  {Systems}}}. \bibinfo{publisher}{ACM}, \bibinfo{pages}{2285--2294}.
\newblock


\bibitem[\protect\citeauthoryear{Dillahunt, Wang, Wheeler, Cheng, Hecht, and
  Zhu}{Dillahunt et~al\mbox{.}}{2017}]%
        {dillahunt_sharing_2017}
\bibfield{author}{\bibinfo{person}{Tawanna~R. Dillahunt},
  \bibinfo{person}{Xinyi Wang}, \bibinfo{person}{Earnest Wheeler},
  \bibinfo{person}{Hao~Fei Cheng}, \bibinfo{person}{Brent Hecht}, {and}
  \bibinfo{person}{Haiyi Zhu}.} \bibinfo{year}{2017}\natexlab{}.
\newblock \showarticletitle{The {Sharing} {Economy} in {Computing}: {A}
  {Systematic} {Literature} {Review}}.
\newblock \bibinfo{journal}{\emph{Proceedings of the ACM on Human-Computer
  Interaction}} \bibinfo{volume}{1}, \bibinfo{number}{CSCW}
  (\bibinfo{date}{Dec.} \bibinfo{year}{2017}), \bibinfo{pages}{1--26}.
\newblock
\showISSN{25730142}
\urldef\tempurl%
\url{https://doi.org/10.1145/3134673}
\showDOI{\tempurl}


\bibitem[\protect\citeauthoryear{Drentea, Clay, Roth, and Mittelman}{Drentea
  et~al\mbox{.}}{2006}]%
        {drentea_predictors_2006}
\bibfield{author}{\bibinfo{person}{Patricia Drentea},
  \bibinfo{person}{Olivio~J. Clay}, \bibinfo{person}{David~L. Roth}, {and}
  \bibinfo{person}{Mary~S. Mittelman}.} \bibinfo{year}{2006}\natexlab{}.
\newblock \showarticletitle{Predictors of improvement in social support:
  {Five}-year effects of a structured intervention for caregivers of spouses
  with {Alzheimer}'s disease}.
\newblock \bibinfo{journal}{\emph{Social Science \& Medicine}}
  \bibinfo{volume}{63}, \bibinfo{number}{4} (\bibinfo{date}{Aug.}
  \bibinfo{year}{2006}), \bibinfo{pages}{957--967}.
\newblock
\showISSN{0277-9536}


\bibitem[\protect\citeauthoryear{DuBenske, Gustafson, Namkoong, Hawkins,
  Atwood, Brown, Chih, McTavish, Carmack, Buss, and {others}}{DuBenske
  et~al\mbox{.}}{2014}]%
        {dubenske_chess_2014}
\bibfield{author}{\bibinfo{person}{Lori~L DuBenske}, \bibinfo{person}{David~H
  Gustafson}, \bibinfo{person}{Kang Namkoong}, \bibinfo{person}{Robert~P
  Hawkins}, \bibinfo{person}{Amy~K Atwood}, \bibinfo{person}{Roger~L Brown},
  \bibinfo{person}{Ming-Yuan Chih}, \bibinfo{person}{Fiona McTavish},
  \bibinfo{person}{Cindy~L Carmack}, \bibinfo{person}{Mary~K Buss}, {and}
  \bibinfo{person}{{others}}.} \bibinfo{year}{2014}\natexlab{}.
\newblock \showarticletitle{{CHESS} improves cancer caregivers' burden and
  mood: {Results} of an {eHealth} {RCT}.}
\newblock \bibinfo{journal}{\emph{Health Psychology}} \bibinfo{volume}{33},
  \bibinfo{number}{10} (\bibinfo{year}{2014}), \bibinfo{pages}{1261}.
\newblock


\bibitem[\protect\citeauthoryear{England}{England}{[n. d.]}]%
        {england_vatican_nodate}
\bibfield{author}{\bibinfo{person}{Rachel England}.} \bibinfo{year}{[n.
  d.]}\natexlab{}.
\newblock \bibinfo{title}{Vatican launches \$110 'click to pray' wearable
  rosary}.
\newblock
\newblock
\urldef\tempurl%
\url{https://www.engadget.com/2019/10/16/vatican-click-to-pray-wearable-rosary/}
\showURL{%
\tempurl}


\bibitem[\protect\citeauthoryear{Ernala, Rizvi, Birnbaum, Kane, and
  De~Choudhury}{Ernala et~al\mbox{.}}{2017}]%
        {ernala_linguistic_2017}
\bibfield{author}{\bibinfo{person}{Sindhu~Kiranmai Ernala},
  \bibinfo{person}{Asra~F. Rizvi}, \bibinfo{person}{Michael~L. Birnbaum},
  \bibinfo{person}{John~M. Kane}, {and} \bibinfo{person}{Munmun De~Choudhury}.}
  \bibinfo{year}{2017}\natexlab{}.
\newblock \showarticletitle{Linguistic {Markers} {Indicating} {Therapeutic}
  {Outcomes} of {Social} {Media} {Disclosures} of {Schizophrenia}}.
\newblock \bibinfo{journal}{\emph{Proc. ACM Hum.-Comput. Interact.}}
  \bibinfo{volume}{1}, \bibinfo{number}{CSCW} (\bibinfo{date}{Dec.}
  \bibinfo{year}{2017}), \bibinfo{pages}{43:1--43:27}.
\newblock
\showISSN{2573-0142}
\urldef\tempurl%
\url{https://doi.org/10.1145/3134678}
\showDOI{\tempurl}


\bibitem[\protect\citeauthoryear{Eschler, Dehlawi, and Pratt}{Eschler
  et~al\mbox{.}}{2015}]%
        {eschler_self-characterized_2015}
\bibfield{author}{\bibinfo{person}{Jordan Eschler}, \bibinfo{person}{Zakariya
  Dehlawi}, {and} \bibinfo{person}{Wanda Pratt}.}
  \bibinfo{year}{2015}\natexlab{}.
\newblock \showarticletitle{Self-{Characterized} {Illness} {Phase} and
  {Information} {Needs} of {Participants} in an {Online} {Cancer} {Forum}}. In
  \bibinfo{booktitle}{\emph{Ninth {International} {AAAI} {Conference} on {Web}
  and {Social} {Media}}}.
\newblock


\bibitem[\protect\citeauthoryear{Eschler and Pratt}{Eschler and Pratt}{2017}]%
        {eschler_im_2017}
\bibfield{author}{\bibinfo{person}{Jordan Eschler} {and} \bibinfo{person}{Wanda
  Pratt}.} \bibinfo{year}{2017}\natexlab{}.
\newblock \showarticletitle{"{I}'m {So} {Glad} {I} {Met} {You}": {Designing}
  {Dynamic} {Collaborative} {Support} for {Young} {Adult} {Cancer}
  {Survivors}}. In \bibinfo{booktitle}{\emph{Proceedings of the 2017 {ACM}
  {Conference} on {Computer} {Supported} {Cooperative} {Work} and {Social}
  {Computing}}} \emph{(\bibinfo{series}{{CSCW} '17})}.
  \bibinfo{publisher}{ACM}, \bibinfo{address}{New York, NY, USA},
  \bibinfo{pages}{1763--1774}.
\newblock
\showISBNx{978-1-4503-4335-0}
\urldef\tempurl%
\url{https://doi.org/10.1145/2998181.2998326}
\showDOI{\tempurl}
\newblock
\shownote{event-place: Portland, Oregon, USA.}


\bibitem[\protect\citeauthoryear{Fast, Chen, and Bernstein}{Fast
  et~al\mbox{.}}{2016}]%
        {fast_empath:_2016}
\bibfield{author}{\bibinfo{person}{Ethan Fast}, \bibinfo{person}{Binbin Chen},
  {and} \bibinfo{person}{Michael~S. Bernstein}.}
  \bibinfo{year}{2016}\natexlab{}.
\newblock \showarticletitle{Empath: {Understanding} {Topic} {Signals} in
  {Large}-{Scale} {Text}}. \bibinfo{publisher}{ACM Press},
  \bibinfo{pages}{4647--4657}.
\newblock
\showISBNx{978-1-4503-3362-7}
\urldef\tempurl%
\url{https://doi.org/10.1145/2858036.2858535}
\showDOI{\tempurl}


\bibitem[\protect\citeauthoryear{Ferrell, Twaddle, Melnick, and Meier}{Ferrell
  et~al\mbox{.}}{2018}]%
        {ferrell_national_2018}
\bibfield{author}{\bibinfo{person}{Betty~R. Ferrell},
  \bibinfo{person}{Martha~L. Twaddle}, \bibinfo{person}{Amy Melnick}, {and}
  \bibinfo{person}{Diane~E. Meier}.} \bibinfo{year}{2018}\natexlab{}.
\newblock \showarticletitle{National {Consensus} {Project} {Clinical}
  {Practice} {Guidelines} for {Quality} {Palliative} {Care} {Guidelines}, 4th
  {Edition}}.
\newblock \bibinfo{journal}{\emph{Journal of Palliative Medicine}}
  (\bibinfo{date}{Sept.} \bibinfo{year}{2018}).
\newblock
\showISSN{1557-7740}
\urldef\tempurl%
\url{https://doi.org/10.1089/jpm.2018.0431}
\showDOI{\tempurl}


\bibitem[\protect\citeauthoryear{Flickinger, DeBolt, Waldman, Reynolds, Cohn,
  Beach, Ingersoll, and Dillingham}{Flickinger et~al\mbox{.}}{2017}]%
        {flickinger_social_2017}
\bibfield{author}{\bibinfo{person}{Tabor~E. Flickinger},
  \bibinfo{person}{Claire DeBolt}, \bibinfo{person}{Ava~Lena Waldman},
  \bibinfo{person}{George Reynolds}, \bibinfo{person}{Wendy~F. Cohn},
  \bibinfo{person}{Mary~Catherine Beach}, \bibinfo{person}{Karen Ingersoll},
  {and} \bibinfo{person}{Rebecca Dillingham}.} \bibinfo{year}{2017}\natexlab{}.
\newblock \showarticletitle{Social {Support} in a {Virtual} {Community}:
  {Analysis} of a {Clinic}-{Affiliated} {Online} {Support} {Group} for
  {Persons} {Living} with {HIV}/{AIDS}}.
\newblock \bibinfo{journal}{\emph{AIDS and Behavior}} \bibinfo{volume}{21},
  \bibinfo{number}{11} (\bibinfo{date}{Nov.} \bibinfo{year}{2017}),
  \bibinfo{pages}{3087--3099}.
\newblock
\showISSN{1573-3254}


\bibitem[\protect\citeauthoryear{Fogel and Nehmad}{Fogel and Nehmad}{2009}]%
        {fogel_internet_2009}
\bibfield{author}{\bibinfo{person}{Joshua Fogel} {and} \bibinfo{person}{Elham
  Nehmad}.} \bibinfo{year}{2009}\natexlab{}.
\newblock \showarticletitle{Internet social network communities: {Risk} taking,
  trust, and privacy concerns}.
\newblock \bibinfo{journal}{\emph{Computers in Human Behavior}}
  \bibinfo{volume}{25}, \bibinfo{number}{1} (\bibinfo{date}{Jan.}
  \bibinfo{year}{2009}), \bibinfo{pages}{153--160}.
\newblock
\showISSN{0747-5632}
\urldef\tempurl%
\url{https://doi.org/10.1016/j.chb.2008.08.006}
\showDOI{\tempurl}


\bibitem[\protect\citeauthoryear{Freeman and Wohn}{Freeman and Wohn}{2017}]%
        {freeman_social_2017}
\bibfield{author}{\bibinfo{person}{Guo Freeman} {and}
  \bibinfo{person}{Donghee~Yvette Wohn}.} \bibinfo{year}{2017}\natexlab{}.
\newblock \showarticletitle{Social {Support} in {eSports}: {Building}
  {Emotional} and {Esteem} {Support} from {Instrumental} {Support}
  {Interactions} in a {Highly} {Competitive} {Environment}}. In
  \bibinfo{booktitle}{\emph{Proceedings of the {Annual} {Symposium} on
  {Computer}-{Human} {Interaction} in {Play}}} \emph{(\bibinfo{series}{{CHI}
  {PLAY} '17})}. \bibinfo{publisher}{ACM}, \bibinfo{address}{New York, NY,
  USA}, \bibinfo{pages}{435--447}.
\newblock
\showISBNx{978-1-4503-4898-0}
\urldef\tempurl%
\url{https://doi.org/10.1145/3116595.3116635}
\showDOI{\tempurl}
\newblock
\shownote{event-place: Amsterdam, The Netherlands.}


\bibitem[\protect\citeauthoryear{Frost and Massagli}{Frost and
  Massagli}{2008}]%
        {frost_social_2008}
\bibfield{author}{\bibinfo{person}{Jeana Frost} {and} \bibinfo{person}{Michael
  Massagli}.} \bibinfo{year}{2008}\natexlab{}.
\newblock \showarticletitle{Social {Uses} of {Personal} {Health} {Information}
  {Within} {PatientsLikeMe}, an {Online} {Patient} {Community}: {What} {Can}
  {Happen} {When} {Patients} {Have} {Access} to {One} {Another}'s {Data}}.
\newblock \bibinfo{journal}{\emph{Journal of Medical Internet Research}}
  \bibinfo{volume}{10}, \bibinfo{number}{3} (\bibinfo{year}{2008}),
  \bibinfo{pages}{e15}.
\newblock
\urldef\tempurl%
\url{https://doi.org/10.2196/jmir.1053}
\showDOI{\tempurl}


\bibitem[\protect\citeauthoryear{Fund}{Fund}{2018}]%
        {fund_worldwide_2012}
\bibfield{author}{\bibinfo{person}{World Cancer~Research Fund}.}
  \bibinfo{year}{2018}\natexlab{}.
\newblock \bibinfo{booktitle}{\emph{Worldwide data {\textbackslash}textbar
  {World} {Cancer} {Research} {Fund} {International}}}.
\newblock
\urldef\tempurl%
\url{https://www.wcrf.org/dietandcancer/cancer-trends/worldwide-cancer-data}
\showURL{%
\tempurl}


\bibitem[\protect\citeauthoryear{Gach and Brubaker}{Gach and Brubaker}{2020}]%
        {gach2020experiences}
\bibfield{author}{\bibinfo{person}{Katie~Z Gach} {and} \bibinfo{person}{Jed~R
  Brubaker}.} \bibinfo{year}{2020}\natexlab{}.
\newblock \showarticletitle{Experiences of Trust in Postmortem Profile
  Management}.
\newblock \bibinfo{journal}{\emph{ACM Transactions on Social Computing}}
  \bibinfo{volume}{3}, \bibinfo{number}{1} (\bibinfo{year}{2020}),
  \bibinfo{pages}{1--26}.
\newblock


\bibitem[\protect\citeauthoryear{Gilbert and Karahalios}{Gilbert and
  Karahalios}{2009}]%
        {gilbert_predicting_2009}
\bibfield{author}{\bibinfo{person}{Eric Gilbert} {and} \bibinfo{person}{Karrie
  Karahalios}.} \bibinfo{year}{2009}\natexlab{}.
\newblock \showarticletitle{Predicting {Tie} {Strength} with {Social} {Media}}.
  In \bibinfo{booktitle}{\emph{Proceedings of the {SIGCHI} {Conference} on
  {Human} {Factors} in {Computing} {Systems}}} \emph{(\bibinfo{series}{{CHI}
  '09})}. \bibinfo{publisher}{ACM}, \bibinfo{address}{New York, NY, USA},
  \bibinfo{pages}{211--220}.
\newblock
\showISBNx{978-1-60558-246-7}
\urldef\tempurl%
\url{https://doi.org/10.1145/1518701.1518736}
\showDOI{\tempurl}
\newblock
\shownote{event-place: Boston, MA, USA.}


\bibitem[\protect\citeauthoryear{Glueckauf, Ketterson, Loomis, and
  Dages}{Glueckauf et~al\mbox{.}}{2004}]%
        {glueckauf_online_2004}
\bibfield{author}{\bibinfo{person}{Robert~L. Glueckauf},
  \bibinfo{person}{Timothy~U. Ketterson}, \bibinfo{person}{Jeffrey~S. Loomis},
  {and} \bibinfo{person}{Pat Dages}.} \bibinfo{year}{2004}\natexlab{}.
\newblock \showarticletitle{Online {Support} and {Education} for {Dementia}
  {Caregivers}: {Overview}, {Utilization}, and {Initial} {Program}
  {Evaluation}}.
\newblock \bibinfo{journal}{\emph{Telemedicine Journal and e-Health}}
  \bibinfo{volume}{10}, \bibinfo{number}{2} (\bibinfo{date}{June}
  \bibinfo{year}{2004}), \bibinfo{pages}{223--232}.
\newblock
\showISSN{1530-5627}
\urldef\tempurl%
\url{https://doi.org/10.1089/tmj.2004.10.223}
\showDOI{\tempurl}


\bibitem[\protect\citeauthoryear{Gray, Fitch, Davis, and Phillips}{Gray
  et~al\mbox{.}}{1996}]%
        {gray_breast_1996}
\bibfield{author}{\bibinfo{person}{Ross Gray}, \bibinfo{person}{Margaret
  Fitch}, \bibinfo{person}{Christine Davis}, {and} \bibinfo{person}{Catherine
  Phillips}.} \bibinfo{year}{1996}\natexlab{}.
\newblock \showarticletitle{Breast cancer and prostate cancer self-help groups:
  {Reflections} on differences}.
\newblock \bibinfo{journal}{\emph{Psycho-Oncology}} \bibinfo{volume}{5},
  \bibinfo{number}{2} (\bibinfo{year}{1996}), \bibinfo{pages}{137--142}.
\newblock
\showISSN{1099-1611}
\urldef\tempurl%
\url{https://doi.org/10.1002/(SICI)1099-1611(199606)5:2<137::AID-PON222>3.0.CO;2-E}
\showDOI{\tempurl}


\bibitem[\protect\citeauthoryear{Hardin}{Hardin}{2001}]%
        {hardin2001conceptions}
\bibfield{author}{\bibinfo{person}{Russell Hardin}.}
  \bibinfo{year}{2001}\natexlab{}.
\newblock \showarticletitle{Conceptions and explanations of trust.}
\newblock  (\bibinfo{year}{2001}).
\newblock


\bibitem[\protect\citeauthoryear{Hartzler, Skeels, Mukai, Powell, Klasnja, and
  Pratt}{Hartzler et~al\mbox{.}}{2011}]%
        {hartzler_sharing_2011}
\bibfield{author}{\bibinfo{person}{Andrea Hartzler},
  \bibinfo{person}{Meredith~McLain Skeels}, \bibinfo{person}{Marlee Mukai},
  \bibinfo{person}{Christopher Powell}, \bibinfo{person}{Predrag Klasnja},
  {and} \bibinfo{person}{Wanda Pratt}.} \bibinfo{year}{2011}\natexlab{}.
\newblock \showarticletitle{Sharing is caring, but not error free: transparency
  of granular controls for sharing personal health information in social
  networks}. In \bibinfo{booktitle}{\emph{{AMIA} {Annual} {Symposium}
  {Proceedings}}}, Vol.~\bibinfo{volume}{2011}. \bibinfo{publisher}{American
  Medical Informatics Association}, \bibinfo{pages}{559}.
\newblock


\bibitem[\protect\citeauthoryear{Helft, Eckles, Johnson-Calley, and
  Daugherty}{Helft et~al\mbox{.}}{2005}]%
        {helft2005use}
\bibfield{author}{\bibinfo{person}{Paul~R Helft}, \bibinfo{person}{Rachael~E
  Eckles}, \bibinfo{person}{Cynthia~Stair Johnson-Calley}, {and}
  \bibinfo{person}{Christopher~K Daugherty}.} \bibinfo{year}{2005}\natexlab{}.
\newblock \showarticletitle{Use of the internet to obtain cancer information
  among cancer patients at an urban county hospital}.
\newblock \bibinfo{journal}{\emph{Journal of Clinical Oncology}}
  \bibinfo{volume}{23}, \bibinfo{number}{22} (\bibinfo{year}{2005}),
  \bibinfo{pages}{4954--4962}.
\newblock


\bibitem[\protect\citeauthoryear{Hlavac}{Hlavac}{2018}]%
        {stargazer_package}
\bibfield{author}{\bibinfo{person}{Marek Hlavac}.}
  \bibinfo{year}{2018}\natexlab{}.
\newblock \bibinfo{booktitle}{\emph{stargazer: Well-Formatted Regression and
  Summary Statistics Tables}}.
\newblock Central European Labour Studies Institute (CELSI), Bratislava,
  Slovakia.
\newblock
\urldef\tempurl%
\url{https://CRAN.R-project.org/package=stargazer}
\showURL{%
\tempurl}
\newblock
\shownote{R package version 5.2.2.}


\bibitem[\protect\citeauthoryear{Hornbaek, Sander, Bargas-Avila, and
  Simonsen}{Hornbaek et~al\mbox{.}}{2014}]%
        {hornbaek_is_2014}
\bibfield{author}{\bibinfo{person}{Kasper Hornbaek}, \bibinfo{person}{Soren~S.
  Sander}, \bibinfo{person}{Javier Bargas-Avila}, {and}
  \bibinfo{person}{Jakob~Grue Simonsen}.} \bibinfo{year}{2014}\natexlab{}.
\newblock \showarticletitle{Is {Once} {Enough}? {On} the {Extent} and {Content}
  of {Replications} in {Human}-{Computer} {Interaction}}.
\newblock  (\bibinfo{year}{2014}).
\newblock
\urldef\tempurl%
\url{https://research.google.com/pubs/pub42512.html}
\showURL{%
\tempurl}


\bibitem[\protect\citeauthoryear{Hsiao, Wong, Miller, Ambs, Goldstein, Smith,
  Ballard-Barbash, Becerra, Cheng, and Wenger}{Hsiao et~al\mbox{.}}{2008}]%
        {hsiao_role_2008}
\bibfield{author}{\bibinfo{person}{An-Fu Hsiao}, \bibinfo{person}{Mitchell~D.
  Wong}, \bibinfo{person}{Melissa~F. Miller}, \bibinfo{person}{Anita~H. Ambs},
  \bibinfo{person}{Michael~S. Goldstein}, \bibinfo{person}{Ashley Smith},
  \bibinfo{person}{Rachel Ballard-Barbash}, \bibinfo{person}{Lida~S. Becerra},
  \bibinfo{person}{Eric~M. Cheng}, {and} \bibinfo{person}{Neil~S. Wenger}.}
  \bibinfo{year}{2008}\natexlab{}.
\newblock \showarticletitle{Role of {Religiosity} and {Spirituality} in
  {Complementary} and {Alternative} {Medicine} {Use} {Among} {Cancer}
  {Survivors} in {California}}.
\newblock \bibinfo{journal}{\emph{Integrative Cancer Therapies}}
  \bibinfo{volume}{7}, \bibinfo{number}{3} (\bibinfo{date}{Sept.}
  \bibinfo{year}{2008}), \bibinfo{pages}{139--146}.
\newblock
\showISSN{1534-7354}
\urldef\tempurl%
\url{https://doi.org/10.1177/1534735408322847}
\showDOI{\tempurl}


\bibitem[\protect\citeauthoryear{Hsieh and Shannon}{Hsieh and Shannon}{2005}]%
        {hsieh_three_2005}
\bibfield{author}{\bibinfo{person}{Hsiu-Fang Hsieh} {and}
  \bibinfo{person}{Sarah~E. Shannon}.} \bibinfo{year}{2005}\natexlab{}.
\newblock \showarticletitle{Three {Approaches} to {Qualitative} {Content}
  {Analysis}}.
\newblock \bibinfo{journal}{\emph{Qualitative Health Research}}
  \bibinfo{volume}{15}, \bibinfo{number}{9} (\bibinfo{date}{Nov.}
  \bibinfo{year}{2005}), \bibinfo{pages}{1277--1288}.
\newblock
\showISSN{1049-7323, 1552-7557}
\urldef\tempurl%
\url{https://doi.org/10.1177/1049732305276687}
\showDOI{\tempurl}


\bibitem[\protect\citeauthoryear{Jacobs, Clawson, and Mynatt}{Jacobs
  et~al\mbox{.}}{2015a}]%
        {jacobs_lessons_2015}
\bibfield{author}{\bibinfo{person}{Maia Jacobs}, \bibinfo{person}{James
  Clawson}, {and} \bibinfo{person}{Elizabeth~D. Mynatt}.}
  \bibinfo{year}{2015}\natexlab{a}.
\newblock \showarticletitle{Lessons learned from a yearlong deployment of
  customizable breast cancer tablet computers}. \bibinfo{publisher}{ACM},
  \bibinfo{pages}{7}.
\newblock


\bibitem[\protect\citeauthoryear{Jacobs, Clawson, and Mynatt}{Jacobs
  et~al\mbox{.}}{2016}]%
        {jacobs_cancer_2016}
\bibfield{author}{\bibinfo{person}{Maia Jacobs}, \bibinfo{person}{James
  Clawson}, {and} \bibinfo{person}{Elizabeth~D. Mynatt}.}
  \bibinfo{year}{2016}\natexlab{}.
\newblock \showarticletitle{A {Cancer} {Journey} {Framework}: {Guiding} the
  {Design} of {Holistic} {Health} {Technology}}. In
  \bibinfo{booktitle}{\emph{Proceedings of the 10th {EAI} {International}
  {Conference} on {Pervasive} {Computing} {Technologies} for {Healthcare}}}
  \emph{(\bibinfo{series}{{PervasiveHealth} '16})}. \bibinfo{publisher}{ICST
  (Institute for Computer Sciences, Social-Informatics and Telecommunications
  Engineering)}, \bibinfo{address}{ICST, Brussels, Belgium, Belgium},
  \bibinfo{pages}{114--121}.
\newblock
\showISBNx{978-1-63190-051-8}


\bibitem[\protect\citeauthoryear{Jacobs, Johnson, and Mynatt}{Jacobs
  et~al\mbox{.}}{2018}]%
        {jacobs_mypath:_2018}
\bibfield{author}{\bibinfo{person}{Maia Jacobs}, \bibinfo{person}{Jeremy
  Johnson}, {and} \bibinfo{person}{Elizabeth~D. Mynatt}.}
  \bibinfo{year}{2018}\natexlab{}.
\newblock \showarticletitle{{MyPath}: {Investigating} {Breast} {Cancer}
  {Patients}' {Use} of {Personalized} {Health} {Information}}.
\newblock \bibinfo{journal}{\emph{Proc. ACM Hum.-Comput. Interact.}}
  \bibinfo{volume}{2}, \bibinfo{number}{CSCW} (\bibinfo{date}{Nov.}
  \bibinfo{year}{2018}), \bibinfo{pages}{78:1--78:21}.
\newblock
\showISSN{2573-0142}
\urldef\tempurl%
\url{https://doi.org/10.1145/3274347}
\showDOI{\tempurl}


\bibitem[\protect\citeauthoryear{Jacobs, Clawson, and Mynatt}{Jacobs
  et~al\mbox{.}}{2015b}]%
        {jacobs_comparing_2015}
\bibfield{author}{\bibinfo{person}{Maia~L. Jacobs}, \bibinfo{person}{James
  Clawson}, {and} \bibinfo{person}{Elizabeth~D. Mynatt}.}
  \bibinfo{year}{2015}\natexlab{b}.
\newblock \showarticletitle{Comparing health information sharing preferences of
  cancer patients, doctors, and navigators}. \bibinfo{publisher}{ACM},
  \bibinfo{pages}{808--818}.
\newblock


\bibitem[\protect\citeauthoryear{Japsen}{Japsen}{[n. d.]}]%
        {japsen_lyft_nodate}
\bibfield{author}{\bibinfo{person}{Bruce Japsen}.} \bibinfo{year}{[n.
  d.]}\natexlab{}.
\newblock \bibinfo{title}{Lyft {Hails} {Major} {Hospital} {Partner} {In}
  {Sutter} {Health}}.
\newblock
\newblock
\urldef\tempurl%
\url{https://www.forbes.com/sites/brucejapsen/2020/01/13/lyft-hails-major-hospital-partner-in-sutter-health/}
\showURL{%
\tempurl}


\bibitem[\protect\citeauthoryear{Jiang and Brubaker}{Jiang and
  Brubaker}{2018}]%
        {jiang_tending_2018}
\bibfield{author}{\bibinfo{person}{Jialun~"Aaron" Jiang} {and}
  \bibinfo{person}{Jed~R. Brubaker}.} \bibinfo{year}{2018}\natexlab{}.
\newblock \showarticletitle{Tending {Unmarked} {Graves}: {Classification} of
  {Post}-mortem {Content} on {Social} {Media}}.
\newblock \bibinfo{journal}{\emph{Proc. ACM Hum.-Comput. Interact.}}
  \bibinfo{volume}{2}, \bibinfo{number}{CSCW} (\bibinfo{date}{Nov.}
  \bibinfo{year}{2018}), \bibinfo{pages}{81:1--81:19}.
\newblock
\showISSN{2573-0142}
\urldef\tempurl%
\url{https://doi.org/10.1145/3274350}
\showDOI{\tempurl}


\bibitem[\protect\citeauthoryear{Joinson and Paine}{Joinson and Paine}{2009}]%
        {joinson_self-disclosure_2009}
\bibfield{author}{\bibinfo{person}{Adam~N. Joinson} {and}
  \bibinfo{person}{Carina~B. Paine}.} \bibinfo{year}{2009}\natexlab{}.
\newblock \showarticletitle{Self-disclosure, {Privacy} and the {Internet}}.
\newblock \bibinfo{journal}{\emph{Oxford Handbook of Internet Psychology}}
  (\bibinfo{date}{Feb.} \bibinfo{year}{2009}).
\newblock
\urldef\tempurl%
\url{https://doi.org/10.1093/oxfordhb/9780199561803.013.0016}
\showDOI{\tempurl}


\bibitem[\protect\citeauthoryear{Kalish}{Kalish}{2012}]%
        {kalish_evidence-based_2012}
\bibfield{author}{\bibinfo{person}{Naomi Kalish}.}
  \bibinfo{year}{2012}\natexlab{}.
\newblock \showarticletitle{Evidence-based spiritual care: a literature
  review}.
\newblock \bibinfo{journal}{\emph{Current Opinion in Supportive and Palliative
  Care}} \bibinfo{volume}{6}, \bibinfo{number}{2} (\bibinfo{date}{June}
  \bibinfo{year}{2012}), \bibinfo{pages}{242}.
\newblock
\showISSN{1751-4258}
\urldef\tempurl%
\url{https://doi.org/10.1097/SPC.0b013e328353811c}
\showDOI{\tempurl}


\bibitem[\protect\citeauthoryear{Kim, Vaccaro, Karahalios, and Hong}{Kim
  et~al\mbox{.}}{2017}]%
        {kim_not_2017}
\bibfield{author}{\bibinfo{person}{Jennifer~G. Kim}, \bibinfo{person}{Kristen
  Vaccaro}, \bibinfo{person}{Karrie Karahalios}, {and} \bibinfo{person}{Hwajung
  Hong}.} \bibinfo{year}{2017}\natexlab{}.
\newblock \showarticletitle{"{Not} by {Money} {Alone}": {Social} {Support}
  {Opportunities} in {Medical} {Crowdfunding} {Campaigns}}. In
  \bibinfo{booktitle}{\emph{Proceedings of the 2017 {ACM} {Conference} on
  {Computer} {Supported} {Cooperative} {Work} and {Social} {Computing}}}
  \emph{(\bibinfo{series}{{CSCW} '17})}. \bibinfo{publisher}{ACM},
  \bibinfo{address}{New York, NY, USA}, \bibinfo{pages}{1997--2009}.
\newblock
\showISBNx{978-1-4503-4335-0}
\urldef\tempurl%
\url{https://doi.org/10.1145/2998181.2998245}
\showDOI{\tempurl}
\newblock
\shownote{event-place: Portland, Oregon, USA.}


\bibitem[\protect\citeauthoryear{Kittur, Suh, and Chi}{Kittur
  et~al\mbox{.}}{2008}]%
        {kittur_can_2008}
\bibfield{author}{\bibinfo{person}{Aniket Kittur}, \bibinfo{person}{Bongwon
  Suh}, {and} \bibinfo{person}{Ed~H. Chi}.} \bibinfo{year}{2008}\natexlab{}.
\newblock \showarticletitle{Can {You} {Ever} {Trust} a {Wiki}?: {Impacting}
  {Perceived} {Trustworthiness} in {Wikipedia}}. In
  \bibinfo{booktitle}{\emph{Proceedings of the 2008 {ACM} {Conference} on
  {Computer} {Supported} {Cooperative} {Work}}} \emph{(\bibinfo{series}{{CSCW}
  '08})}. \bibinfo{publisher}{ACM}, \bibinfo{address}{New York, NY, USA},
  \bibinfo{pages}{477--480}.
\newblock
\showISBNx{978-1-60558-007-4}
\urldef\tempurl%
\url{https://doi.org/10.1145/1460563.1460639}
\showDOI{\tempurl}
\newblock
\shownote{event-place: San Diego, CA, USA.}


\bibitem[\protect\citeauthoryear{Lamberg}{Lamberg}{2003}]%
        {lamberg_online_2003}
\bibfield{author}{\bibinfo{person}{Lynne Lamberg}.}
  \bibinfo{year}{2003}\natexlab{}.
\newblock \showarticletitle{Online {Empathy} for {Mood} {Disorders}}.
\newblock \bibinfo{journal}{\emph{JAMA}} \bibinfo{volume}{289},
  \bibinfo{number}{23} (\bibinfo{date}{June} \bibinfo{year}{2003}),
  \bibinfo{pages}{3073--3077}.
\newblock
\showISSN{0098-7484}
\urldef\tempurl%
\url{https://doi.org/10.1001/jama.289.23.3073}
\showDOI{\tempurl}


\bibitem[\protect\citeauthoryear{Langford, Li, and Strehl}{Langford
  et~al\mbox{.}}{2007}]%
        {langford_vowpal_2007}
\bibfield{author}{\bibinfo{person}{John Langford}, \bibinfo{person}{Lihong Li},
  {and} \bibinfo{person}{Alex Strehl}.} \bibinfo{year}{2007}\natexlab{}.
\newblock \bibinfo{booktitle}{\emph{Vowpal wabbit online learning project}}.
\newblock \bibinfo{publisher}{Technical report, http://hunch.net}.
\newblock


\bibitem[\protect\citeauthoryear{LaRock}{LaRock}{2009}]%
        {larock_lyft_2009}
\bibfield{author}{\bibinfo{person}{Zoe LaRock}.}
  \bibinfo{year}{2009}\natexlab{}.
\newblock \bibinfo{title}{Lyft is widening its health stance with grocery
  rides}.
\newblock
\newblock
\urldef\tempurl%
\url{https://www.businessinsider.com/lyft-expanding-discounted-rides-to-grocery-stores-2019-4}
\showURL{%
\tempurl}


\bibitem[\protect\citeauthoryear{Leskovec, Huttenlocher, and
  Kleinberg}{Leskovec et~al\mbox{.}}{2010}]%
        {leskovec_signed_2010}
\bibfield{author}{\bibinfo{person}{Jure Leskovec}, \bibinfo{person}{Daniel
  Huttenlocher}, {and} \bibinfo{person}{Jon Kleinberg}.}
  \bibinfo{year}{2010}\natexlab{}.
\newblock \showarticletitle{Signed {Networks} in {Social} {Media}}. In
  \bibinfo{booktitle}{\emph{Proceedings of the {SIGCHI} {Conference} on {Human}
  {Factors} in {Computing} {Systems}}} \emph{(\bibinfo{series}{{CHI} '10})}.
  \bibinfo{publisher}{ACM}, \bibinfo{address}{New York, NY, USA},
  \bibinfo{pages}{1361--1370}.
\newblock
\showISBNx{978-1-60558-929-9}
\urldef\tempurl%
\url{https://doi.org/10.1145/1753326.1753532}
\showDOI{\tempurl}
\newblock
\shownote{event-place: Atlanta, Georgia, USA.}


\bibitem[\protect\citeauthoryear{Levonian, Erikson, Luo, Narayanan, Rubya,
  Vachher, Terveen, and Yarosh}{Levonian et~al\mbox{.}}{2020}]%
        {levonianbridging}
\bibfield{author}{\bibinfo{person}{Zachary Levonian},
  \bibinfo{person}{Drew~Richard Erikson}, \bibinfo{person}{Wenqi Luo},
  \bibinfo{person}{Saumik Narayanan}, \bibinfo{person}{Sabirat Rubya},
  \bibinfo{person}{Prateek Vachher}, \bibinfo{person}{Loren Terveen}, {and}
  \bibinfo{person}{Svetlana Yarosh}.} \bibinfo{year}{2020}\natexlab{}.
\newblock \showarticletitle{Bridging Qualitative and Quantitative Methods for
  User Modeling: Tracing Cancer Patient Behavior in an Online Health
  Community}.
\newblock \bibinfo{journal}{\emph{ICWSM}} (\bibinfo{year}{2020}).
\newblock


\bibitem[\protect\citeauthoryear{Ma, Smith, He, Narayanan, Giaquinto, Evans,
  Hanson, and Yarosh}{Ma et~al\mbox{.}}{2017}]%
        {ma_write_2017}
\bibfield{author}{\bibinfo{person}{Haiwei Ma}, \bibinfo{person}{C.~Estelle
  Smith}, \bibinfo{person}{Lu He}, \bibinfo{person}{Saumik Narayanan},
  \bibinfo{person}{Robert~A. Giaquinto}, \bibinfo{person}{Roni Evans},
  \bibinfo{person}{Linda Hanson}, {and} \bibinfo{person}{Svetlana Yarosh}.}
  \bibinfo{year}{2017}\natexlab{}.
\newblock \showarticletitle{Write for {Life}: {Persisting} in {Online} {Health}
  {Communities} {Through} {Expressive} {Writing} and {Social} {Support}}.
\newblock \bibinfo{journal}{\emph{Proc. ACM Hum.-Comput. Interact.}}
  \bibinfo{volume}{1}, \bibinfo{number}{CSCW} (\bibinfo{date}{Dec.}
  \bibinfo{year}{2017}), \bibinfo{pages}{73:1--73:24}.
\newblock
\showISSN{2573-0142}
\urldef\tempurl%
\url{https://doi.org/10.1145/3134708}
\showDOI{\tempurl}


\bibitem[\protect\citeauthoryear{MacLeod, Bastin, Liu, Siek, and
  Connelly}{MacLeod et~al\mbox{.}}{2017}]%
        {macleod_be_2017}
\bibfield{author}{\bibinfo{person}{Haley MacLeod}, \bibinfo{person}{Grace
  Bastin}, \bibinfo{person}{Leslie~S. Liu}, \bibinfo{person}{Katie Siek}, {and}
  \bibinfo{person}{Kay Connelly}.} \bibinfo{year}{2017}\natexlab{}.
\newblock \showarticletitle{"{Be} {Grateful} {You} {Don}'{T} {Have} a {Real}
  {Disease}": {Understanding} {Rare} {Disease} {Relationships}}. In
  \bibinfo{booktitle}{\emph{Proceedings of the 2017 {CHI} {Conference} on
  {Human} {Factors} in {Computing} {Systems}}} \emph{(\bibinfo{series}{{CHI}
  '17})}. \bibinfo{publisher}{ACM}, \bibinfo{address}{New York, NY, USA},
  \bibinfo{pages}{1660--1673}.
\newblock
\showISBNx{978-1-4503-4655-9}
\urldef\tempurl%
\url{https://doi.org/10.1145/3025453.3025796}
\showDOI{\tempurl}
\newblock
\shownote{event-place: Denver, Colorado, USA.}


\bibitem[\protect\citeauthoryear{Mankoff, Kuksenok, Kiesler, Rode, and
  Waldman}{Mankoff et~al\mbox{.}}{2011}]%
        {mankoff_competing_2011}
\bibfield{author}{\bibinfo{person}{Jennifer Mankoff}, \bibinfo{person}{Kateryna
  Kuksenok}, \bibinfo{person}{Sara Kiesler}, \bibinfo{person}{Jennifer~A.
  Rode}, {and} \bibinfo{person}{Kelly Waldman}.}
  \bibinfo{year}{2011}\natexlab{}.
\newblock \showarticletitle{Competing {Online} {Viewpoints} and {Models} of
  {Chronic} {Illness}}. In \bibinfo{booktitle}{\emph{Proceedings of the
  {SIGCHI} {Conference} on {Human} {Factors} in {Computing} {Systems}}}
  \emph{(\bibinfo{series}{{CHI} '11})}. \bibinfo{publisher}{ACM},
  \bibinfo{address}{New York, NY, USA}, \bibinfo{pages}{589--598}.
\newblock
\showISBNx{978-1-4503-0228-9}
\urldef\tempurl%
\url{https://doi.org/10.1145/1978942.1979027}
\showDOI{\tempurl}


\bibitem[\protect\citeauthoryear{Martins, Carilho, Schnell, Duarte, Couto,
  Carrico, and Guerreiro}{Martins et~al\mbox{.}}{2014}]%
        {martins_friendsourcing_2014}
\bibfield{author}{\bibinfo{person}{Joao Martins}, \bibinfo{person}{Jose
  Carilho}, \bibinfo{person}{Oliver Schnell}, \bibinfo{person}{Carlos Duarte},
  \bibinfo{person}{Francisco~M Couto}, \bibinfo{person}{Luis Carrico}, {and}
  \bibinfo{person}{Tiago Guerreiro}.} \bibinfo{year}{2014}\natexlab{}.
\newblock \showarticletitle{Friendsourcing the unmet needs of people with
  dementia}. In \bibinfo{booktitle}{\emph{Proceedings of the 11th {Web} for
  {All} {Conference}}}. \bibinfo{publisher}{ACM}, \bibinfo{pages}{35}.
\newblock


\bibitem[\protect\citeauthoryear{Mattson and Hall}{Mattson and Hall}{2011}]%
        {mattson_health_2011}
\bibfield{author}{\bibinfo{person}{Marifran Mattson} {and}
  \bibinfo{person}{Jennifer~Gibb Hall}.} \bibinfo{year}{2011}\natexlab{}.
\newblock \bibinfo{booktitle}{\emph{Health as {Communication} {Nexus} : {A}
  {Service}-{Learning} {Approach}}}.
\newblock \bibinfo{publisher}{Kendall Hunt Publishing Company}.
\newblock
\urldef\tempurl%
\url{https://www.google.com/search?q=Health+as+Communication+Nexus+%3A+A+Service-Learning+Approach.&oq=Health+as+Communication+Nexus+%3A+A+Service-Learning+Approach.&aqs=chrome..69i57.266j0j7&sourceid=chrome&ie=UTF-8}
\showURL{%
\tempurl}


\bibitem[\protect\citeauthoryear{McGraw and Wong}{McGraw and Wong}{1992}]%
        {mcgraw_common_1992}
\bibfield{author}{\bibinfo{person}{Kenneth~O. McGraw} {and}
  \bibinfo{person}{S.~P. Wong}.} \bibinfo{year}{1992}\natexlab{}.
\newblock \showarticletitle{A common language effect size statistic}.
\newblock \bibinfo{journal}{\emph{Psychological Bulletin}}
  \bibinfo{volume}{111}, \bibinfo{number}{2} (\bibinfo{date}{March}
  \bibinfo{year}{1992}), \bibinfo{pages}{361--365}.
\newblock
\urldef\tempurl%
\url{https://doi.org/10.1037/0033-2909.111.2.361}
\showDOI{\tempurl}


\bibitem[\protect\citeauthoryear{Mcknight and Chervany}{Mcknight and
  Chervany}{1996}]%
        {mcknight_meanings_1996}
\bibfield{author}{\bibinfo{person}{D.~Harrison Mcknight} {and}
  \bibinfo{person}{Norman~L. Chervany}.} \bibinfo{year}{1996}\natexlab{}.
\newblock \bibinfo{booktitle}{\emph{The {Meanings} of {Trust}}}.
\newblock \bibinfo{type}{{T}echnical {R}eport}.
\newblock


\bibitem[\protect\citeauthoryear{McLachlan, Opila, Shah, Sun, and
  Naaman}{McLachlan et~al\mbox{.}}{2016}]%
        {mclachlan_you_2016}
\bibfield{author}{\bibinfo{person}{Ross McLachlan}, \bibinfo{person}{Claire
  Opila}, \bibinfo{person}{Neha Shah}, \bibinfo{person}{Emily Sun}, {and}
  \bibinfo{person}{Mor Naaman}.} \bibinfo{year}{2016}\natexlab{}.
\newblock \showarticletitle{You {Can}'{T} {Always} {Get} {What} {You} {Want}:
  {Challenges} in {P}2P {Resource} {Sharing}}. In
  \bibinfo{booktitle}{\emph{Proceedings of the 2016 {CHI} {Conference}
  {Extended} {Abstracts} on {Human} {Factors} in {Computing} {Systems}}}
  \emph{(\bibinfo{series}{{CHI} {EA} '16})}. \bibinfo{publisher}{ACM},
  \bibinfo{address}{New York, NY, USA}, \bibinfo{pages}{1301--1307}.
\newblock
\showISBNx{978-1-4503-4082-3}
\urldef\tempurl%
\url{https://doi.org/10.1145/2851581.2892358}
\showDOI{\tempurl}


\bibitem[\protect\citeauthoryear{Meier, Lyons, Frydman, Forlenza, and
  Rimer}{Meier et~al\mbox{.}}{2007}]%
        {meier_how_2007}
\bibfield{author}{\bibinfo{person}{Andrea Meier}, \bibinfo{person}{Elizabeth
  Lyons}, \bibinfo{person}{Gilles Frydman}, \bibinfo{person}{Michael Forlenza},
  {and} \bibinfo{person}{Barbara Rimer}.} \bibinfo{year}{2007}\natexlab{}.
\newblock \showarticletitle{How {Cancer} {Survivors} {Provide} {Support} on
  {Cancer}-{Related} {Internet} {Mailing} {Lists}}.
\newblock \bibinfo{journal}{\emph{Journal of Medical Internet Research}}
  \bibinfo{volume}{9}, \bibinfo{number}{2} (\bibinfo{year}{2007}),
  \bibinfo{pages}{e12}.
\newblock


\bibitem[\protect\citeauthoryear{Mikolov, Chen, Corrado, and Dean}{Mikolov
  et~al\mbox{.}}{2013a}]%
        {mikolov_distributed_2013}
\bibfield{author}{\bibinfo{person}{Tomas Mikolov}, \bibinfo{person}{Kai Chen},
  \bibinfo{person}{Greg Corrado}, {and} \bibinfo{person}{Jeffrey Dean}.}
  \bibinfo{year}{2013}\natexlab{a}.
\newblock \showarticletitle{Distributed {Representations} of {Words} and
  {Phrases} and their {Compositionality}}.
\newblock \bibinfo{journal}{\emph{NIPS}} (\bibinfo{year}{2013}),
  \bibinfo{pages}{1--9}.
\newblock
\showISSN{10495258}
\urldef\tempurl%
\url{https://doi.org/10.1162/jmlr.2003.3.4-5.951}
\showDOI{\tempurl}


\bibitem[\protect\citeauthoryear{Mikolov, Chen, Corrado, and Dean}{Mikolov
  et~al\mbox{.}}{2013b}]%
        {mikolov_efficient_2013}
\bibfield{author}{\bibinfo{person}{Tomas Mikolov}, \bibinfo{person}{Kai Chen},
  \bibinfo{person}{Greg Corrado}, {and} \bibinfo{person}{Jeffrey Dean}.}
  \bibinfo{year}{2013}\natexlab{b}.
\newblock \showarticletitle{Efficient {Estimation} of {Word} {Representations}
  in {Vector} {Space}}.
\newblock \bibinfo{journal}{\emph{Arxiv}} (\bibinfo{year}{2013}),
  \bibinfo{pages}{1--12}.
\newblock
\showISSN{15324435}
\urldef\tempurl%
\url{https://doi.org/10.1162/153244303322533223}
\showDOI{\tempurl}


\bibitem[\protect\citeauthoryear{Moadel, Morgan, Fatone, Grennan, Carter,
  Laruffa, Skummy, and Dutcher}{Moadel et~al\mbox{.}}{1999}]%
        {moadel_seeking_1999}
\bibfield{author}{\bibinfo{person}{Alyson Moadel}, \bibinfo{person}{Carole
  Morgan}, \bibinfo{person}{Anne Fatone}, \bibinfo{person}{Jennifer Grennan},
  \bibinfo{person}{Jeanne Carter}, \bibinfo{person}{Gia Laruffa},
  \bibinfo{person}{Anne Skummy}, {and} \bibinfo{person}{Janice Dutcher}.}
  \bibinfo{year}{1999}\natexlab{}.
\newblock \showarticletitle{Seeking meaning and hope: self-reported spiritual
  and existential needs among an ethnically-diverse cancer patient population}.
\newblock \bibinfo{journal}{\emph{Psycho-Oncology: Journal of the
  Psychological, Social and Behavioral Dimensions of Cancer}}
  \bibinfo{volume}{8}, \bibinfo{number}{5} (\bibinfo{year}{1999}),
  \bibinfo{pages}{378--385}.
\newblock


\bibitem[\protect\citeauthoryear{Mueller, Sedley, and Ferrall-Nunge}{Mueller
  et~al\mbox{.}}{2014}]%
        {mueller_survey_2014}
\bibfield{author}{\bibinfo{person}{Hendrik Mueller}, \bibinfo{person}{Aaron
  Sedley}, {and} \bibinfo{person}{Elizabeth Ferrall-Nunge}.}
  \bibinfo{year}{2014}\natexlab{}.
\newblock \showarticletitle{Survey {Research} in {HCI}}.
\newblock In \bibinfo{booktitle}{\emph{Ways of {Knowing} in {HCI}}}.
  \bibinfo{publisher}{Springer, New York, NY}, \bibinfo{pages}{229--266}.
\newblock
\showISBNx{978-1-4939-0377-1 978-1-4939-0378-8}
\urldef\tempurl%
\url{https://doi.org/10.1007/978-1-4939-0378-8_10}
\showDOI{\tempurl}


\bibitem[\protect\citeauthoryear{Muller and Kuhn}{Muller and Kuhn}{1993}]%
        {muller_participatory_1993}
\bibfield{author}{\bibinfo{person}{Michael~J. Muller} {and}
  \bibinfo{person}{Sarah Kuhn}.} \bibinfo{year}{1993}\natexlab{}.
\newblock \showarticletitle{Participatory {Design}}.
\newblock \bibinfo{journal}{\emph{Commun. ACM}} \bibinfo{volume}{36},
  \bibinfo{number}{6} (\bibinfo{date}{June} \bibinfo{year}{1993}),
  \bibinfo{pages}{24--28}.
\newblock
\showISSN{0001-0782}


\bibitem[\protect\citeauthoryear{Narayanan and Shmatikov}{Narayanan and
  Shmatikov}{2010}]%
        {narayanan_myths_2010}
\bibfield{author}{\bibinfo{person}{Arvind Narayanan} {and}
  \bibinfo{person}{Vitaly Shmatikov}.} \bibinfo{year}{2010}\natexlab{}.
\newblock \showarticletitle{Myths and {Fallacies} of "{Personally}
  {Identifiable} {Information}"}.
\newblock \bibinfo{journal}{\emph{Commun. ACM}} \bibinfo{volume}{53},
  \bibinfo{number}{6} (\bibinfo{date}{June} \bibinfo{year}{2010}),
  \bibinfo{pages}{24--26}.
\newblock
\showISSN{0001-0782}
\urldef\tempurl%
\url{https://doi.org/10.1145/1743546.1743558}
\showDOI{\tempurl}


\bibitem[\protect\citeauthoryear{Newman, Lauterbach, Munson, Resnick, and
  Morris}{Newman et~al\mbox{.}}{2011}]%
        {newman_its_2011}
\bibfield{author}{\bibinfo{person}{Mark~W. Newman}, \bibinfo{person}{Debra
  Lauterbach}, \bibinfo{person}{Sean~A. Munson}, \bibinfo{person}{Paul
  Resnick}, {and} \bibinfo{person}{Margaret~E. Morris}.}
  \bibinfo{year}{2011}\natexlab{}.
\newblock \showarticletitle{It's {Not} {That} {I} {Don}'{T} {Have} {Problems},
  {I}'{M} {Just} {Not} {Putting} {Them} on {Facebook}: {Challenges} and
  {Opportunities} in {Using} {Online} {Social} {Networks} for {Health}}. In
  \bibinfo{booktitle}{\emph{Proceedings of the {ACM} 2011 {Conference} on
  {Computer} {Supported} {Cooperative} {Work}}} \emph{(\bibinfo{series}{{CSCW}
  '11})}. \bibinfo{publisher}{ACM}, \bibinfo{address}{New York, NY, USA},
  \bibinfo{pages}{341--350}.
\newblock
\showISBNx{978-1-4503-0556-3}
\urldef\tempurl%
\url{https://doi.org/10.1145/1958824.1958876}
\showDOI{\tempurl}


\bibitem[\protect\citeauthoryear{Nicholas, Chahauver, Hetherington, McNeill,
  and Bouffet}{Nicholas et~al\mbox{.}}{2012}]%
        {nicholas_evaluation_2012}
\bibfield{author}{\bibinfo{person}{David~B. Nicholas}, \bibinfo{person}{Anu
  Chahauver}, \bibinfo{person}{Ross Hetherington}, \bibinfo{person}{Ted
  McNeill}, {and} \bibinfo{person}{Eric Bouffet}.}
  \bibinfo{year}{2012}\natexlab{}.
\newblock \showarticletitle{Evaluation of an {Online} {Peer} {Support}
  {Network} for {Fathers} of a {Child} {With} a {Brain} {Tumor}}.
\newblock \bibinfo{journal}{\emph{Social Work in Health Care}}
  \bibinfo{volume}{51}, \bibinfo{number}{3} (\bibinfo{date}{March}
  \bibinfo{year}{2012}), \bibinfo{pages}{232--245}.
\newblock
\showISSN{0098-1389}
\urldef\tempurl%
\url{https://doi.org/10.1080/00981389.2011.631696}
\showDOI{\tempurl}


\bibitem[\protect\citeauthoryear{of~Medicine}{of~Medicine}{2015}]%
        {institute_of_medicine_dying_2015}
\bibfield{author}{\bibinfo{person}{Institute of Medicine}.}
  \bibinfo{year}{2015}\natexlab{}.
\newblock \bibinfo{booktitle}{\emph{Dying in {America}: {Improving} quality and
  honoring individual preferences near the end of life}}.
\newblock \bibinfo{publisher}{National Academies Press},
  \bibinfo{address}{Washington, D.C.}
\newblock


\bibitem[\protect\citeauthoryear{Olson, Grudin, and Horvitz}{Olson
  et~al\mbox{.}}{2005}]%
        {olson_study_2005}
\bibfield{author}{\bibinfo{person}{Judith~S. Olson}, \bibinfo{person}{Jonathan
  Grudin}, {and} \bibinfo{person}{Eric Horvitz}.}
  \bibinfo{year}{2005}\natexlab{}.
\newblock \showarticletitle{A {Study} of {Preferences} for {Sharing} and
  {Privacy}}. In \bibinfo{booktitle}{\emph{{CHI} '05 {Extended} {Abstracts} on
  {Human} {Factors} in {Computing} {Systems}}} \emph{(\bibinfo{series}{{CHI}
  {EA} '05})}. \bibinfo{publisher}{ACM}, \bibinfo{address}{New York, NY, USA},
  \bibinfo{pages}{1985--1988}.
\newblock
\showISBNx{978-1-59593-002-6}
\urldef\tempurl%
\url{https://doi.org/10.1145/1056808.1057073}
\showDOI{\tempurl}
\newblock
\shownote{event-place: Portland, OR, USA.}


\bibitem[\protect\citeauthoryear{Peng, Guo, Tsang, and Ma}{Peng
  et~al\mbox{.}}{2020}]%
        {pengexploring}
\bibfield{author}{\bibinfo{person}{Zhenhui Peng}, \bibinfo{person}{Qingyu Guo},
  \bibinfo{person}{Ka~Wing Tsang}, {and} \bibinfo{person}{Xiaojuan Ma}.}
  \bibinfo{year}{2020}\natexlab{}.
\newblock \showarticletitle{Exploring the Effects of Technological Writing
  Assistance for Support Providers in Online Mental Health Community}.
\newblock  (\bibinfo{year}{2020}).
\newblock


\bibitem[\protect\citeauthoryear{Phelps, Lauderdale, Alcorn, Dillinger,
  Balboni, Van~Wert, VanderWeele, and Balboni}{Phelps et~al\mbox{.}}{2012}]%
        {phelps_addressing_2012}
\bibfield{author}{\bibinfo{person}{Andrea~C. Phelps},
  \bibinfo{person}{Katharine~E. Lauderdale}, \bibinfo{person}{Sara Alcorn},
  \bibinfo{person}{Jennifer Dillinger}, \bibinfo{person}{Michael~T. Balboni},
  \bibinfo{person}{Michael Van~Wert}, \bibinfo{person}{Tyler~J. VanderWeele},
  {and} \bibinfo{person}{Tracy~A. Balboni}.} \bibinfo{year}{2012}\natexlab{}.
\newblock \showarticletitle{Addressing {Spirituality} {Within} the {Care} of
  {Patients} at the {End} of {Life}: {Perspectives} of {Patients} {With}
  {Advanced} {Cancer}, {Oncologists}, and {Oncology} {Nurses}}.
\newblock \bibinfo{journal}{\emph{Journal of Clinical Oncology}}
  \bibinfo{volume}{30}, \bibinfo{number}{20} (\bibinfo{date}{July}
  \bibinfo{year}{2012}), \bibinfo{pages}{2538--2544}.
\newblock
\showISSN{0732-183X}
\urldef\tempurl%
\url{https://doi.org/10.1200/JCO.2011.40.3766}
\showDOI{\tempurl}


\bibitem[\protect\citeauthoryear{Puchalski and Ferrell}{Puchalski and
  Ferrell}{2011}]%
        {puchalski2011making}
\bibfield{author}{\bibinfo{person}{Christina Puchalski} {and}
  \bibinfo{person}{Betty Ferrell}.} \bibinfo{year}{2011}\natexlab{}.
\newblock \bibinfo{booktitle}{\emph{Making health care whole: Integrating
  spirituality into patient care}}.
\newblock \bibinfo{publisher}{Templeton Foundation Press}.
\newblock


\bibitem[\protect\citeauthoryear{Puchalski, Ferrell, Virani, Otis-Green, Baird,
  Bull, Chochinov, Handzo, Nelson-Becker, Prince-Paul, Pugliese, and
  Sulmasy}{Puchalski et~al\mbox{.}}{2009}]%
        {puchalski_improving_2009}
\bibfield{author}{\bibinfo{person}{Christina Puchalski}, \bibinfo{person}{Betty
  Ferrell}, \bibinfo{person}{Rose Virani}, \bibinfo{person}{Shirley
  Otis-Green}, \bibinfo{person}{Pamela Baird}, \bibinfo{person}{Janet Bull},
  \bibinfo{person}{Harvey Chochinov}, \bibinfo{person}{George Handzo},
  \bibinfo{person}{Holly Nelson-Becker}, \bibinfo{person}{Maryjo Prince-Paul},
  \bibinfo{person}{Karen Pugliese}, {and} \bibinfo{person}{Daniel Sulmasy}.}
  \bibinfo{year}{2009}\natexlab{}.
\newblock \showarticletitle{Improving the {Quality} of {Spiritual} {Care} as a
  {Dimension} of {Palliative} {Care}: {The} {Report} of the {Consensus}
  {Conference}}.
\newblock \bibinfo{journal}{\emph{Journal of Palliative Medicine}}
  \bibinfo{volume}{12}, \bibinfo{number}{10} (\bibinfo{date}{Oct.}
  \bibinfo{year}{2009}), \bibinfo{pages}{885--904}.
\newblock
\showISSN{1096-6218}
\urldef\tempurl%
\url{https://doi.org/10.1089/jpm.2009.0142}
\showDOI{\tempurl}


\bibitem[\protect\citeauthoryear{{Qualtrics}}{{Qualtrics}}{2018}]%
        {qualtrics_qualtrics_2018}
\bibfield{author}{\bibinfo{person}{{Qualtrics}}.}
  \bibinfo{year}{2018}\natexlab{}.
\newblock \bibinfo{booktitle}{\emph{Qualtrics {Software}}}.
\newblock
\urldef\tempurl%
\url{https://www.qualtrics.com}
\showURL{%
\tempurl}


\bibitem[\protect\citeauthoryear{Rahman, Addo, and Ahamed}{Rahman
  et~al\mbox{.}}{2014}]%
        {rahman_prisn:_2014}
\bibfield{author}{\bibinfo{person}{Farzana Rahman}, \bibinfo{person}{Ivor~D.
  Addo}, {and} \bibinfo{person}{Sheikh~I. Ahamed}.}
  \bibinfo{year}{2014}\natexlab{}.
\newblock \showarticletitle{{PriSN}: {A} {Privacy} {Protection} {Framework} for
  {Healthcare} {Social} {Networking} {Sites}}. In
  \bibinfo{booktitle}{\emph{Proceedings of the 2014 {Conference} on {Research}
  in {Adaptive} and {Convergent} {Systems}}} \emph{(\bibinfo{series}{{RACS}
  '14})}. \bibinfo{publisher}{ACM}, \bibinfo{address}{New York, NY, USA},
  \bibinfo{pages}{66--71}.
\newblock
\showISBNx{978-1-4503-3060-2}
\urldef\tempurl%
\url{https://doi.org/10.1145/2663761.2664199}
\showDOI{\tempurl}


\bibitem[\protect\citeauthoryear{Rains and Wright}{Rains and Wright}{2016}]%
        {rains_social_2016}
\bibfield{author}{\bibinfo{person}{Stephen~A. Rains} {and}
  \bibinfo{person}{Kevin~B. Wright}.} \bibinfo{year}{2016}\natexlab{}.
\newblock \showarticletitle{Social {Support} and {Computer}-{Mediated}
  {Communication}: {A} {State}-of-the-{Art} {Review} and {Agenda} for {Future}
  {Research}}.
\newblock \bibinfo{journal}{\emph{Annals of the International Communication
  Association}} \bibinfo{volume}{40}, \bibinfo{number}{1} (\bibinfo{date}{Jan.}
  \bibinfo{year}{2016}), \bibinfo{pages}{175--211}.
\newblock
\showISSN{2380-8985}
\urldef\tempurl%
\url{https://doi.org/10.1080/23808985.2015.11735260}
\showDOI{\tempurl}


\bibitem[\protect\citeauthoryear{Rho, Haimson, Andalibi, Mazmanian, and
  Hayes}{Rho et~al\mbox{.}}{2017}]%
        {rho_class_2017}
\bibfield{author}{\bibinfo{person}{Eugenia Ha~Rim Rho},
  \bibinfo{person}{Oliver~L. Haimson}, \bibinfo{person}{Nazanin Andalibi},
  \bibinfo{person}{Melissa Mazmanian}, {and} \bibinfo{person}{Gillian~R.
  Hayes}.} \bibinfo{year}{2017}\natexlab{}.
\newblock \showarticletitle{Class {Confessions}: {Restorative} {Properties} in
  {Online} {Experiences} of {Socioeconomic} {Stigma}}. In
  \bibinfo{booktitle}{\emph{Proceedings of the 2017 {CHI} {Conference} on
  {Human} {Factors} in {Computing} {Systems}}} \emph{(\bibinfo{series}{{CHI}
  '17})}. \bibinfo{publisher}{ACM}, \bibinfo{address}{New York, NY, USA},
  \bibinfo{pages}{3377--3389}.
\newblock
\showISBNx{978-1-4503-4655-9}
\urldef\tempurl%
\url{https://doi.org/10.1145/3025453.3025921}
\showDOI{\tempurl}
\newblock
\shownote{event-place: Denver, Colorado, USA.}


\bibitem[\protect\citeauthoryear{Rhue and Robert}{Rhue and Robert}{2018}]%
        {rhue_emotional_2018}
\bibfield{author}{\bibinfo{person}{Lauren Rhue} {and}
  \bibinfo{person}{Lionel~P. Robert}.} \bibinfo{year}{2018}\natexlab{}.
\newblock \showarticletitle{Emotional {Delivery} in {Pro}-social {Crowdfunding}
  {Success}}. In \bibinfo{booktitle}{\emph{Extended {Abstracts} of the 2018
  {CHI} {Conference} on {Human} {Factors} in {Computing} {Systems}}}
  \emph{(\bibinfo{series}{{CHI} {EA} '18})}. \bibinfo{publisher}{ACM},
  \bibinfo{address}{New York, NY, USA}, \bibinfo{pages}{LBW019:1--LBW019:6}.
\newblock
\showISBNx{978-1-4503-5621-3}
\urldef\tempurl%
\url{https://doi.org/10.1145/3170427.3188534}
\showDOI{\tempurl}
\newblock
\shownote{event-place: Montreal QC, Canada.}


\bibitem[\protect\citeauthoryear{Rubya and Yarosh}{Rubya and Yarosh}{2017}]%
        {rubya_video-mediated_2017}
\bibfield{author}{\bibinfo{person}{Sabirat Rubya} {and}
  \bibinfo{person}{Svetlana Yarosh}.} \bibinfo{year}{2017}\natexlab{}.
\newblock \showarticletitle{Video-{Mediated} {Peer} {Support} in an {Online}
  {Community} for {Recovery} from {Substance} {Use} {Disorders}}. In
  \bibinfo{booktitle}{\emph{Proceedings of the 2017 {ACM} {Conference} on
  {Computer} {Supported} {Cooperative} {Work} and {Social} {Computing}}}
  \emph{(\bibinfo{series}{{CSCW} '17})}. \bibinfo{publisher}{ACM},
  \bibinfo{address}{New York, NY, USA}, \bibinfo{pages}{1454--1469}.
\newblock
\showISBNx{978-1-4503-4335-0}
\urldef\tempurl%
\url{https://doi.org/10.1145/2998181.2998246}
\showDOI{\tempurl}


\bibitem[\protect\citeauthoryear{Schorch, Wan, Randall, and Wulf}{Schorch
  et~al\mbox{.}}{2016}]%
        {schorch_designing_2016}
\bibfield{author}{\bibinfo{person}{Maren Schorch}, \bibinfo{person}{Lin Wan},
  \bibinfo{person}{David~William Randall}, {and} \bibinfo{person}{Volker
  Wulf}.} \bibinfo{year}{2016}\natexlab{}.
\newblock \showarticletitle{Designing for {Those} who are {Overlooked}:
  {Insider} {Perspectives} on {Care} {Practices} and {Cooperative} {Work} of
  {Elderly} {Informal} {Caregivers}}. In \bibinfo{booktitle}{\emph{Proceedings
  of the 19th {ACM} {Conference} on {Computer}-{Supported} {Cooperative} {Work}
  \& {Social} {Computing}}}. \bibinfo{publisher}{ACM},
  \bibinfo{pages}{787--799}.
\newblock


\bibitem[\protect\citeauthoryear{Schulz and Beach}{Schulz and Beach}{1999}]%
        {schulz_caregiving_1999}
\bibfield{author}{\bibinfo{person}{Richard Schulz} {and}
  \bibinfo{person}{Scott~R. Beach}.} \bibinfo{year}{1999}\natexlab{}.
\newblock \showarticletitle{Caregiving as a {Risk} {Factor} for {Mortality}:
  {The} {Caregiver} {Health} {Effects} {Study}}.
\newblock \bibinfo{journal}{\emph{JAMA}} \bibinfo{volume}{282},
  \bibinfo{number}{23} (\bibinfo{date}{Dec.} \bibinfo{year}{1999}),
  \bibinfo{pages}{2215--2219}.
\newblock
\showISSN{0098-7484}
\urldef\tempurl%
\url{https://doi.org/10.1001/jama.282.23.2215}
\showDOI{\tempurl}


\bibitem[\protect\citeauthoryear{Sharma and De~Choudhury}{Sharma and
  De~Choudhury}{2018}]%
        {sharma_mental_2018}
\bibfield{author}{\bibinfo{person}{Eva Sharma} {and} \bibinfo{person}{Munmun
  De~Choudhury}.} \bibinfo{year}{2018}\natexlab{}.
\newblock \showarticletitle{Mental {Health} {Support} and {Its} {Relationship}
  to {Linguistic} {Accommodation} in {Online} {Communities}}. In
  \bibinfo{booktitle}{\emph{Proceedings of the 2018 {CHI} {Conference} on
  {Human} {Factors} in {Computing} {Systems}}} \emph{(\bibinfo{series}{{CHI}
  '18})}. \bibinfo{publisher}{ACM}, \bibinfo{address}{New York, NY, USA},
  \bibinfo{pages}{641:1--641:13}.
\newblock
\showISBNx{978-1-4503-5620-6}
\urldef\tempurl%
\url{https://doi.org/10.1145/3173574.3174215}
\showDOI{\tempurl}
\newblock
\shownote{event-place: Montreal QC, Canada.}


\bibitem[\protect\citeauthoryear{Shaw, Han, Kim, Gustafson, Hawkins, Cleary,
  McTavish, Pingree, Eliason, and Lumpkins}{Shaw et~al\mbox{.}}{2007}]%
        {shaw2007effects}
\bibfield{author}{\bibinfo{person}{Bret Shaw}, \bibinfo{person}{Jeong~Yeob
  Han}, \bibinfo{person}{Eunkyung Kim}, \bibinfo{person}{David Gustafson},
  \bibinfo{person}{Robert Hawkins}, \bibinfo{person}{James Cleary},
  \bibinfo{person}{Fiona McTavish}, \bibinfo{person}{Suzanne Pingree},
  \bibinfo{person}{Patricia Eliason}, {and} \bibinfo{person}{Crystal
  Lumpkins}.} \bibinfo{year}{2007}\natexlab{}.
\newblock \showarticletitle{Effects of prayer and religious expression within
  computer support groups on women with breast cancer}.
\newblock \bibinfo{journal}{\emph{Psycho-Oncology: Journal of the
  Psychological, Social and Behavioral Dimensions of Cancer}}
  \bibinfo{volume}{16}, \bibinfo{number}{7} (\bibinfo{year}{2007}),
  \bibinfo{pages}{676--687}.
\newblock


\bibitem[\protect\citeauthoryear{Shumaker and Brownell}{Shumaker and
  Brownell}{1984}]%
        {shumaker_toward_1984}
\bibfield{author}{\bibinfo{person}{Sally~A Shumaker} {and}
  \bibinfo{person}{Arlene Brownell}.} \bibinfo{year}{1984}\natexlab{}.
\newblock \showarticletitle{Toward a theory of social support: {Closing}
  conceptual gaps}.
\newblock \bibinfo{journal}{\emph{Journal of social issues}}
  \bibinfo{volume}{40}, \bibinfo{number}{4} (\bibinfo{year}{1984}),
  \bibinfo{pages}{11--36}.
\newblock


\bibitem[\protect\citeauthoryear{Siler, Mamier, Winslow, and Ferrell}{Siler
  et~al\mbox{.}}{2019}]%
        {siler_interprofessional_2019}
\bibfield{author}{\bibinfo{person}{S. Siler}, \bibinfo{person}{I. Mamier},
  \bibinfo{person}{B.~W. Winslow}, {and} \bibinfo{person}{B.~R. Ferrell}.}
  \bibinfo{year}{2019}\natexlab{}.
\newblock \showarticletitle{Interprofessional {Perspectives} on {Providing}
  {Spiritual} {Care} for {Patients} {With} {Lung} {Cancer} in {Outpatient}
  {Settings}.}
\newblock \bibinfo{journal}{\emph{Oncology nursing forum}}
  \bibinfo{volume}{46}, \bibinfo{number}{1} (\bibinfo{date}{Jan.}
  \bibinfo{year}{2019}), \bibinfo{pages}{49--58}.
\newblock
\showISSN{0190-535X}
\urldef\tempurl%
\url{https://doi.org/10.1188/19.ONF.49-58}
\showDOI{\tempurl}


\bibitem[\protect\citeauthoryear{Simoni, Franks, Lehavot, and Yard}{Simoni
  et~al\mbox{.}}{2011}]%
        {simoni_peer_2011}
\bibfield{author}{\bibinfo{person}{Jane~M. Simoni}, \bibinfo{person}{Julie~C.
  Franks}, \bibinfo{person}{Keren Lehavot}, {and} \bibinfo{person}{Samantha~S.
  Yard}.} \bibinfo{year}{2011}\natexlab{}.
\newblock \showarticletitle{Peer interventions to promote health: {Conceptual}
  considerations}.
\newblock \bibinfo{journal}{\emph{American Journal of Orthopsychiatry}}
  \bibinfo{volume}{81}, \bibinfo{number}{3} (\bibinfo{year}{2011}),
  \bibinfo{pages}{351--359}.
\newblock
\showISSN{1939-0025(Electronic),0002-9432(Print)}
\urldef\tempurl%
\url{https://doi.org/10.1111/j.1939-0025.2011.01103.x}
\showDOI{\tempurl}


\bibitem[\protect\citeauthoryear{Skeels}{Skeels}{2010}]%
        {skeels_sharing_2010}
\bibfield{author}{\bibinfo{person}{Meredith~McLain Skeels}.}
  \bibinfo{year}{2010}\natexlab{}.
\newblock \emph{\bibinfo{title}{Sharing by design: {Understanding} and
  supporting personal health information sharing and collaboration within
  social networks}}.
\newblock Ph.{D}. \bibinfo{school}{University of Washington},
  \bibinfo{address}{Washington, United States}.
\newblock


\bibitem[\protect\citeauthoryear{Skeels, Unruh, Powell, and Pratt}{Skeels
  et~al\mbox{.}}{2010}]%
        {skeels_catalyzing_2010}
\bibfield{author}{\bibinfo{person}{Meredith~McLain Skeels},
  \bibinfo{person}{Kenton~T. Unruh}, \bibinfo{person}{Christopher Powell},
  {and} \bibinfo{person}{Wanda Pratt}.} \bibinfo{year}{2010}\natexlab{}.
\newblock \showarticletitle{Catalyzing {Social} {Support} for {Breast} {Cancer}
  {Patients}}. In \bibinfo{booktitle}{\emph{Proceedings of the {SIGCHI}
  {Conference} on {Human} {Factors} in {Computing} {Systems}}}
  \emph{(\bibinfo{series}{{CHI} '10})}. \bibinfo{publisher}{ACM},
  \bibinfo{address}{New York, NY, USA}, \bibinfo{pages}{173--182}.
\newblock
\showISBNx{978-1-60558-929-9}


\bibitem[\protect\citeauthoryear{Sonderen}{Sonderen}{1993}]%
        {sonderen_het_1993}
\bibfield{author}{\bibinfo{person}{F.L.P.~van Sonderen}.}
  \bibinfo{year}{1993}\natexlab{}.
\newblock \bibinfo{booktitle}{\emph{Het meten van sociale steun met de
  {Sociale} {Steun} {Lijst}-{Interacties} ({SSL}-{I}) en {Sociale} {Steun}
  {Lijst}-{Discrepanties} ({SSL}-{D}): een handleiding [{Assessing} social
  support with the {Social} {Support} {List}-{Interactions} ({SSL}-{I}) and the
  social support list-discrepancies, a manual]}}.
\newblock \bibinfo{publisher}{Noordelijk Centrum voor Gezondheidsvraagstukken,
  Rijksuniversiteit Groningen}, \bibinfo{address}{Groningen}.
\newblock
\showISBNx{978-90-72156-59-4}


\bibitem[\protect\citeauthoryear{Stokes}{Stokes}{1983}]%
        {stokes_predicting_1983}
\bibfield{author}{\bibinfo{person}{Joseph~P. Stokes}.}
  \bibinfo{year}{1983}\natexlab{}.
\newblock \showarticletitle{Predicting satisfaction with social support from
  social network structure}.
\newblock \bibinfo{journal}{\emph{American Journal of Community Psychology}}
  \bibinfo{volume}{11}, \bibinfo{number}{2} (\bibinfo{year}{1983}),
  \bibinfo{pages}{141--152}.
\newblock


\bibitem[\protect\citeauthoryear{Taylor and Mamier}{Taylor and Mamier}{2005}]%
        {taylor_spiritual_2005}
\bibfield{author}{\bibinfo{person}{Elizabeth~Johnston Taylor} {and}
  \bibinfo{person}{Iris Mamier}.} \bibinfo{year}{2005}\natexlab{}.
\newblock \showarticletitle{Spiritual care nursing: what cancer patients and
  family caregivers want}.
\newblock \bibinfo{journal}{\emph{Journal of Advanced Nursing}}
  \bibinfo{volume}{49}, \bibinfo{number}{3} (\bibinfo{year}{2005}),
  \bibinfo{pages}{260--267}.
\newblock
\showISSN{1365-2648}
\urldef\tempurl%
\url{https://doi.org/10.1111/j.1365-2648.2004.03285.x}
\showDOI{\tempurl}


\bibitem[\protect\citeauthoryear{Thebault-Spieker, Terveen, and
  Hecht}{Thebault-Spieker et~al\mbox{.}}{2017}]%
        {thebault-spieker_toward_2017}
\bibfield{author}{\bibinfo{person}{Jacob Thebault-Spieker},
  \bibinfo{person}{Loren Terveen}, {and} \bibinfo{person}{Brent Hecht}.}
  \bibinfo{year}{2017}\natexlab{}.
\newblock \showarticletitle{Toward a {Geographic} {Understanding} of the
  {Sharing} {Economy}: {Systemic} {Biases} in {UberX} and {TaskRabbit}}.
\newblock \bibinfo{journal}{\emph{ACM Trans. Comput.-Hum. Interact.}}
  \bibinfo{volume}{24}, \bibinfo{number}{3} (\bibinfo{date}{April}
  \bibinfo{year}{2017}), \bibinfo{pages}{21:1--21:40}.
\newblock
\showISSN{1073-0516}


\bibitem[\protect\citeauthoryear{Thebault-Spieker, Terveen, and
  Hecht}{Thebault-Spieker et~al\mbox{.}}{2015}]%
        {thebault-spieker_avoiding_2015}
\bibfield{author}{\bibinfo{person}{Jacob Thebault-Spieker},
  \bibinfo{person}{Loren~G Terveen}, {and} \bibinfo{person}{Brent Hecht}.}
  \bibinfo{year}{2015}\natexlab{}.
\newblock \showarticletitle{Avoiding the south side and the suburbs: {The}
  geography of mobile crowdsourcing markets}. In
  \bibinfo{booktitle}{\emph{Proceedings of the 18th {ACM} {Conference} on
  {Computer} {Supported} {Cooperative} {Work} \& {Social} {Computing}}}.
  \bibinfo{publisher}{ACM}, \bibinfo{pages}{265--275}.
\newblock


\bibitem[\protect\citeauthoryear{Thong, Kaptein, Krediet, Boeschoten, and
  Dekker}{Thong et~al\mbox{.}}{2007}]%
        {thong_social_2007}
\bibfield{author}{\bibinfo{person}{Melissa S.~Y. Thong},
  \bibinfo{person}{Adrian~A. Kaptein}, \bibinfo{person}{Raymond~T. Krediet},
  \bibinfo{person}{Elisabeth~W. Boeschoten}, {and} \bibinfo{person}{Friedo~W.
  Dekker}.} \bibinfo{year}{2007}\natexlab{}.
\newblock \showarticletitle{Social support predicts survival in dialysis
  patients}.
\newblock \bibinfo{journal}{\emph{Nephrology Dialysis Transplantation}}
  \bibinfo{volume}{22}, \bibinfo{number}{3} (\bibinfo{date}{March}
  \bibinfo{year}{2007}), \bibinfo{pages}{845--850}.
\newblock
\showISSN{0931-0509}


\bibitem[\protect\citeauthoryear{Umberson and Karas~Montez}{Umberson and
  Karas~Montez}{2010}]%
        {umberson_social_2010}
\bibfield{author}{\bibinfo{person}{Debra Umberson} {and}
  \bibinfo{person}{Jennifer Karas~Montez}.} \bibinfo{year}{2010}\natexlab{}.
\newblock \showarticletitle{Social {Relationships} and {Health}: {A}
  {Flashpoint} for {Health} {Policy}}.
\newblock \bibinfo{journal}{\emph{Journal of Health and Social Behavior}}
  \bibinfo{volume}{51}, \bibinfo{number}{1\_suppl} (\bibinfo{date}{March}
  \bibinfo{year}{2010}), \bibinfo{pages}{S54--S66}.
\newblock
\showISSN{0022-1465}
\urldef\tempurl%
\url{https://doi.org/10.1177/0022146510383501}
\showDOI{\tempurl}


\bibitem[\protect\citeauthoryear{Vallurupalli, Lauderdale, Balboni, Phelps,
  Block, Ng, Kachnic, VanderWeele, and Balboni}{Vallurupalli
  et~al\mbox{.}}{2012}]%
        {vallurupalli_role_2012}
\bibfield{author}{\bibinfo{person}{Ms~Mounica Vallurupalli},
  \bibinfo{person}{Ms~Katharine Lauderdale}, \bibinfo{person}{Michael~J
  Balboni}, \bibinfo{person}{Andrea~C Phelps}, \bibinfo{person}{Susan~D Block},
  \bibinfo{person}{Andrea~K Ng}, \bibinfo{person}{Lisa~A Kachnic},
  \bibinfo{person}{Tyler~J VanderWeele}, {and} \bibinfo{person}{Tracy~A
  Balboni}.} \bibinfo{year}{2012}\natexlab{}.
\newblock \showarticletitle{The role of spirituality and religious coping in
  the quality of life of patients with advanced cancer receiving palliative
  radiation therapy}.
\newblock \bibinfo{journal}{\emph{The journal of supportive oncology}}
  \bibinfo{volume}{10}, \bibinfo{number}{2} (\bibinfo{year}{2012}),
  \bibinfo{pages}{81}.
\newblock


\bibitem[\protect\citeauthoryear{Vitak and Kim}{Vitak and Kim}{2014}]%
        {vitak_``you_2014}
\bibfield{author}{\bibinfo{person}{Jessica Vitak} {and}
  \bibinfo{person}{Jinyoung Kim}.} \bibinfo{year}{2014}\natexlab{}.
\newblock \showarticletitle{``{You} {Can}'t {Block} {People} {Offline}'':
  {Examining} {How} {Facebook}'s {Affordances} {Shape} the {Disclosure}
  {Process}}. In \bibinfo{booktitle}{\emph{Proceedings of the 17th {ACM}
  {Conference} on {Computer} {Supported} {Cooperative} {Work} \& {Social}
  {Computing}}} \emph{(\bibinfo{series}{{CSCW} '14})}.
  \bibinfo{publisher}{ACM}, \bibinfo{address}{New York, NY, USA},
  \bibinfo{pages}{461--474}.
\newblock
\showISBNx{978-1-4503-2540-0}
\urldef\tempurl%
\url{https://doi.org/10.1145/2531602.2531672}
\showDOI{\tempurl}
\newblock
\shownote{event-place: Baltimore, Maryland, USA.}


\bibitem[\protect\citeauthoryear{Vlahovic, Wang, Kraut, and Levine}{Vlahovic
  et~al\mbox{.}}{2014}]%
        {vlahovic2014support}
\bibfield{author}{\bibinfo{person}{Tatiana~A Vlahovic},
  \bibinfo{person}{Yi-Chia Wang}, \bibinfo{person}{Robert~E Kraut}, {and}
  \bibinfo{person}{John~M Levine}.} \bibinfo{year}{2014}\natexlab{}.
\newblock \showarticletitle{Support matching and satisfaction in an online
  breast cancer support community}. In \bibinfo{booktitle}{\emph{Proceedings of
  the SIGCHI Conference on Human Factors in Computing Systems}}.
  \bibinfo{pages}{1625--1634}.
\newblock


\bibitem[\protect\citeauthoryear{Walther and Boyd}{Walther and Boyd}{2002}]%
        {walther2002attraction}
\bibfield{author}{\bibinfo{person}{Joseph~B Walther} {and}
  \bibinfo{person}{Shawn Boyd}.} \bibinfo{year}{2002}\natexlab{}.
\newblock \showarticletitle{Attraction to computer-mediated social support}.
\newblock \bibinfo{journal}{\emph{Communication technology and society:
  Audience adoption and uses}}  \bibinfo{volume}{153188}
  (\bibinfo{year}{2002}).
\newblock


\bibitem[\protect\citeauthoryear{Wang, Zhao, and Street}{Wang
  et~al\mbox{.}}{2014}]%
        {wang_social_2014}
\bibfield{author}{\bibinfo{person}{Xi Wang}, \bibinfo{person}{Kang Zhao}, {and}
  \bibinfo{person}{Nick Street}.} \bibinfo{year}{2014}\natexlab{}.
\newblock \showarticletitle{Social support and user engagement in online health
  communities}. In \bibinfo{booktitle}{\emph{International {Conference} on
  {Smart} {Health}}}. \bibinfo{publisher}{Springer}, \bibinfo{pages}{97--110}.
\newblock


\bibitem[\protect\citeauthoryear{Wang, Kraut, and Levine}{Wang
  et~al\mbox{.}}{2012}]%
        {wang_stay_2012}
\bibfield{author}{\bibinfo{person}{Yi-Chia Wang}, \bibinfo{person}{Robert
  Kraut}, {and} \bibinfo{person}{John~M Levine}.}
  \bibinfo{year}{2012}\natexlab{}.
\newblock \showarticletitle{To stay or leave?: the relationship of emotional
  and informational support to commitment in online health support groups}. In
  \bibinfo{booktitle}{\emph{Proceedings of the {ACM} 2012 conference on
  {Computer} {Supported} {Cooperative} {Work}}}. \bibinfo{publisher}{ACM},
  \bibinfo{pages}{833--842}.
\newblock


\bibitem[\protect\citeauthoryear{Wills}{Wills}{1985}]%
        {wills_supportive_1985}
\bibfield{author}{\bibinfo{person}{Thomas~Ashby Wills}.}
  \bibinfo{year}{1985}\natexlab{}.
\newblock \showarticletitle{Supportive functions of inter-personal
  relationships.}
\newblock In \bibinfo{booktitle}{\emph{Social support and health.}}
  \bibinfo{publisher}{Academic Press}, \bibinfo{address}{San Diego, CA, US},
  \bibinfo{pages}{61--82}.
\newblock


\bibitem[\protect\citeauthoryear{Wohn, Freeman, and McLaughlin}{Wohn
  et~al\mbox{.}}{2018}]%
        {wohn_explaining_2018}
\bibfield{author}{\bibinfo{person}{Donghee~Yvette Wohn}, \bibinfo{person}{Guo
  Freeman}, {and} \bibinfo{person}{Caitlin McLaughlin}.}
  \bibinfo{year}{2018}\natexlab{}.
\newblock \showarticletitle{Explaining {Viewers}' {Emotional}, {Instrumental},
  and {Financial} {Support} {Provision} for {Live} {Streamers}}. In
  \bibinfo{booktitle}{\emph{Proceedings of the 2018 {CHI} {Conference} on
  {Human} {Factors} in {Computing} {Systems}}} \emph{(\bibinfo{series}{{CHI}
  '18})}. \bibinfo{publisher}{ACM}, \bibinfo{address}{New York, NY, USA},
  \bibinfo{pages}{474:1--474:13}.
\newblock
\showISBNx{978-1-4503-5620-6}
\urldef\tempurl%
\url{https://doi.org/10.1145/3173574.3174048}
\showDOI{\tempurl}
\newblock
\shownote{event-place: Montreal QC, Canada.}


\bibitem[\protect\citeauthoryear{Wright}{Wright}{2000}]%
        {wright2000perceptions}
\bibfield{author}{\bibinfo{person}{Kevin Wright}.}
  \bibinfo{year}{2000}\natexlab{}.
\newblock \showarticletitle{Perceptions of on-line support providers: An
  examination of perceived homophily, source credibility, communication and
  social support within on-line support groups}.
\newblock \bibinfo{journal}{\emph{Communication Quarterly}}
  \bibinfo{volume}{48}, \bibinfo{number}{1} (\bibinfo{year}{2000}),
  \bibinfo{pages}{44--59}.
\newblock


\bibitem[\protect\citeauthoryear{Wu and Harden}{Wu and Harden}{2015}]%
        {wu_symptom_2015}
\bibfield{author}{\bibinfo{person}{Horng-Shiuann Wu} {and}
  \bibinfo{person}{Janet~K. Harden}.} \bibinfo{year}{2015}\natexlab{}.
\newblock \showarticletitle{Symptom {Burden} and {Quality} of {Life} in
  {Survivorship}: {A} {Review} of the {Literature}}.
\newblock \bibinfo{journal}{\emph{Cancer Nursing}} \bibinfo{volume}{38},
  \bibinfo{number}{1} (\bibinfo{date}{Feb.} \bibinfo{year}{2015}),
  \bibinfo{pages}{E29}.
\newblock
\showISSN{0162-220X}
\urldef\tempurl%
\url{https://doi.org/10.1097/NCC.0000000000000135}
\showDOI{\tempurl}


\bibitem[\protect\citeauthoryear{Wyche and Grinter}{Wyche and Grinter}{2009}]%
        {wyche_extraordinary_2009}
\bibfield{author}{\bibinfo{person}{Susan~P. Wyche} {and}
  \bibinfo{person}{Rebecca~E. Grinter}.} \bibinfo{year}{2009}\natexlab{}.
\newblock \showarticletitle{Extraordinary {Computing}: {Religion} {As} a {Lens}
  for {Reconsidering} the {Home}}. In \bibinfo{booktitle}{\emph{Proceedings of
  the {SIGCHI} {Conference} on {Human} {Factors} in {Computing} {Systems}}}
  \emph{(\bibinfo{series}{{CHI} '09})}. \bibinfo{publisher}{ACM},
  \bibinfo{address}{New York, NY, USA}, \bibinfo{pages}{749--758}.
\newblock
\showISBNx{978-1-60558-246-7}


\bibitem[\protect\citeauthoryear{Wyche, Hayes, Harvel, and Grinter}{Wyche
  et~al\mbox{.}}{2006}]%
        {wyche_technology_2006}
\bibfield{author}{\bibinfo{person}{Susan~P. Wyche}, \bibinfo{person}{Gillian~R.
  Hayes}, \bibinfo{person}{Lonnie~D. Harvel}, {and} \bibinfo{person}{Rebecca~E.
  Grinter}.} \bibinfo{year}{2006}\natexlab{}.
\newblock \showarticletitle{Technology in {Spiritual} {Formation}: {An}
  {Exploratory} {Study} of {Computer} {Mediated} {Religious} {Communications}}.
  In \bibinfo{booktitle}{\emph{Proceedings of the 2006 20th {Anniversary}
  {Conference} on {Computer} {Supported} {Cooperative} {Work}}}
  \emph{(\bibinfo{series}{{CSCW} '06})}. \bibinfo{publisher}{ACM},
  \bibinfo{address}{New York, NY, USA}.
\newblock
\showISBNx{978-1-59593-249-5}


\bibitem[\protect\citeauthoryear{Xing, Guo, Bai, Qian, and Chen}{Xing
  et~al\mbox{.}}{2018}]%
        {xing_are_2018}
\bibfield{author}{\bibinfo{person}{Lu Xing}, \bibinfo{person}{Xiujing Guo},
  \bibinfo{person}{Lu Bai}, \bibinfo{person}{Jiahui Qian}, {and}
  \bibinfo{person}{Jing Chen}.} \bibinfo{year}{2018}\natexlab{}.
\newblock \showarticletitle{Are spiritual interventions beneficial to patients
  with cancer?}
\newblock \bibinfo{journal}{\emph{Medicine}} \bibinfo{volume}{97},
  \bibinfo{number}{35} (\bibinfo{date}{Aug.} \bibinfo{year}{2018}).
\newblock
\showISSN{0025-7974}
\urldef\tempurl%
\url{https://doi.org/10.1097/MD.0000000000011948}
\showDOI{\tempurl}


\bibitem[\protect\citeauthoryear{Yang, Kraut, and Levine}{Yang
  et~al\mbox{.}}{2017}]%
        {yang_commitment_2017}
\bibfield{author}{\bibinfo{person}{Diyi Yang}, \bibinfo{person}{Robert Kraut},
  {and} \bibinfo{person}{John~M. Levine}.} \bibinfo{year}{2017}\natexlab{}.
\newblock \showarticletitle{Commitment of {Newcomers} and {Old}-timers to
  {Online} {Health} {Support} {Communities}}. In
  \bibinfo{booktitle}{\emph{Proceedings of the 2017 {CHI} {Conference} on
  {Human} {Factors} in {Computing} {Systems}}} \emph{(\bibinfo{series}{{CHI}
  '17})}. \bibinfo{publisher}{ACM}, \bibinfo{address}{New York, NY, USA},
  \bibinfo{pages}{6363--6375}.
\newblock
\showISBNx{978-1-4503-4655-9}
\urldef\tempurl%
\url{https://doi.org/10.1145/3025453.3026008}
\showDOI{\tempurl}


\bibitem[\protect\citeauthoryear{Zafar, Peppercorn, Schrag, Taylor, Goetzinger,
  Zhong, and Abernethy}{Zafar et~al\mbox{.}}{2013}]%
        {zafar_financial_2013}
\bibfield{author}{\bibinfo{person}{S.~Yousuf Zafar},
  \bibinfo{person}{Jeffrey~M. Peppercorn}, \bibinfo{person}{Deborah Schrag},
  \bibinfo{person}{Donald~H. Taylor}, \bibinfo{person}{Amy~M. Goetzinger},
  \bibinfo{person}{Xiaoyin Zhong}, {and} \bibinfo{person}{Amy~P. Abernethy}.}
  \bibinfo{year}{2013}\natexlab{}.
\newblock \showarticletitle{The {Financial} {Toxicity} of {Cancer} {Treatment}:
  {A} {Pilot} {Study} {Assessing} {Out}-of-{Pocket} {Expenses} and the
  {Insured} {Cancer} {Patient}'s {Experience}}.
\newblock \bibinfo{journal}{\emph{The Oncologist}} \bibinfo{volume}{18},
  \bibinfo{number}{4} (\bibinfo{date}{April} \bibinfo{year}{2013}),
  \bibinfo{pages}{381--390}.
\newblock
\showISSN{1083-7159, 1549-490X}


\end{thebibliography}
